\documentclass[useAMS,usenatbib]{mn2e}
\usepackage{graphicx}
\usepackage{amsmath}

\newcommand{\degree}{$^{\circ}$}
\def\vhel{\ifmmode{V_{{\rm HEL}}}\else{$V_{{\rm HEL}}$}\fi}
\def\vsys{\ifmmode{V_{\rm sys}}\else{$V_{\rm sys}$}\fi}
\def\kms{\ifmmode{~{\rm km\,s}^{-1}}\else{~km~s$^{-1}$}\fi}
\def\vlsr{\ifmmode{v_{\rm lsr}}\else{$v_{\rm lsr}$}\fi}

\title[Optical multi--band polarimetry of post-AGB stars]
{Multi--band polarimetry of post-asymptotic giant branch stars stars ~- I. Optical measurements.} 

\author[Akras et al.] 
{S. Akras$^{1,2}\thanks{e-mail:stavrosakras@on.br}$, J. C. Ram\'irez V\'elez$^{3}$, N. Nanouris$^{4}$, G. Ramos--Larios$^{5}$, J. M. L\'opez$^{6}$,
\newauthor
D. Hiriart$^{6}$, D. Panoglou$^{7}$\\
$^{1}$Observat\'orio Nacional/MCTIC, Rua Gen. Jos\'{e} Cristino, 77, 20921-400, Rio de Janeiro, Brazil\\
$^{2}$Observat\'orio do Valongo, Universidade Federal do Rio de Janeiro, Ladeira Pedro Antonio 43, 20080-090, Rio de Janeiro, Brazil\\
$^{3}$Instituto de Astronom\'ia, Universidad Nacional Aut\'onoma de M\'exico, Apartado Postal 70-264, M\'exico, D.F. 04510, Mexico\\
$^{4}$IAASARS, National Observatory of Athens, I. Metaxa \& V. Pavlou, Penteli, GR--15236, Athens, Greece\\
$^{5}$Instituto de Astronom\'ia y Meteorolog\'ia, Av. Vallarta No. 2602, Col. Arcos Vallarta, CP, 44130, Guadalajara, Jalisco, Mexico\\
$^{6}$Instituto de Astronom\'ia, Universidad Nacional Aut\'onoma de M\'exico, Ensenada 22800, Baja California, Mexico\\
$^{7}$Instituto de Astronomia, Geof\'{ı}sica e Ci\^{e}ncias Atmosf\'ericas, Universidade de S\~ao Paulo, SP 05508-900, Brazil}
 
\begin{document}  

\date{Received **insert**; Accepted **insert**} 

\pagerange{\pageref{firstpage}--\pageref{lastpage}}

\maketitle
\label{firstpage}

\begin{abstract}

We present new optical broad--band ({\it UBVRI}) aperture polarimetric observations of 53 post-asymptotic giant branch (AGB) stars selected to exhibit a large near--infrared excess. 24 out of the 53 stars (45\% of our sample) are presented for the first time. A statistical analysis shows four distinctive groups of polarized post-AGB stars: unpolarized or very lowly polarized  (degree of polarization or DoP$<$1\%), lowly polarized (1\% $<$ DoP$ <$ 4\%), moderately polarized (4\% $<$ DoP $<$ 8\%) and highly polarized  (DoP $>$ 8\%). 23 out of the 53 (66\%) belong to the first group,  10 (19\%) to the second, five (9\%) to the third and only three (6\%) to the last group. 
Approximately, 34\% of our sample was found to be unpolarized objects, which is close to the percentage of round planetary nebulae. On average, the low and moderate groups show a wavelength-dependent polarization that increases towards shorter wavelength, implying an intrinsic origin of the polarization, which signifies a Rayleigh-like scattering spectrum typical for non-symmetrical envelopes composed principally of small dust grains. The moderately polarized stars exhibit higher {\it K-W3} and {\it W1-W3} colour indices compared with the group of lowly polarized stars suggesting a possible relation between DoP and mass-loss rate. Moreover, they are found to be systematically colder (redder in {\it B-V}), which may be associated with the condensation process close to these stars that results in higher degree of polarization. We also provide evidence that multiple scattering in optically thin polar outflows is the mechanism that gives high DoP in post-AGB stars with a bipolar or multi-polar envelopes. 

\noindent
\end{abstract}


\section{Introduction}

One of the most challenging problems in the field of post-asymptotic giant branch (AGB) stars and Planetary Nebulae (PNe), which still continues to intrigue and perplex astronomers, is the mechanism responsible for the formation of complex morphologies. It is known from optical and near--infrared (IR) imaging surveys that the majority of post-AGB stars and PNe show an aspherical geometry (elliptical, bipolar, multi--polar; Manchado et al. 1996, 2011; Parker et al. 2006; Boumis et al. 2003, 2006; Sahai et al. 2011; Sabin et al. 2014).

Post-AGB stars represent the phase of stellar evolution where the circumstellar envelope (CSE) has not expanded enough to become optically thin and, thus, observations in the optical are impeded. These CSEs are very thick due to the large amount of mass ejected during the AGB phase (10$^{-4}$--10$^{-5}$M$_{\odot}$ yr$^{-1}$; Bujarrabal et al. 2001). The central stars evolve towards higher effective temperatures, reappearing again in the optical range as a consequence of the dilution of the CSE on a time-scale that is mainly dependent on the initial mass of the progenitor star (Bl\"{o}cker 1995). However, the rapid evolution of the most massive post-AGB stars does not provide enough time for the CSE to expand and become optically thin. Evidence of this evolutionary path is provided by young PNe whose very thick envelopes result in extremely high internal extinctions that prevent their detection in the optical range (Ramos-Larios et al. 2012).

The PN phase, on the other hand, is the consequent result of the interaction between stellar winds from low--intermediate mass stars (1--8 $\rm{M_{\odot}}$) during the AGB and post-AGB phases. The star is hot and luminous enough to ionize the gas, resulting in the formation of a colourful nebula. This complex structure around the central star undergoes spectacular changes in the structure and geometry during the transition from the AGB to the post-AGB, and, finally, to the PNe phase. 

Optical and IR imaging surveys of post-AGB stars have also revealed very complex CSEs, with a wide variety of shapes and forms (e.g.~Trammell et al. 1994; Meixner et al. 1999; Ueta et al. 2000; Gledhill et al. 2001; Gledhill 2005; Parthasarathy et al. 2005; Lagadec et al. 2011).  It is suggested that the spherical symmetry of the AGB stellar wind breaks somewhere between the very late AGB mass--loss phase (Trammell et al. 1994) and the early post-AGB phase (Gledhill et al. 2001). However, the exact mechanism responsible for this deviation is still poorly known (see e.g. Balick \& Frank 2002 and reference therein).

The presence of an equatorial density enhancement, such as a disk or torus around these stars, has been  credibly proposed by several authors  to be the mechanism responsible for the formation of aspherical PNe (e.g. Livio \& Soker 1988; Soker \& Livio 1994; Soker 2002; Frank \& Blackman 2004 among others). To enable the formation of highly collimated, fast--moving jets or outflows, a large amount of orbiting mass is needed, in accordance to the magnetohydrodynamic model of Frank \& Blackman (2004).  This mass can be provided through the mass--exchange interaction  of close binary systems (e.g. Soker \& Livio 1994; Rasio \& Livio 1996; see also the review by De Marco 2009). The presence of these dusty circumstellar disks has been confirmed in some proto--PNe (PPNe) with highly collimated outflows such as M~2--9 (Lykou et al. 2011), Mz3 (Chesneau et al. 2007), Red Rectangle and 89~Her (Bujarrabal et al. 2005, 2007).

The presence of  such a large amount of dust in CSEs results in: (i) a large near--IR excess in the spectral energy distribution (SED) (e.g. De Ruyter et al. 2006), and (ii) the production of significant amount of linear polarization due to stellar light scattering by large ($>$1~$\mu$m) or small ($\sim$0.2~$\mu$m) dust grains, and free electrons. The grains size may be responsible for the different wavelength dependences of polarization (Gledhill \& Yates 2003). CSEs  have been found to be richer in small grains.
 
Optical and near--IR polarization imaging along with spectro--polarimetric observations have become very powerful techniques for investigating dust properties (i.e. distribution and composition), as well as the geometry of CSEs around AGB and post-AGB stars (e.g. Johnson \& Jones 1991; Parthasarathy et al. 2005; Ueta et al. 2000, 2005, 2007; Meixner et al. 2000; Gledhill et al. 2001; Gledhill 2005). 

Studying the polarization characteristics of several stars from the red giant to the PNe phase, Johnson \& Jones (1991)  concluded that the degree of polarization increases through the transition from the AGB to the post-AGB phase, and decreases again at the PNe phase. This result implies that a spherically symmetric AGB wind  may break somewhere between the late AGB mass--loss and early post-AGB phases, whereas the dust formation and mass--loss  rates are relatively high (Soker 2000). Moreover, the expansion of CSEs over time leads to a gradually decrease of dust density and results in a lower degree of polarization. L\'opez \& Hiriart (2011a \& 2011b) reported a linear relation between the degree of polarization and the mass--loss rate in evolved carbon stars.

Based on polarimetric images of post-AGB and PPNe with NICMOS on the {\it Hubble Space Telescope (HST)}, Gledhill et al. (2001) claimed that the degree of linear polarization is strongly related to the light scattered from disk--like structures, oriented at different inclination angles. Later, Ueta et al. (2005, 2007) proposed that besides the effect of orientation, the optical depth is also a crucial parameter that reflects the polarization characteristics of DUPLEX (dust--prominent longitudinally extended; elliptical PPNe) and SOLE (star--obvious low--level elongated; toroidal PPNe) nebulae, in agreement with the theoretical predictions of radiative transfer models (Ueta \& Meixner 2003).

Brown et al. (1978) derived a general expression for the linear polarization in optically thin CSEs around binary systems, providing a method that allows us to determine the geometric characteristics of these CSEs from the observed Stokes parameters. Posterior investigation of the effect of multi-scattering polarization in axisymmetric geometries, such as equatorial disks, ellipsoidal envelopes and polar jets, has shown that multi--scattering results in a higher degree of polarization than single scattering in CSEs with high optical depths (Wood et al. 1996a,b). Finally, a more comprehensive study on the effect of multi--scattering in circumstellar disks around binary stars was performed by Hoffman et al. (2003).

In this paper, we present new optical broad--band and multi--band ({\it U, V, B, R} and {\it I}) linear aperture polarimetric measurements of a sample of 53 post-AGB stars, out of which 24 (or 45\%) are presented for the first time. The remaining 29 sources (55\%) have been already observed either in the optical or near-IR wavelengths. Based on the degree of polarization in the {\it V}-band, we classify the programme stars from unpolarized to highly polarized. This study follows the same technique conducted by Bieging et al. (2006) and Lopez \& Hiriart (2011a) to identify possible correlations between the degree of polarization with near-IR colour indices.

In Section 2, we present the selection criteria of our sample, while the observations and data analysis are described in Section 3. Our results and distinct comments on each programme star are presented in Section 4. We discuss our results in Section 5 and we wrap up with the conclusions in Section 6.  It is noteworthy that the current paper is the first out of a sequence focusing on the optical polarimetric data. Furthermore, new near--IR ({\it J, H, K}) polarimetric observations will also be obtained and presented in a forthcoming paper.

\section{Sample Selection}

A sample of 53 post-AGB stars was selected from De Ruyter et al. (2006) and the Toru\'n catalogue of Galactic post-AGB and relative objects (Szczerba et al. 2007, 2012) by applying the following criteria to all the programme stars: (i) they exhibit a near--IR excess in their SED profiles, which may be associated with the thermal emission of hot dust grains (Evans 1985) and provides a signature of a dusty circumstellar disk/torus (De Ruyter et al. 2006); and ii) they are bright enough at optical wavelengths to perform accurate linear polarimetry from the 0.84~m telescope at the Observatorio Astron\'omico Nacional in Sierra de San Pedro M\'artir, Mexico. In Table \ref{table1}, we list for the stars of our sample, their coordinates, their classification type and information about their binarity.

Our sample consists of 29 post-AGB or likely post-AGB stars (three PPNe: Red Rectangle, Frosty Leo and Minkowski Footprint), 23 RV Tauri stars and one yellow hypergiant star (YHG). In particular, the RV Tauri stars define a subgroup of pulsating post-AGB stars whose IRAS fluxes are located in a specific area of the IRAS  colour--colour diagram, known as the RV Tauri box (1.0$<[12]$--$[25]<$1.55 \& 0.2 $< [25]$--$[60] <$ 1.0; Evans 1985). Among the 53 objects, seven are known binary systems with a range of orbital periods between 100 and 1000~d (see De Ruyter et al. 2006 for more details), 15 are candidate binary systems, and 6 are confirmed single stars.

\begin{table*}
\centering
\caption{Log file of the observations. The IRAS number, the  name, the Galactic coordinates {\it l} and {\it b},  the equatorial coordinates RA and Dec, the type, a note of binarity and the reference from our sample are tabulated below in the same order.} 
\label{table1}
\begin{tabular}{cccccccccccccccc}
\hline
IRAS & Name & lll.l $\pm$ bb.b & RA(J2000) & Dec(J2000) & Type & Binarity & Reference \\ 
\hline

Z02229+6208 & $-$ & 133.7 +01.5 & 02 26 41.7 & +62 21 22 & post-AGB & $-$ & $-$\\
04166+5719  & TW Cam & 148.3 +05.3 & 04 20 47.6 & +57 26 28.5 & RV Tauri & candidate & 1,6\\  
04296+3429  & GLMP 74 & 166.2 -09.1 & 04 32 56.9 & +34 36 12.4 & post-AGB & no & 12\\ 
04440+2605  & RV Tau & 174.8 -12.2 & 04 47 06.7 & +26 10 45.6 & RV Tauri & candidate & 1 \\  
05040+4820  & LS V +48 26 & 159.8 +04.8 & 05 07 50.3 & +48 24 09.4 & post-AGB & $-$ & $-$\\
05113+1347  & GLMP 88 & 188.9 -14.3 & 05 14 07.7 & +13 50 28.3 & post-AGB & $-$ & $-$\\
05208-2035  & BD -20 1073 & 222.8 -28.3 & 05 22 59.4 & -20 32 53.0 & post-AGB & candidate & 1,6 \\
05280+3817  & V428 Aur & 170.5 +02.5 & 05 31 26.7 & +38 19 10.5 & RV Tauri & $-$ & $-$\\
05381+1012  & GLMP 177 & 195.5 -10.6 & 05 40 57.0 & +10 14 25 & post-AGB & $-$ & $-$\\
06072+0953  & CT Ori & 199.4 -04.6 & 06 09 57.9 & +09 52 31.8 & RV Tauri & candidate & 1\\
06108+2743  & SU Gem & 184.2 +04.8 & 06 14 00.8 & +27 42 12.2 & RV Tauri & candidate & 1,6\\
06160-1701  & UY Cma & 224.8 -14.9 & 06 18 16.4 & -17 02 34.7 & RV Tauri & candidate & 1\\
06176-1036  & Rec Rectangle & 218.9 -11.8 & 06 19 58 & -10 38 15 & post-AGB & yes & 2\\
06338+5333  & V382 Aur & 161.9 +19.6 & 06 37 52.4 & +53 31 02 & post-AGB & yes & 3\\  
07008+1050  & PS Gem & 204.7 +07.6 & 07 03 39.6 & +10 46 13 & post-AGB & yes & 6 \\
07134+1005  & LS V +10 15 & 206.7 +10.0 & 07 16 10.3 & +09 59 47.9 & post-AGB & no & 4,12\\
07331+0021  & AL Cmi & 217.8 +09.9 & 07 35 41.2 & +00 14 58 & RV Tauri & $-$ & $-$\\ 
07430+1115  & GLMP 192 & 208.9 +17.1 & 07 45 51.4 & +11 08 19.6 & post-AGB & $-$ & $-$\\  
08187-1905  & V552 Pup & 240.6 +09.8 & 08 20 57.1 & -19 15 03.4 & post-AGB & $-$ & $-$\\
09371+1212  & Frosty Leo & 221.9 +42.7 & 09 39 53.9 & 11 58 52.3 & post-AGB & $-$ & $-$\\
11157+5257  & DZ Uma & 150.7 +59.1 & 11 18 33.6 & +52 40 54.6 & RV Tauri & $-$ & $-$\\
11472-0800  & AF Crt & 277.9 +51.6 & 11 49 48 & -08 17 20.5 & Rv Tauri & candidate & 5 \\
13467-0141  & CE Vir & 330.8 +57.8 & 13 49 17.1 & -01 55 44.9 & RV Tauri & $-$ & $-$ \\
17436+5003  & V814 Her & 077.1 +30.9 & 17 44 55.5 & +50 02 39.5 & post-AGB & no & 4\\ 
F17495+0757 & TX Oph & 033.3 +16.8 & 17 52 01.1 & +07 56 29.3 & RV Tauri & $-$ & $-$\\
17534+2603  & 89 Her & 051.4 +23.2 & 17 55 25.2 & +26 02 59.9 & post-AGB & yes & 7 \\
18095+2704  & V887 Her & 053.8 +20.2 & 18 11 30.6 & +27 05 15.6 & post-AGB & $-$ & $-$\\
18123+0511  & $-$ & 033.4 +10.5 & 18 14 49.4 & +05 12 55.7 & RV Tauri & candidate & 1\\
18281+2149  & AC Her & 050.5 +14.2 & 18 30 16.2 & 21 52 00.6 & RV Tauri & yes & 10 \\ 
18564-0814  & AD Aql & 026.5 -05.4 & 18 59 08.7 & -08 10 14.1 & RV Tauri & candidate & 1\\
19090+3829  & EG Lyr & 069.8 +13.0 & 19 10 48.7 & +38 34 25 & RV Tauri & $-$ & $-$\\
19114+0002  & V1427 Aql & 035.6 -04.9 & 19 13 58.6 & +00 07 31.9 & YHG & $-$ & $-$ \\ 
19125+0343  & $-$ & 039.0 -03.5 & 19 15 01.2 & +03 48 42.7 & RV Tauri & candidate & 1,6\\  
19163+2745  & EP Lyr & 060.7 +06.9 & 19 18 19.5 & +27 51 03.2 & RV Tauri & candidate & 1,6\\
19199+3950  & HP Lyr & 072.0 +11.7 & 19 21 39.1 & +39 56 08.0 & RV Tauri & yes & 8 \\
19343+2926  & Min Footprint & 064.1 +04.3 & 19 36 18.9 & +29 32 50 & post-AGB & yes & 9 \\
19386+0155  & V1648 Aql & 040.5 -10.1 & 19 41 08.3 & +02 02 31.2 & post-AGB & $-$ & $-$\\ 
19475+3119  & LS II +31 9 & 067.2 +02.7 & 19 49 29.6 & +31 27 16 & post-AGB & no & 5\\ 
19486+1350  & TW Aql & 052.2 -06.4 & 19 51 00.9 & +13 58 15.6 & RV Tauri & $-$ & $-$ \\
19500-1709  & V5112 Sqr & 023.9 -21.0 & 19 52 52.7 & -17 01 50.3 & post-AGB & no & 5,12\\
20000+3239  & GLMP 963 & 069.8 +01.2 & 20 01 52.5 & +32 47 32.9 & post-AGB & $-$ & $-$\\
20004+2955  & V1027 Cyg & 067.4 -00.4 & 20 02 27.4 & +30 04 25.5 & post-AGB & $-$ & $-$\\
20056+1834  & QY Sqe & 058.4 -07.5 & 20 07 54.6 & +18 42 54.5 & RV Tauri & candidate & 1,6 \\
20117+1634  & R Sge & 057.5 -09.8 & 20 14 03.7 & +16 43 35.1 & RV Tauri & candidate & 1 \\
20160+2734  & AU Vul & 067.3 -04.5 & 20 18 05.9 & +27 44 04 & post-AGB & $-$ & $-$\\
20343+2625  & V Vul & 068.8 -08.5 & 20 36 32.0 & +26 36 14.5 & RV Tauri & candidate & 1 \\
21546+4721  & GLMP 1047 & 095.0 -05.6 & 21 56 32.9 & +47 36 12.8 & post-AGB & $-$ & $-$\\
22023+5249  & LS III +52 24 & 099.3 -02.0 & 22 04 12.3 & +53 04 01.4 & post-AGB & $-$ & $-$ \\
22223+4327  & BD +42 4388 & 096.8 -11.6 & 22 24 31.4 & +43 43 10.9 & post-AGB & no & 5,12\\ 
22223+5556  & TX Oph & 103.4 -01.0 & 22 24 13.7 & +56 11 33.3 & RV Tauri & $-$ & $-$\\
22272+5435  & V354 Lac & 103.3 -02.5 & 22 29 10.4 & +54 51 06 & post-AGB & candidate & 5 \\
22327-1731  & HD 213985 & 043.2 -57.1 & 22 35 27.5 & -17 15 26.9 & post-AGB & yes & 2,6\\  
$-$         & BD +39\degree 4926 & 098.4 -16.7 & 22 46 11.2 & +40 06 29.3 & post-AGB & yes & 11 \\    

\hline
\end{tabular}                                                                                                                            
\medskip{}
\begin{flushleft}      
References: (1) De Ruyter et al. (2006), (2) Van Winckel et al. (1995), (3) Hrivnak et al. (2008), 
(4) Hrivnak et al. (2011), (5) van Winckel et al. (2012), (6) Gielen et al. (2008), (7) Waters et al. (1993), (8) Kreiner (2004), (9) Castro--Carrizo et al. (2012), (10) Van Winckel et al. (1998), (11) Kodaira et al. (1970), (12) Reyniers (2002)
\end{flushleft}
\end{table*}

\section{Observations}

\subsection{Reduction process}

Linear polarization measurements were performed with the 0.84~m, f/15, telescope of the Observatorio 
Astron\'omico Nacional in the Sierra de San Pedro M\'artir, Mexico, using the POLIMA polarimeter (Hiriart et al. 2005) in conjunction with a 2048 $\times$ 2048 CCD camera (13$\times$13~$\mu$m pixels), which results in a scale of 0.44~arcsec~$\rm{pixel^{-1}}$. The polarimeter consists of a rotating Glan--Taylor prism driven by a stepper motor. As the polarimeter is a single--beam device with a very slow modulation, good photometric conditions are required for accurate polarimetry. The observations were performed between 2012 January and December at/or near the new Moon phase. The broad-band filters {\it U, B, V, R} and {\it I} were used with exposure times from 5 to 180~s.   

\begin{figure*}
\begin{center}
\includegraphics[width=18cm]{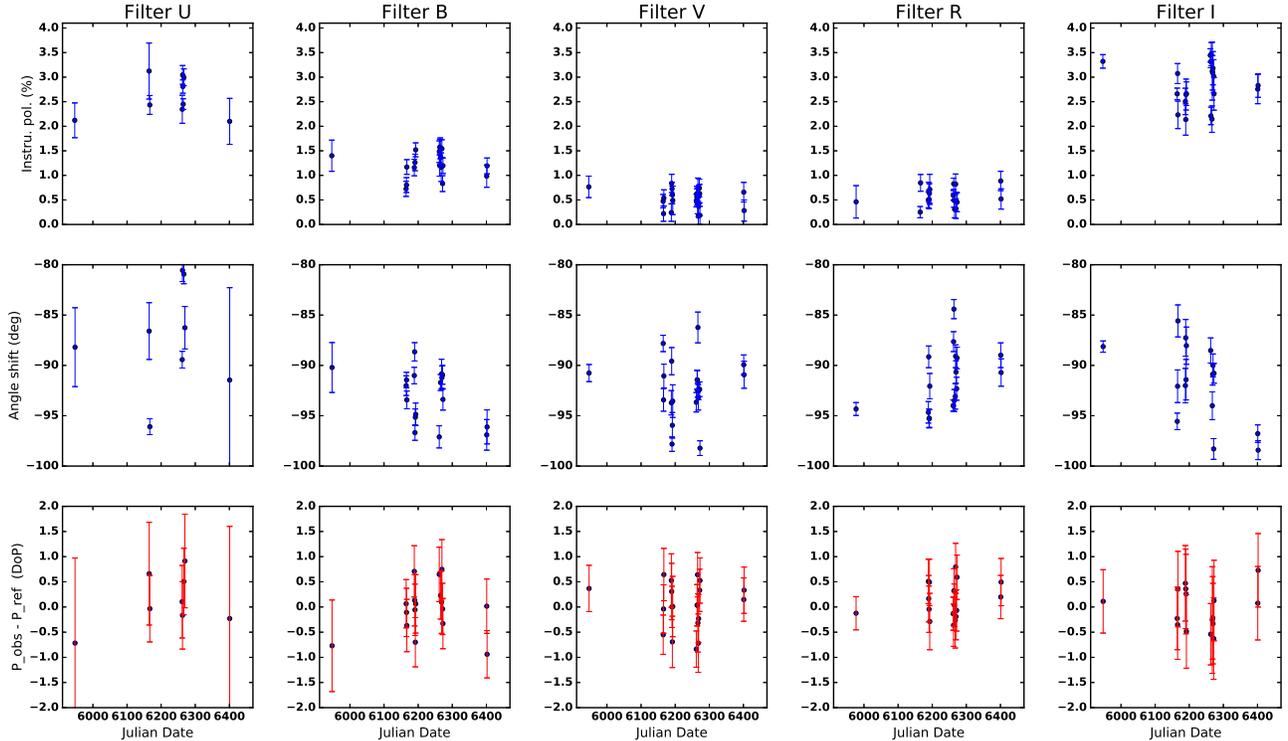}
 \caption{ The upper panels correspond to the observations of the unpolarized stars, i.e. the instrumental polarization in each band, while  the middle panels correspond to the angle shift of the polarimeter with respect to north. In the lower panels, we show the difference between our polarization measurements and the values reported by other authors for the standard polarized stars. The error bars in the upper two panels represent only the obervational error (in blue), while in the lower panels the error bars (in red) represent the observational plus the instrumental errors (see Table \ref{table1}). In the $\chi$-axis, the observational dates are given in JD-2450000.}
\label{fig_2}
\end{center}
\end{figure*}

Separate images were obtained for all bands and at relative position angles of the prism of 0\degree, 90\degree\ and 45\degree, 135\degree, in order to reduce the time between the two measurements when evaluating each Stokes parameter. The instrumental polarization as well as the rotation angle of the polarimeter were estimated by observing a number of unpolarized and polarized standard stars (Schmidt et al. 1992) during our campaigns. In Fig. \ref{fig_2} we show the observations of the polarized and unpolarized standard stars. Five unpolarized and six polarized standard stars were observed. The total number of observations in all bands for the standars stars, including both polarized and unpolarized ones, is 168. In Table \ref{table_polima}, we present the instrumental degree of polarization and rotation angle for each filter. Our values in the {\it V} filter are in very good agreement with the values presented by Lopez \& Hiriart (2011a, 2011b). Despite the relatively high instrumental polarization in the {\it U} and {\it I} filters, the relative difference between our measurements and those from the literature of the polarized standard stars is not higher than 8\%.

The data reduction was performed with the POLIMA pipeline developed by Hiriart et al. (2005). The code applies a standard reduction technique which includes removal of cosmic rays, subtraction of the dark current and bias, as well as flat--field normalization. Multiple flat--field frames were taken at each of the four positions of the polarizer and for all the filters at dusk and dawn. The flat--field frames were combined to form a normalized frame for each filter and prism position. Finally, all individual science frames for each filter and prism position were bias subtracted and flat--fielded.

The normalized Stokes parameters $q$ and $u$ for each object were calculated as 

\begin{equation}
\it{ q = \frac{F(0^\circ) - F(90^\circ)}{F(0^\circ) + F(90^\circ)}} \; ,
\end{equation}
and
\begin{equation}
\it{ u = \frac{F(45^\circ) - F(135^\circ)}{F(45^\circ) + F(135^\circ)}} \; ,
\end{equation}

where {\it F($\phi$)} is the flux at polarizer position $\phi$. The $q$ and $u$ Stokes parameters are related to the total linear polarization $p_L$ and the PA of polarization $\theta$ by

\begin{equation}
\it{p_L=\sqrt{q^2 + u^2}}  \; ,
\label{eq_P}
\end{equation}
and
\begin{equation}
\it{\theta = \frac{1}{2} \arctan \frac{u}{q}}  \; .
\end{equation}

Since we did not perform measurements of the circular polarization, we will hereafter 
refer to the total linear polarization, given by equation~(3), as the degree of polarization (DoP).

Given that POLIMA is a single--beam polarimeter, sky variations during the exposures may induce false polarized signals. In general, the total flux 
$F$ is equal to either $F_1$=$F$(0\degree)+$F$(90\degree) and $F_2$=$F$(45\degree)+$F$(135\degree) (e.g. Shahzamanian et al. 2015). Hence, the difference between the two (parallel and perpendicular) fluxes $F_1$ and $F_2$ indicates the sky stability during the observations. This method allows us to reject measurements with a high flux difference.

\begin{figure}
\begin{center}
  \includegraphics[width=9cm]{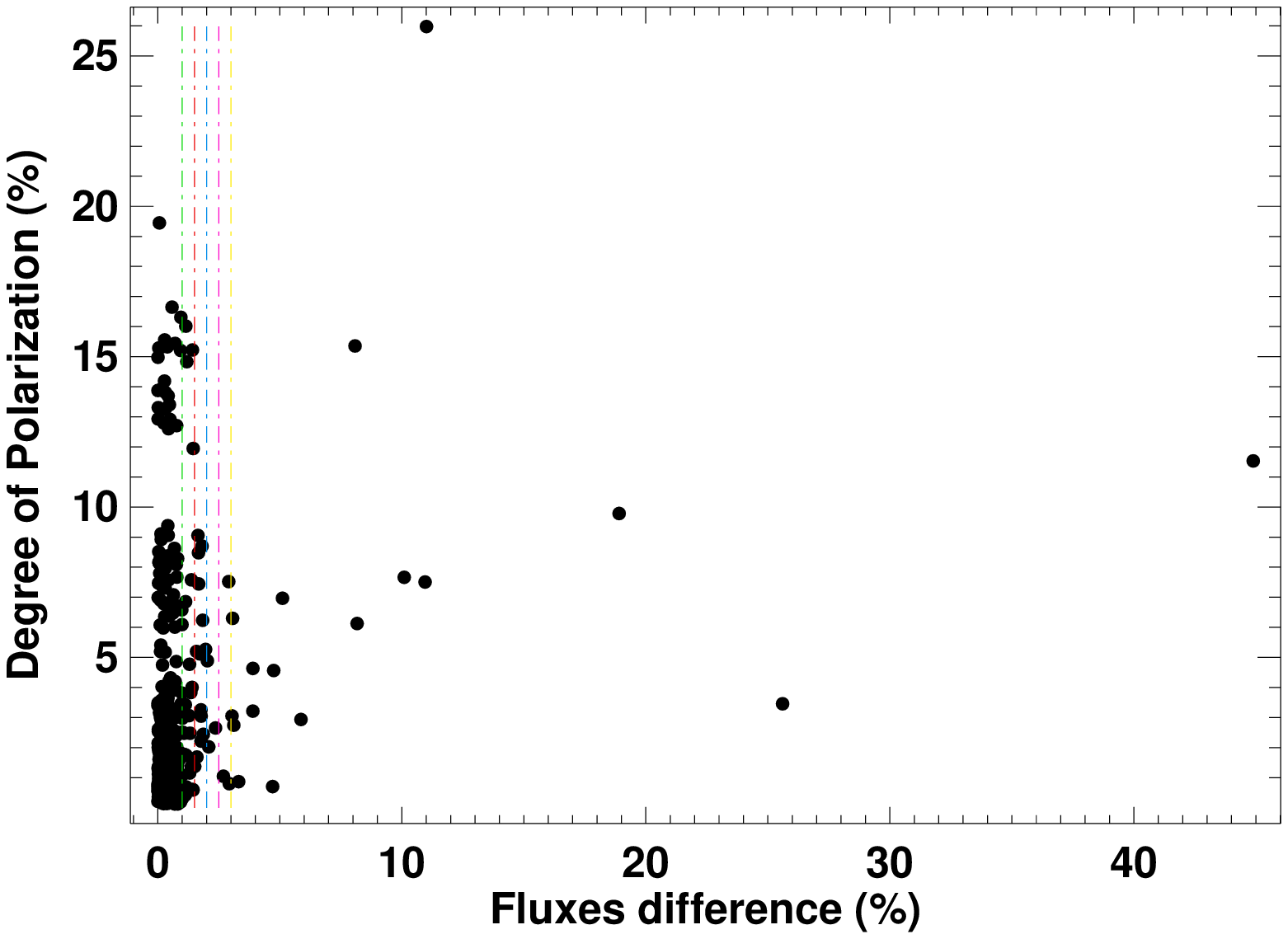}
 \includegraphics[width=9cm]{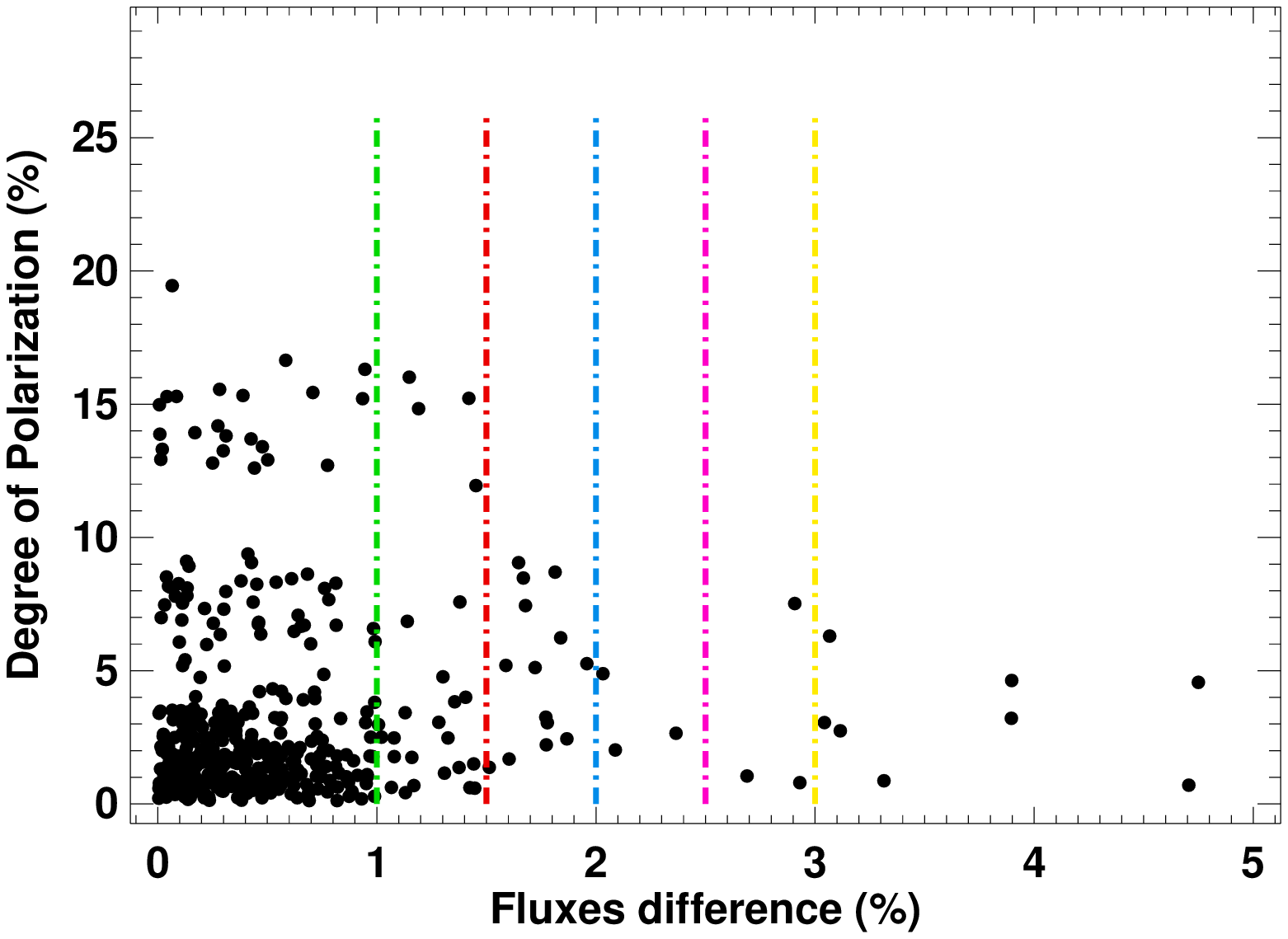}
 \caption{Flux difference between the parallel and perpendicular polarizer angles versus DoP. Upper panel: DOP values for all  available measurements (447). Lower panel: DoP values with flux difference less than 5\%. The vertical lines indicate flux differences from 1\% to 3\% in steps of 0.5\% (see text for details).}
 \label{fig1}
\end{center}
\end{figure}

Fig.~\ref{fig1} shows the DoP as a function of flux difference. The upper panel shows the 447 individual measurements, while the lower panel displays only the measurements with flux difference $<$5\%. The vertical lines indicate flux differences from 1\% to 3\% in steps of 0.5\%. Most of the high 
DoP measurements ($>$10\%) show a flux difference lower than 1.0\% and only four of them exhibit a flux difference between 1.0\% and 1.5\%. Apparently, high DoP values  may be recorded because of a significantly high flux difference due to sky variability and mistakenly be considered as real (Fig.~1, upper panel). Hence, we decide to accept and publish only those measurements with a flux difference lower than 3\%, thus excluding 18 of the total 447 measurements ($\sim$4\% of the whole sample).

\begin{table*}
\centering
\caption{ Instrumental degree of polarization, rotation angle of the polarimeter with respect to north and the relative difference in percentage between our measurements of the polarized standard stars and their values in the literature.}
\label{table_polima}
\begin{tabular}{cccccccccccccccc}
\hline
Filter                & U & B & V & R & I \\ 
\hline
\rm{P$_{\rm{instrumental}}$} ($\%$) & 2.58$\pm$0.14 & 1.24$\pm$0.07 & 0.51$\pm$0.05 & 0.57$\pm$0.05 & 2.81$\pm$0.11 \\ 
Rotation angle (\degree)  & -87.44$\pm$2.2& -93.04$\pm$0.62 & -92.3$\pm$0.67 & -91.58$\pm$0.69 & -91.74$\pm$1.06 \\ 
Rel. dif. ($\%$)    & 8.07$\pm$3.38 & 6.64$\pm$2.21 & 6.8$\pm$1.86 & 5.87$\pm$1.6 & 7.96$\pm$2.15 \\
\hline
\end{tabular}                                                                                                                            
\medskip{}
\begin{flushleft}  
\end{flushleft}
\end{table*}

\subsection{Interstellar polarization}

As the stellar light propagates through the interstellar medium (ISM), it is subject to emission, absorption and scattering mechanisms produced by the magnetically aligned dust grains that further polarize the light at all wavelength bands. In the visible (e.g. Reiz \& Franco 1998; Fosalba et al. 2002) and near-IR regimes (e.g. Jones 1989), the interstellar polarization (ISP) shows a strong dependence on the interstellar extinction. The ISP degree-extinction relation is found to be almost linear in the nearby (d$<$2~kpc) and close to the Galactic plane ($|$b$|<$10\degree) area. As a result, the observed Stokes parameters at a wavelength $\lambda$, $\it{q(\lambda)}$ and $\it{u(\lambda)}$, are a combination of the intrinsic Stokes parameters of the object, $\it{Q_{\star}(\lambda)}$ and $\it{U_{\star}(\lambda)}$, 
and those induced from the interstellar medium, $\it{q_{ISM}}(\lambda)$ and $\it{u_{ISM}(\lambda)}$. Thus, we have:

\begin{equation}
\rm{ q(\lambda) = Q_{\star}(\lambda) + q_{ISM}(\lambda)}
\end{equation}

\begin{equation}  
\rm{ u(\lambda) = U_{\star}(\lambda) + u_{ISM}(\lambda)}
\end{equation}

To disentangle starlight polarization from the ISM contamination, the $\rm{q_{ISM}}(\lambda)$ and $\rm{u_{ISM}(\lambda)}$ values must be estimated along the line of sight of each programme star. To deal with the proper estimation for each of the five available photometric bands, we adopt the empirical law of Serkowski (1971) which models the ISP as a function of wavelength:

\begin{equation}
\rm{ISP(\lambda)=p_{max,ISM}\times \exp{[-K \ln^2{\frac{\lambda_{max}}{\lambda}}]}} \ ; 
\end{equation}

where $\rm {\lambda_{max}}$ is the wavelength at which ISP reaches its maximum value $\rm{p_{max,ISM}}$, while K is a constant parameter that remains constant irrespective of the examined wavelength (e.g. Serkowski 1971; Coyne et al. 1974; Serkowski et al. 1975; Codina-Landaberry \& Magalhaes 1976).

Therefore, the properties of a linear ISP can be fully determined for any wavelength provided the quantities $\rm{\lambda_{max}}$, $\rm{p_{max,ISM}}$ and $\rm{\theta_{ISM}}$ are available, and the ISP vector neither rotates with the wavelength nor varies throughout time (PA may vary if measurements are obtained at different times; Coyne 1974). We set $\rm {\lambda_{max}}$=544~nm as the average wavelength at which the ISP is maximized, while K = 1.15 is adopted as the optimal value  (Serkowski et al. 1975).

In an attempt to estimate the parameters $\rm{p_{max,ISM}}$ and $\rm{\theta_{ISM}}$ towards the programme stars, we apply the star field method using the catalogue of Heiles (2000). The catalogue lists only stellar polarization measurements in the {\it V} band that are considered to be a firm representative of the  maximum ISP. The Stokes parameters $\it{q_j}$ and $\it{u_j}$ are then determined for each neighbour star $j$ located inside a circle of angular radius $\rm{r_c}$ = 5\degree\  centred on our programme star. Extreme values (outliers) are carefully removed from the analysis by performing robust statistics. Likewise, stars with values deviating by at least 4$\sigma$ from the median are excluded from the procedure. A distance-dependent weight $\rm{w_j}$ is then assigned to the selected stars following the suggestion of Bastien (1985):

\begin{equation}
\rm{ w_j = 1 - \frac{r_j}{r_c}} , 
\end{equation}

with $\rm{r_{j}}$ being the angular distance from the programme star to the $\rm{j^{th}}$  catalogue neighbour star. Finally, the distance-weighted average of the Stokes parameters accounting for all neighbours are calculated through the formulae:

\begin{equation}
\rm{\overline{q}_{ISM}=\frac{\sum^{n}_{j=1} q_j\cdot\ w_j}{\sum^{n}_{j=1} w_j}} \,
\end{equation}
and
\begin{equation}
\rm{\overline{u}_{ISM}=\frac{\sum^{n}_{j=1} u_j\cdot\ w_j}{\sum^{n}_{j=1} w_j}} \ ,
\end{equation}

where n is the number of  neighbour stars that do not fulfil our statistical criterion for outlier detection. The ISP and the PA towards each of our programme stars are obtained by employing equations~(3) and (4). The intrinsic Stokes parameters, $\it{Q_{\star}}$ and $\it{U_{\star}}$, are finally  calculated by subtracting the ISP contribution by means of equations~(5) and (6). In six cases (IRAS 06338+5333, IRAS 07134+1005, IRAS 07430+1115, IRAS 09371+1212, IRAS 11472+0800 and IRAS 17436+5003), a circle with a radius of 10\degree\ was selected, since no stars were found within the circle of 5\degree.

\begin{figure}
\begin{center}
  \includegraphics[scale=0.30]{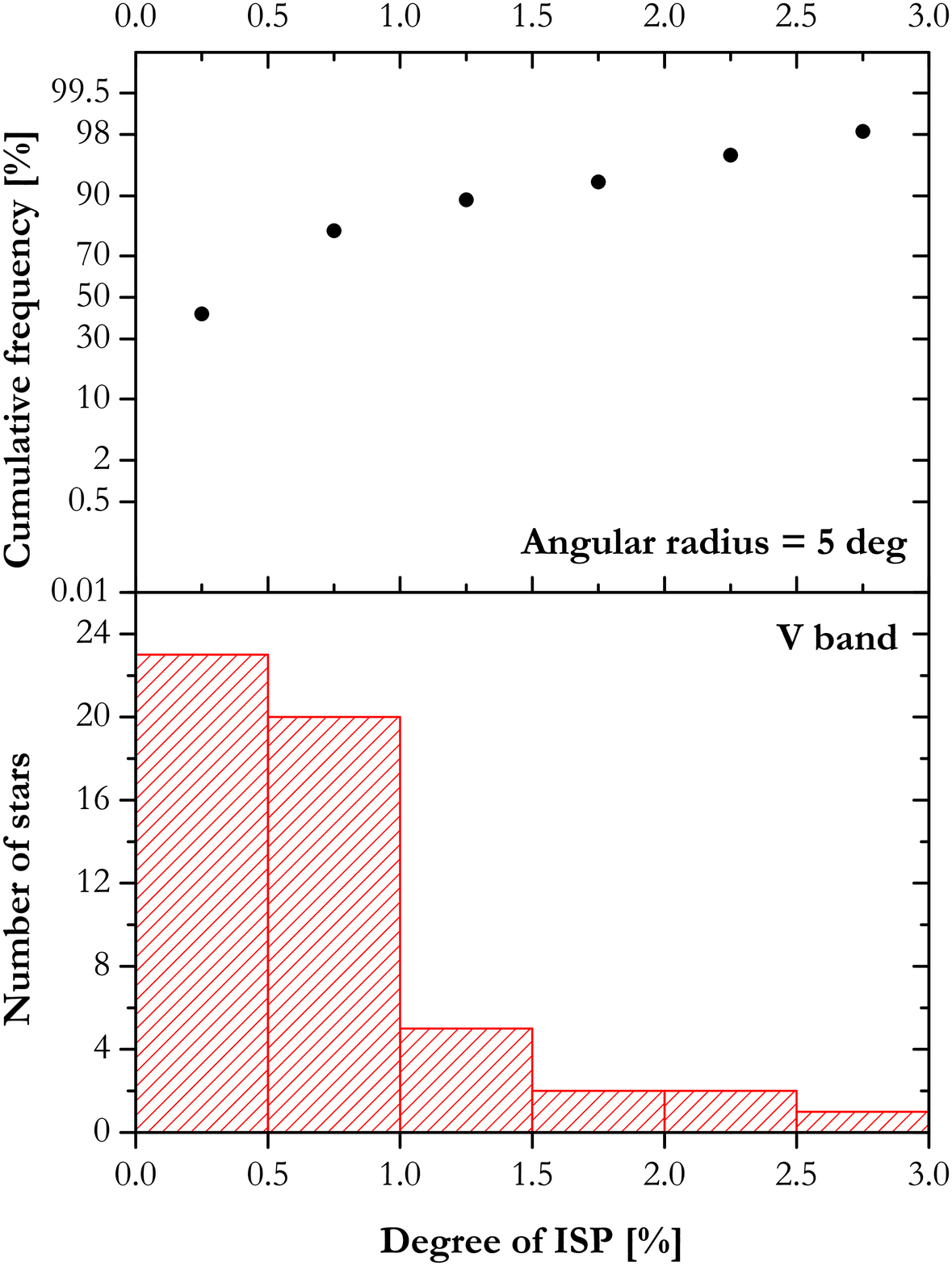}
  \includegraphics[scale=0.05]{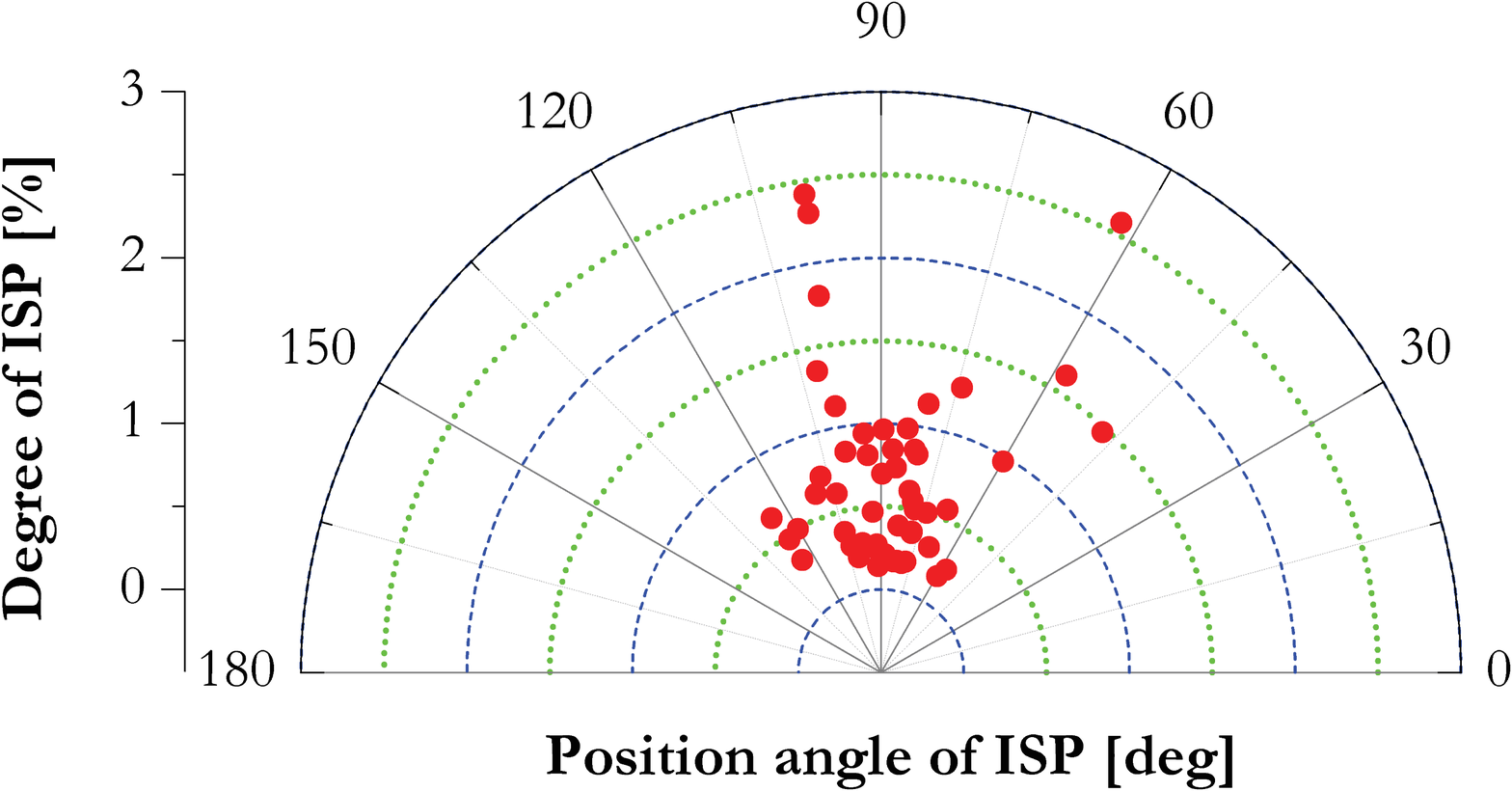}
 \caption{Top/center panels: Distribution of ISP degree toward the programme stars (top: cumulative frequency of stars; centre: number of stars). The size of DoP bins is 0.5\%.  Bottom panel: Distribution of ISP vector parameters in polar coordinates. The blue and green circles correspond to DoP values which differ by 0.5\%.}
\label{fig_IP}
\end{center}
\end{figure}

Once $\rm{p_{max,ISM}}$ and $\rm{\theta_{ISM}}$ are defined, then the ISP values in the remaining {\it U, B, R} and {\it I} bands  are calculated through equation~(7) by setting as effective wavelength the values of 360, 430, 640, and 890~nm, respectively. Note that the same method for estimating the ISP has been used in many other similar works as well (e.g. Yudin et al. 2003; Lopez \& Hiriart 2011a, 2011b; Malatesta et al. 2013).

The upper panel in Fig. \ref{fig_IP} displays the distribution of ISP in the {\it V} band, and reveals that 90\% of the estimated values are less than 1.5\%. Additionally, the lower panel shows the same distribution depicted in a polar diagram with the PA set as the angular coordinate and the radial axis representing the ISP degree. The majority of the estimated PAs are found to be distributed around 90\degree ($\pm$15\degree), as expected from large-scale ISP statistics (see Fosalba et al. 2002). Only  16 measurements are found to be out of this range of PAs, most of which have an ISP degree lower than 0.5\%.  Only 10\% of these measurements exhibit DoP higher than 1\%, indicating a significant contribution of the ISM (upper panel, Fig. \ref{fig_IP}). Moreover, it is worth mentioning that most of the ISP measurements with DoP higher than 1\% are found to exhibit the same PA$\sim$100\degree, and $\sim$60\degree. This indicates that the alignment of dust grains in  those directions is strong, probably due to the Galactic magnetic field.  Conclusively, this analysis provides a boost of confidence that our ISP estimates are reliable within their uncertainty levels.	

\section{Results}
 In the following sections, the results for each object are described and compared to previously published works. The intrinsic DoPs as well as the PAs for each star are given in Table~B1, and the plots of the intrinsic DoP as a function of wavelength ({\it U, B, V, R} and {\it I bands}) are shown in Appendix~A (Fig.~A1). The effective wavelengths for the {\it U, B, V, R} and {\it I bands} filters are 0.36, 0.43, 0.54, 0.64 and 0.89 $\mu$m, respectively.

In the subsequent analysis, we refer to an object as unpolarized or very lowly polarized for DoP$\leq$1\%, lowly polarized for 1\%$<$DoP$\leq$4\%, moderately polarized for 4\%$<$DoP$\leq$8\%, and highly polarized for DoP$>$8\%. Some of the very lowly polarized object cannot be distinguished from the unpolarized ones because of the relatively high uncertainty of DoP due to the interstellar polarization.

It should be noted that polarization measurements ($\rm{DoP=\sqrt(u^2+q^2)}$) are biased at low signal-to-noise ratios (S/N) and further correction is required (see, e.g., Simmons \& Stewart 1985; Quinn 2012 and references therein). The Modified ASyptotic Estimator (MAS; Plaszczynski et al. 2014) is used for estimating the bias-noise for each measurement. The corrected $\rm{DoP_{MAS}}$ values are also given in Table~B1. In  the analysis of each individual object below, the uncorrected DoP for the low S/N bias-noise are used since this correction had not been applied in previous studies.

\subsection{Comments on Individual Stars}

\subsubsection{IRAS Z02229+6208}
 The elliptical nebula in IRAS Z02229+6208 is oriented at PA=30--35\degree\ with a size of 2.1 $\times$ 1.3~arcsec (Ueta et al. 2000, 2007). High--resolution  polarimetric maps, obtained with NICMOS on board the {\it HST},have  revealed high DoP up to 50\%--60\% in the north-western and south-eastern lobes, respectively, whereas it is lower (30\%) along the direction of PA$\sim$150\degree\ (Ueta et al. 2005, 2007). Ground--based mid--IR images have also shown a seemingly spherically symmetric envelope with a diameter of 4~arcsec (Meixner et al. 1999). The shift in PA between the DoP and mid--IR emission maps is attributed to a possible precession motion of the torus (Ueta et al. 2005, 2007) .

Our multi--band polarimetric observations show that the polarization is wavelength-independent within the errors in contrast to the spectro-polarimetric data from Beiging et al. (2006). On average, our DoP$\sim$7\% and PA$\sim$130\degree\ agree with the average values derived by Bieging et al. (2006) and aperture polarimetry (Ueta et al. 2007). The difference between our measurements and those from Bieging et al. (2006) may be the correction of bias-noise applied to our data. Despite the very highly polarized lobes (Ueta et al. 2005, 2007), the net aperture DoP of IRAS Z02229+6208 is significantly lower. The reason for this is twofold: (i) the cancellation of perpendicular polarization vectors, and (ii) the dilution of polarized flux due to the unpolarized stellar light. IRAS Z02229+6208 is clearly a moderately polarized source.

\subsubsection{IRAS 04166+5719}
IRAS 04166+5719 or TW Cam is a variable RV Tauri star, whose DoP and PA are wavelength-independent. A decrease in DoP towards longer wavelengths has been reported by Nook, Cardelli \& Nordsieck (1990). Our values are significantly higher than those quoted by Nook et al. (1990). Using the Julian Dates of the observations and the pulsation period of TW Cam (85.6~d; Evans 1985), we conclude that our observations were performed near the minimum magnitude phase, in contrast to those obtained by Nook  et al. (1990). This implies that  the polarization measurements by Nook et al. (1990) are probable diluted by the unpolarized stellar light.

The interstellar light may also affect the results. Nook et al. (1990) calculated the ISP in the direction of IRAS 04166+5719 close to 2.97\%  (PA=135.8\degree) within a circle of 10\degree, whereas a lower value of 1.8\% and PA=99\degree\ is estimated in this work based on the recently published catalogue of Heiles (2000).

\subsubsection{IRAS 04296+3429}	
HST images of IRAS 04296+3429 have revealed a multi--polar structure, with two pairs 
of highly collimated outflows oriented at PAs of $\sim$25\degree\ and $\sim$100\degree, 
respectively (Sahai 1999, Ueta et al. 2000, 2005, 2007). Both outflows are found to be 
highly polarized (DoP$\geq$30\%) in contrast to the lowly polarized inner region. 

Spectro--polarimetric data have shown that the DoP increases to the blue part of spectrum (Trammell et al. 1994, Oppenheimer et al. 2005). Our data show a roughly constant DoP$\sim$5\% with a possible increase to the {\it B} band (see  Table~B1). 
This value of $\sim$5\% is close to the aperture polarization measurement in the near-IR regime (Ueta et al. 2007). 
The PA measurements are in good agreement with both spectro-polarimetric studies, showing a rotation from 175 to 135\degree. Multi--scattering in these optically thin outflows may explain the high DoP values (Wood et al. 1996a,b).

\subsubsection{IRAS 04440+2605}
IRAS 04440+2605 shows a strong IR excess in the SED attributed to hot--dust emission, owing to a circumstellar disk around a binary system (De Ruyter et al. 2006). We present here the first polarimetric observations of this pulsating RV Tauri star. We find a moderate time variation in DoP and PA, for which a close binary system may be responsible. IRAS 04440+2605 is classified as a lowly polarized object .

\subsubsection{IRAS 05040+4820}
A nearly constant DoP$\sim$2\% and PA$\sim$140\degree\ are found for IRAS 05040+4820. No indication of time variability of polarization is found. Our observed values, before the correction of bias noise and ISP, are in very good agreement with those quoted  by Parthasarathy et al. (2005). IRAS 05040+4820 is apparently a lowly polarized object.  

\subsubsection{IRAS 05113+1347}
Despite that we observed IRAS 05113+1347 only in the {\it B, R} and {\it I} bands, we are able to deduce some general results. Its DoP increases towards shorter wavelengths from $\sim$1.2\% in the {\it I} band to $\sim$2.5\% in the {\it B} band, whereas its PA varies between $\sim$100\degree\ and $\sim$125\degree. Our results agree with those derived from spectro--polarimetric data (Bieging et al. 2006).

\subsubsection{IRAS 05208--2035}
The first polarimetric observations of IRAS 05208--2035 are presented here. Despite some possible time variations in polarization, IRAS 05208--2035 is most probable an unpolarized source.

\subsubsection{IRAS 05280+3817}
We present here the first polarimetric observations of IRAS 05280+3817. The DoP is from 0.3 to 0.5\%. Because of the high uncertainties, it is classified as an unpolarized object.

\subsubsection{IRAS 05381+1012}
The first polarimetric observations of IRAS 05381+1012 indicate it is an unpolarized object (0.1\%$<$DoP$<$0.25\% and a constant PA at 175--180\degree). Even after the correction of ISP, IRAS 05381+1012 exhibits a wavelength-dependent polarization that resembles the Serkwoski's law profile, indicating a possible interstellar origin. The reddening E(B-V)=0.29$\pm$0.03~mag of IRAS 05381+1012 (Vickers et al. 2015) implies an ISP of 1.3\%, as inferred from the ISM  extinction--polarization relation by Fosalba et al. (2002). This value is almost 5 times higher than our value, estimated with the field star method, namely 0.26\%  in the {\it V} band. It should be noted that the former equation is an empirical formula that provides the ISP as a function of reddening, but the spread of ISP measurements introduce a high uncertainty in the final result. We classify IRAS 05381+1012 as an unpolarized object.

\subsubsection{IRAS 06072+0953}
IRAS 06072+0953 is a variable RV Tauri star with a pulsation period of 67.3~d (Kiss et al. 2007). Its SED exhibits a strong IR excess associated with the emission from hot-dust grains in the envelope. Based on the SED, De Ruyter et al. (2006) claim that IRAS 06072+0953 is a candidate binary system, surrounded by a dusty disk. However, there is no observational confirmation of the disk or the binary companion.

We present here the first polarimetric observations of this RV-Tauri star. It was observed during three nights, with some evidence of possible time variability. However, a constant DoP$\sim$0.5\% and PA$\sim$ 140\degree\ seems very likely. In conclusion, IRAS 06072+0953 is classified as a very lowly polarized object despite its axisymmetric CSE. This suggests that the direct unpolarized light from the central star significantly dilutes the polarized light from the CSE, and results in a very low net DoP. 

\subsubsection{IRAS 06108+2743} 
The SED of IRAS 06108+2743 shows a strong IR--excess (De Ruyter et al. 2006). The first polarimetric observations of IRAS 06108+2743 reveal a very lowly polarized object with DoP$<$0.5\% and 105$<$PA$<$140\degree. Despite the correction for ISP, the polarization of IRAS 06108+2743 displays a Serkwoski profile, implying an interstellar origin. The large reddening E(B-V)=0.8$\pm$0.3~mag (De Ruyter et al. 2006) yields an ISP degree of 2.92$\pm$0.88\%, which is almost 3-4 times higher than our value. This further supports the interstellar origin for its intrinsic polarization. We, therefore, classify this object as an unpolarized one.

\subsubsection{IRAS 06160--1701}
IRAS 06160--1701 was also observed for the first time. It is found to be a very lowly polarized or even unpolarized object with probably a spherically symmetric CSE. Its strong IR excess implies the presence of a dusty disk around a binary system (De Ruyter et al. 2006). If IRAS 06160--1701 is a binary system, it must be seen face--on because of the very low DoP.

\subsubsection{IRAS 06176--1036}
The IRAS source 06176--1036 (also known as the Red Rectangle), is a PPN with a bi--conical morphology and a binary system (Waelkens et al. 1996). The first indications of an equatorial disk came from the optical images by Ostebart et al. (1997) and the high-angular-resolution near--IR images by M\'ekarnai et al. (1998), which resolved the central region and unveiled two lobes divided by a thin dark lane. Interferometric CO observations confirmed the presence of a molecular disk in Keplerian rotation perpendicular to the axis of the lobes (Jura 1997, Bujarrabal et al. 2003, 2007). 

Spectro--polarimetric data by Schmidt, Cohen \& Margon (1980) demonstrated that the DoP of this object increases towards shorter wavelengths (from $\sim$6\% to $\sim$2\%) while it has a nearly constant PA at $\sim$75\%. Cohen et al. (1975) showed, however, that the DoP decreases towards shorter wavelengths from 1.30\%$\pm$0.08\% in the {\it I} band to 0.97\%$\pm$0.03\% in the {\it U} band, and the PA is significantly lower between 28.3\degree$\pm$2.2\degree and 43.3\degree$\pm$1.4\degree. 

Our aperture polarization measurements show that IRAS 06176--1036 is a very lowly polarized object with a DoP nearly constant at ~1\%, whereas the PA rotates from 30\degree\ to 80\degree. Our observed values agree with those derived by Reese \& Sitko (1996) of DoP=2.2\% and PA=40\degree\ in the optical regime. 

\subsubsection{IRAS 06338+5333}
IRAS 06338+5333 is a known binary system with a period of 600$\pm$2~d (Hrivnak et al., 2008) and a strong IR excess, as the results of the presence of a dusty circumstellar/circumbinary disk. The intrinsic DoP shows a very lowly polarized object (0.32$<$DoP$<$0.66\%). Our values, before the correction for the bias noise and ISP, are in good agreement with previous studies (Trammell et al. 1994; Parthasarathy et al. 2005).

\subsubsection{IRAS 07008+1050}
IRAS 07008+1050 or PS Gem is a metal--poor ([Fe/H]=-4.8) RV Tauri star (Waelkens et al. 1991). The binary nature of all the metal--poor stars included in our sample, such as IRAS 06338+5333, IRAS 06167-1036, BD +39\degree 4926 and IRAS 07331+0021\footnote{The binary nature of this metal--poor star has not been confirmed yet. However, the recent detection of narrow CO lines (Bujarrabal et al. 2013) suggests the presence of a Keplerian disk which is attributed to a binary system.}, is well established (see Table~1). IRAS 07008+1050 is, indeed, a binary star with an orbital period of 1310$\pm$8~d (Van Winckel et al. 1999, Gielen et al. 2008). Interferometric observations have confirmed the presence of a compact circumbinary disk with a radius $\leq$0.0055~arcsec (Deroo et al. 2006).

Recent spectroscopic observations shown a DoP up to $\sim$1\% with possible variability in time (Harrington \& Kuhn 2009). 
The DoP increases toward shorter wavelengths (from 0.4-0.5\% in the {\it I} band to $\sim$1.1-1.2\% in the  {\it U} band). 
Our values are very close to those reported by Parthasarathy et al. (2005). The PA is also found to vary between 5\degree\ 
and 60\degree. The high DoP towards the blue part of the spectrum supports an intrinsic origin of the polarization. 
IRAS 07008+1050 was observed during three nights  with no significant evidence of time variability on this time-scale.

\subsubsection{IRAS 07134+1005}
IRAS 07134+1005 is a pulsating supergiant star with a period of 36.8~d (Barth\'es et al. 2000). It displays an ellipsoidal hollow shell with a size of 5$\times$4~arcsec and the major axis aligned to PA=25\degree\ (Ueta et al. 2000, 2005). IRAS 07134+1005 has highly polarized outer edges (average value of 55$\pm$16\%) and significantly low-polarization inner region (DoP$<$10\%; Ueta et al. 2005). 

This star was observed during three nights and all observations indicate a very lowly polarized or most likely an unpolarized object with DoP$\sim$0.2\%. Our values agree with those reported by Trammell et al. (1994) and Parthasarathy et al. (2005) but are slightly lower than the value recently reported by Bieging et al. (2006). The very low aperture net DoP found in IRAS 07134+1005, despite the highly polarized outer edges, can be explained by the cancellation of the polarization vectors due to the centrosymmetric pattern, as well as the dilution of scattering light from the CSE. 

\subsubsection{IRAS 07331+0021} 
IRAS 07331+0021 is a metal-poor supergiant star (Klochkova \& Panchuk 1998). We find some possible evidence of time variations in DoP consistence with Parthasarathy et al. (2005). The DoP varies between 0.1\% and 0.8\%, whereas the PA increases from $\sim$125\degree\ in the {\it U} band to $\sim$150\degree in the {\it I} band. 

Recent CO mapping data of IRAS 07331+0021 revealed narrow CO lines like those in the Red Rectangle and 89 Her, suggesting a possible emission from a dusty Keplerian disk (Bujarrabal et al. 2013). The presence of a binary system may be responsible for the variation in polarization. 

\subsubsection{IRAS 07430+1115}
IRAS 07430+1115 was observed only one night in the {\it B, V} and {\it R} bands. The DoP is found to decrease towards longer wavelength from 6\% in the {\it B} band to 1.77\% in the {\it R} band, whereas the PA varies from 135 to 145\degree. Near--IR polarimetry measurements in the {\it J} (DoP=1.2\%, PA=135\degree) and {\it K} bands (DoP=0.8\% and PA=135\degree) reported by Gledhill (2005)  are consistent with our wavelength dependence. The polarized flux image in the {\it J} band unveils the presence of a ring--like structure with a diameter of 0.4~arcsec along the north-east--south-west direction (Gledhill 2005). IRAS 07430+1115 is classified as a lowly polarized object.
 
\subsubsection{IRAS 08187--1905} 
We present the first broad-band polarimetric observations for this pulsating post-AGB star. 
Both DoP and PA (see Table~B1) show  time variations that may be associated with the pulsation of the star like IRAS 07331+0021. Further observations are required. The wavelength dependence of both DoP and PA suggest an intrinsic origin of polarization.

\subsubsection{IRAS 09371+1212 }
IRAS 09371+1212, also known as Frosty Leo, is a well--known and studied PPN. Optical and near-IR images display a complex bipolar or multi-polar structure (Langill, Kwok \& Hrivnak 1994; Scarrott \& Scarrott 1994; Roddier et al. 1995; Sahai et al. 2000) aligned at PA$\sim$110\degree, and an equatorial density enhancement seen almost edge--on. Roddier et al. (1995) claimed the detection of a binary companion at 0.18~arcsec\ from the central star. However, the identical spectral type of both stars makes this detection less probably and the binarity of Frosty Leo still remains controversial. Molecular gas has also been detected in a small inner region with an angular radius less than 6~arcsec, expanding with velocities up to 50\kms\ (Sahai et al. 2000; Castro--Carrizo et al. 2005). 

Optical and near--IR polarimetric observations of Frosty Leo were obtained first by Scarrott \& Scarrott (1994) and more recently by Murakawa et al. (2008). A high DoP up to 60\% has been found mainly in the outer blobs. According to the Mie scattering theory, Scarrott \& Scarrott (1994) argued that dust grains with radii $\alpha$ $\sim$0.1$\micron$ and a steep grain-size distribution are required to explain their polarization results. Although, Murakawa et al. (2008) were able to reproduce the observed DoP considering different size distributions for the disk (0.005$\micron$ $\leq\alpha\leq$2.0$\micron$) and the bipolar lobes (0.005$\micron$ $\leq\alpha\leq$0.7$\micron$). 

IRAS 09371+1212 is one of the three highly polarized objects in our sample. We find a wavelength--dependent DoP that increases towards longer wavelengths from 11\% to 19\% in the {\it B} and {\it I} bands, respectively, which is consistent with scattering by small dust grains (Murakawa et al 2008). The PA rotates from 55\degree\ to 75\degree, almost perpendicular to the symmetric axis of the lobes. This is consistent with multiple scattering of photons in optically thin regions.

\subsubsection{IRAS 11157+5257}
IRAS 11157+5257 or DZ Uma is classified as a RV Tauri star (Szczerba et al. 2007). The first polarization measurements of this RV Tauri star show an unpolarized object (see Table B1).

\subsubsection{IRAS 11472--0800}
IRAS 11472--0800 or AF Crt, is a pulsating RV Tauri star with a period of 31.16~d (Kiss et al. 2007, Van Winckel et al. 2012). The presence of a binary companion with an orbital period of 14~yr or longer has been proposed by Van Winckel et al. (2012), which is approximately 2 times longer than the maximum orbital periods in binary post-AGB stars (2000~d; Van Winckel et al. 2009).

Near--IR polarimetric images of IRAS 11472--0800 present a highly polarized source with an average DoP of $\sim$6\% in both {\it J} and {\it K} bands and maximum values of 10\% and 7.7\%, respectively (Gledhill 2005). The polarization vectors are found to be perfectly aligned along the north-west--south-east direction (PA=135\degree), perpendicular to the axisymmetric structure seen in the {\it K} band image (Gledhill 2005). This high degree of alignment may be associated with multiple scattering along the polar--axis (Wood et al. 1996a,b; Hoffman et al. 2003). 

We find a wavelength-dependent DoP that gradually increases from 2.8 to 5\%, and a PA that rotates from 165\degree\ to 135\degree. Both agree with the near--IR polarimetric measurements of Gledhill (2005). The ISP of IRAS 11472--0800 was estimated to be 0.15\% within circles of 5\degree\ and 10\degree. This further supports the previous estimation of ISP by Mathewson \& Ford (1970).

\subsubsection{IRAS 13467--0141}
The first aperture polarimetric data of IRAS 13467--0141 is presented here. A constant DoP at  0.5\% and PA between 70\degree\ and 87\degree\ are found. IRAS 13467--0141 is assigned to the group of very lowly polarized sources.

\subsubsection{IRAS 17436+5003}
IRAS 17436+5003 has been extensively studied by several groups (Joshi et al. 1987; Trammell et al. 1994; Ueta et al. 2000, Gledhill et al. 2001; Gledhill \& Yates 2003, Parthasarathy et al. 2005). HST optical images (Ueta et al. 2000) and near-IR polarimetric images (Gledhill et al. 2001) display an elliptical envelope aligned at PA=10\degree. It is found to be more polarized along the north--south direction than along the west--east direction, with a maximum DoP value of 20\% in both {\it J} and {\it K} bands (Gledhill et al. 2001).

Our observed values, before the interstellar and bias-noise correction, agree with those quoted by Parthasarathy et al. (2005). The field star method gives an ISP degree of 1.72\% in the {\it V} band, which does not agree with the very low reddening of this object (0.03$\pm$0.01~mag; Vickers et al. 2015). This reddening value yields ISP= 0.2\%. The correction for the bias noise has resulted in a DoP$\sim$0.4-0.5\% (see Table~B1). IRAS 17436+5003 is classified as an unpolarized object despite the value of 0.67$\pm$0.38\% in the {\it B} band.

\subsubsection{IRAS F17495+0757}
The first polarimetric measurements of IRAS F17495+0757 show a constant DoP and PA within the errors ( DoP$\sim$0.3\%, PA$\sim$80\degree). IRAS F17495+0757 is considered as an unpolarized source.

\subsubsection{IRAS 17534+2603}
IRAS 17534+2603 (or 89~Her) is a pulsating star, extensively studied over the last decade. The CO emission reveals an hourglass morphology, seen almost pole--on (15\degree\ with respect to the line of sight) expanding with velocities of 5--10\kms, and a more compact Keplerian rotating disk (Bujarrabal et al. 2007, 2013). This disk is consistent with the strong IR excess in the SED (De Ruyter et al. 2006) and the binary central star (Waters et al. 1993). 

Its DoP measurements indicate an unpolarized source (see Table~B1). In particular, we estimate the DoP between 0.1 and 0.2\%. This value is lower than those quoted from previous polarimetric and spectro--polarimetric studies (Joshi et al. 1987; Trammell et al. 1994; Parthasarathy et al. 2005).  However, it should be noted that the DoPs derived by Joshi et al. (1987) and Parthasarathy et al. (2005) are not corrected for the ISM, which we found to be 0.33$\pm$0.77\% in the {\it V} band, very close to the value  of 0.29\% derived by Trammell et al. (1994). By applying this ISP to Joshi et al.'s (1987) and Parthasarathy et al.'s (2005) data, we obtain intrinsic DoP values very close to our values.

According to this analysis, we do not find time variability in the polarization, as previously suggested by Trammell et al. (1994), based on the data from Polyakova (1987) and Joshi et al. (1987). We have to mention that (i) the data of Polyakova refer to BL Her ({\it l}=045\degree.16, {\it b}=19\degree.46), not to 89~Her (l=051\degree.43, b=23\degree.19); and (ii) Joshi et al.'s polarization measurements are not corrected for the ISP, as mentioned before.

The null polarization of 89~Her indicates that it is seen almost pole--on with an inclination angle lower than 15\degree\ (Bujarrabal et al. 2007). Hillen et al. (2013) also came to a similar conclusion. This means that the central star is directly visible and the unpolarized stellar light dilutes the scattered light from the envelope.

\subsubsection{IRAS 18095+2704}
IRAS 18095+2704 is an F--type post-AGB star (Hrivnak, Kwok \& Volk 1989), surrounded by a bipolar nebula oriented at PA=100\degree\ (Ueta et al. 2000). Near-IR polarization imaging reveals an alignment of the polarization vectors towards the north-west--south-east direction (PA=23\degree, Gledhill et al. 2001), almost perpendicular to the bipolar axis, indicating multiple scattering. 

We find a wavelength--dependent DoP that increases from 0.9\% ({in the \it I}) to 6.7\% ({in the \it U}). Our results agree with previous studies (Trammell et al. 1994; Parthasarathy et al. 2005; Bieging et al. 2006). The high DoP towards the blue part of the spectrum confirms the intrinsic nature of the polarization. Moreover, the DoP values in both {\it H} and {\it K} bands of 2.4$\pm$0.8\% and 2.4$\pm$0.3\%, respectively (Gledhill et al. 2001), indicate an increase of polarization toward the near-IR regime. The PA is found to be nearly constant at $\sim$17\degree\ and is significantly lower than those derived by Trammell et al. (1994) and Parthasarathy et al. (2005), of 131\degree\ and 125\degree\ respectively, but closer to the values derived from Gledhill et al. (2001) and Bieging et al. (2006), 23\degree and 16.4\degree, respectively. 

\subsubsection{IRAS 18123+0511}
The first polarimetric observations of IRAS 18123+0511 are presented here. The intrinsic polarization of this object resembles the Serkwoski's law profile, even after the correction of the ISM, and in conjunction with the constant PA$\sim$32\degree, we presume a likely interstellar origin of its polarization. The reddening of 0.3~mag (De Ruyter et al. 2006) implies an ISP of 1.34\%, approximately 3.5 times higher than the value adopted here (0.37\%).  The polarization measurements, corrected for the bias noise, are found to be very close or slightly higher than the ISP from the reddening method. This implies that we may overestimate the intrinsic polarization of IRAS 18123+0511. We classify this object as a very lowly polarized source.

Recently, Bujarrabal et al. (2013) detected intensive CO molecular lines attributed to a compact rotating disk with an expanding component like those found in the Red Rectangle and 89 Her. 

\subsubsection{IRAS 18281+2149}
IRAS 18281+2149, also known as AC Her, is a confirmed binary system with a period of 
1196$\pm$6~d (Van Winckel et al. 1998). The presence of a dusty circumbinary disk has been speculated due to its strong IR excess (De Ruyter et al. 2006). CO mapping data of IRAS 18181+2149 show the presence of very narrow CO lines indicative of a compact rotating disk (Bujarrabal et al. 2013).  The low reddening of AC Her, E(B-V)=0.17, implies that the disk is seen nearly face--on.

Our data show that AC Her is an unpolarized object and we confirm the results from Nook et al. (1990). Given that this object is a known binary system, the null polarization is consistent with a face-on disk. The cancellation of the perpendicular polarization vectors results in zero or very low DoP. Moreover, scattered light from the envelope is significantly diluted by the direct unpolarized stellar light, given that the central star is very bright ({\it V}=7.01 mag; Ducati 2002). Adaptive optics images at mid-IR do not show any evidence of an extended envelope around AC Her, whereas there is a circumstellar disk with a size lower than 0.2~arcsec\ or 75~AU, adopting a distance of 0.75~kpc (Close et al. 2003).

\subsubsection{IRAS 18564--0814}
The first polarimetric observation of IRAS 18564--0814 show a wavelength--dependent DoP. 
It decreases from  2\% in the {\it U} band to $\sim$0\% in the {\it V}  band, and then increases again to  0.7\% in the 
{\it I}  band. The increase of DoP toward the blue part of wavelength indicates an intrinsic origin of its polarization. 
The PA rotates from 105\degree\ to 150\degree.

\subsubsection{IRAS 19090+3829} 
We present the first polarimetric observations of IRAS 19090+3829. It is found to be a very lowly polarized or unpolarized object with DoP$<$1\%. However, due to relatively high errors, a constant DoP of $\sim$0.5\% cannot be ruled out. 

\subsubsection{IRAS 19114+0002}
IRAS 19114+0002 (HD 179821 or AFGL 2343) has a controversial classification. The postulate of this star was for it be a luminous and massive YHG (Kastner and Weintraub 1995; Castro--Carrizo et al. 2007; Quintana--Lacaci et al. 2007; Quintana-Lacaci, Bujarrabal \& Castro-Carrizo 2008). However, there are some indications, such as the depletion of s--process elements, that indicate it is a low-mass post-AGB star (Th\'evenin , Parthasarathy \& Jasniewicz 2000; Kipper 2008). An aspherical inner region has been unveiled from high-angular-resolution {\it HST} images (Ueta et al. 2000) and interferometric observations of molecular gas ( Bujarrabal et al. 1992; Castro--Carrizo et al. 2007; Quintana--Lacaci et al. 2007, 2008). A spherically symmetric halo extends up to 5-6~arcsec from the central star. 

IRAS 19114+0002 exhibits a double-peak SED with a large near-IR excess, possible associated with emission from hot dust in a circumstellar disk. Nevertheless, radiative transfer models have show that the near-IR excess of IRAS 19114+0002 can also be interpreted as the result of a spherically symmetric dust shell, instead of a dusty disk (Nordhaus et al. 2008). Interferometric observations support this scenario, as they do not confirm the presence of any compact disk around the star (Castro--Carrizo et al. 2007; Quintana--Lacaci et al. 2007, 2008). This analysis points out that strong near-IR excesses are not necessarily the result of a dusty circumstellar disk, but also of a dusty shell. 

We estimate the DoP between  0.9\% and 2\% and a PA nearly constant at 45\degree. Our results agree with those derived by Trammell et al. (1994), Parthasarathy et al. (2005) and Patel et al. (2008). Its intrinsic polarization can be described by the Serkwoski law. The ISP is estimated around 0.27\% using the star field method, which does not agree with the high reddening E(B-V)=0.96$\pm$0.19~mag (Vickers et al. 2015) or equivalently, ISP=3.38\%. On the other hand, a significantly lower ISP=0.7\% (PA=42.4\degree), much closer to our value, has been reported by Trammell et al. (1994). We classify IRAS 19114+0002 as a lowly polarized object.

\subsubsection{IRAS 19125+0343}
IRAS 19125+0343 or BD +03\degree 3950 is a confirmed binary system with a period of 517~d (Gielen et al. 2008). Bujarrabal et al. (2013)  report the detection of narrow CO lines with intense line wings, possibly, associated with an extended rotating disk and low-mass bipolar outflows (5--10\kms).

The first polarimetric observations of this binary post-AGB star are presented here. The intrinsic polarization clearly resembles the Serkwoski's law profile, and in conjunction with the constant PA ($\sim$25\degree), we claim a possible interstellar origin. The very large reddening of 1.2~mag (De Ruyter et al. 2006) gives ISP=4.05\%, which is significantly higher than our value  (0.21\%). From the current data, we cannot deduce whether IRAS 19125+0343 is an unpolarized or a very lowly polarized object.

\subsubsection{IRAS 19163+2745}
IRAS 19163+2745 was also observed for the first time. The data show a lowly polarized object with DoP between  0.3\% and 0.9\% and PA between  35\degree\ and 65\degree. 

The strong IR excess of this object is also attributed to emission from a hot dusty disk (De Ruyter et al. 2006; Gielen et al. 2009). Our ISP is 4 times lower than the value of 2\% derived from the reddening formula [E(B-V)=0.5~mag; De Ruyter et al. 2006]. Based on this analysis and the high uncertainties, we argue that IRAS 19163+2745 is more likely a very lowly polarized object.

\subsubsection{IRAS 19199+3950}
The first polarimetric observations of the eclipsing binary IRAS 19199+3950, with an orbital period of 138.6~d (Kreiner 2004), are presented here. Our data clearly show an unpolarized object, probably seen face-on.

\subsubsection{IRAS 19343+2926}
IRAS 19343+2926, also known as Minkowski Footprint nebula or M 1-92, is an O--rich bipolar PPN consisting of an equatorial dusty torus or disk. Interferometric observations of $^{13}$CO in the J=2--1 line confirmed the presence of a thin equatorial disk (Alcolea, Neri \& Bujarrabal 2007).

IRAS 19343+2926 is a highly polarized object. The DoP increases towards longer wavelengths from 6\% to 16\%, whereas the PA is nearly constant (40\degree) and almost perpendicular to the bipolar axis defined by the lobes (PA=131\degree; Ueta et al. 2007). Such a high DoP signifies significant multiple scattering in the optically thin polar lobes. Wood et al. (1996a,b) and Hoffman et al. (2003) showed that multiple scattering results in higher DoP by reducing the total number of photons that escape from the optically thick equatorial region.

\subsubsection{IRAS 19386+0155}
Optical and IR spectroscopic analysis of IRAS 19386+0155 have shown that both the star and the envelope are O-rich (Pereira,
Lorenz-Martins \& Machado 2004). The authors claim that the SED of IRAS 19386+0155 cannot be reproduced by a spherically symmetric envelope, and a dusty equatorial disk is required. Recent mid-IR images of this IRAS source by Lagadec et al. (2011) unveiled an axisymmetric structure along the south-east--north-west. The {\it J}-band polarimetric image displays an alignment of polarization vectors along the east--west direction that gradually rotate to the south-east--north-west direction in the outer envelope (Gledhill 2005).

We find high DoP at short wavelengths ( 5\% in the {\it U} band) that gradually decreases at longer wavelengths ( 0.6\% in the {\it I} band). This wavelength-dependent polarization is consistent with the very low DoP found in the {\it J} band (Gledhill 2005). Furthermore, the increase of DoP towards shorter wavelengths strongly supports an intrinsic origin for its polarization.

\subsubsection{IRAS 19475+3119}
High-angular resolution {\it HST} images of IRAS 19475+3119 display a quadra-polar PPN with a bipolar outflow towards PA=90 and 145\degree, respectively (Sahai et al. 2007a,b, Si\'odmiak et al. 2008). Molecular CO emission has also been detected and studied by several authors (Likkel et al. 1991, Hrivnak \& Bieging 2005, S\'anchez Contreras et al. 2006, Castro--Carrizo et al. 2010, Hsu \& Lee 2011). In accordance with these works, the fast-moving molecular outflows with projected velocities of 20--30~\kms, trace the optical outflow. The presence of a dusty toroidal-like structure has been confirmed by Hsu \& Lee (2011).

The polarization vectors exhibit a centrosymmetric pattern mainly in the central region, while they have a preferential orientation along the north-east--south-west direction at the outer envelope. The maximum DoP is approximately 20.6$\pm$4.6\% and 11.3$\pm$2.5\% in the {\it J} and {\it K} bands, respectively (Gledhill et al. 2001). Our aperture polarimetric measurements show a DoP between  0.9 and 1.4\%. This low intrinsic polarization suggests
that the star is seen directly, and its stellar unpolarized light dilutes the polarized light from the CSE. The PA is found to vary between 25\degree and 40\degree, which is almost perpendicular to the symmetric axes of the nebula and consistent with the near--IR results of Gledhill et al. (2001).

\subsubsection{IRAS 19486+1350}
We confirm the very low polarization of this low-mass post-AGB star reported by Parthasarathy et al. (2005). In particular, we find a nearly constant DoP at $\sim$0.3\%, whereas its PA rotates from 35\degree\ to 75\degree.

\subsubsection{IRAS 19500--1709}
IRAS 19500--1709 has been classified as a carbon-rich post-AGB F-type supergiant star. Unfortunately, only two measurements in the {\it U} and {\it V} bands can be used for this object. However,  IRAS 19500--1709 appears to be intrinsically polarized with a DoP around 1.5\% and PA=165\degree\ in agreement with the previous work by Parthasarathy et al. (2005).

Based on spectro--polarimetric observations, Trammell et al. (1994) claimed that IRAS 19500--1709 is not intrinsically polarized because of the high ISP (2.3\%). We have to point out that our estimation of the ISP is significantly lower (0.3\%). However, polarimetric images in both {\it J} and {\it K} bands show significantly high DoP of 5.7\% and 6.1\%, respectively (Gledhill et al. 2001). The polarization map in the {\it J} band reveals a seemingly bipolar structure with a symmetric axis towards PA=100\degree. We reckon that IRAS 19500--1709 is polarized but we cannot deduce any robust result for its intrinsic polarization due to the pure estimation of ISP. We classified it as a lowly polarized object.

\subsubsection{IRAS 20000+3239}
The star associated with the IRAS source 20000+3239, is found to be a C-rich G8 Ia type star (Hrivnak, Kwok \& Volk 2000). Optical {\it HST} images show an ellipsoidal structure along PA=120\degree\ (Sahai et al. 2007a, Si\'odmiak et al. 2008). Based on polarimetric maps in the {\it J} and {\it K} bands, Gledhill et al. (2001) deduced that the CSE envelope around this star has a bipolar structure with an equatorial dense torus or disk.
 
According to our observations, IRAS 20000+3239 is a very lowly polarized object with a nearly wavelength-independent polarization ($\sim$1\% and PA=120\degree). Our results do not exhibit a Serkowski profile, as reported by Trammell et al. (1994), while our observed data, before interstellar correction, are the same with those from  Trammell et al. The difference in the intrinsic polarization may be associated with the ISP. In particular, Trammell et al. (1994) report an ISP=3.04 (PA=32\degree) which is almost 4 times higher than our value of 0.74\% (PA=86\degree). Although, we have to mention that near-IR polarization images in the {\it J} and {\it K} bands have shown DoP up to 6.1$\pm$1.5\% and 3.4$\pm$2.1\% respectively (Gledhill et al. 2001). The polarization vectors are found to be aligned along a PA=120\degree\ very close to our PA measurement. IRAS 20000+3239 is classified as a very lowly polarized source but we cannot rule out the possibility that it is intrinsically unpolarized.

\subsubsection{IRAS 20004+2955}
IRAS 20004+2955 has been classified as a PPN candidate (Hrivnak et al. 1989). We find a wavelength-dependent DoP varying from 3.9\% to 6.7\%. The PA increases gradually to longer wavelengths from 60\degree\ to 70\degree. Our results agree very well with those reported by Trammell et al. (1994), unlike those of Parthasarathy et al. (2005).

\subsubsection{IRAS 20056+1834}
IRAS 20056+1834 is a variable RV Tauri star with a period of 60~d (Kiss et al. 2007). Rao, Goswami \& Lambert (2002) claim that this star is highly obscured because of the direct view of a dense torus or disk component, and it is mainly seen by scattered light. This is consistent with the strong near-IR excess (De Ruyter et al. 2006). 

We present here the first broad-band polarimetric observations of IRAS 20056+1834, revealing a highly polarized object with DoP decreasing gradually towards shorter wavelengths. Despite that our results are consistent with those from Rao et al. (2002), this object requires further investigation due to the significant difference in DoP within a period of 80~d.
Moreover, a significantly lower DoP, by a factor of 3--4, as well as an opposite polarization wavelength dependence has been reported by Trammell et al. (1994) .

Our results are consistent with the more recent near-IR polarimetric observation of Gledhill et al. (2001). The authors estimated the maximum DoP in the {\it J} and {\it H} bands of 14.2\% and 7.4\% respectively, as expected from the wavelength dependence found here. Furthermore, the polarization vectors are aligned at PA=135\degree, very close to our value of 140\degree.

\subsubsection{IRAS 20117+1634}
IRAS 20117+1634 is found to be a  very lowly polarized object with a DoP$\sim$0.7\%. This value is in agreement with the value determined by Yudin et al. (2003).  Although the strong IR-excess in the SED implies the presence of a hot dusty circumstellar disk (De Ruyter et al. 2006), no CO emission associated with a compact disk has been detected (Bujarrabal et al. 2013).

\subsubsection{IRAS 20160+2734}
We present the first polarimetric observations of this F3Ie-type, post-AGB star 
(Su\'arez et al. 2006). We find a wavelength dependence for both DoP and PA, increasing towards shorter wavelengths. Similar wavelength--dependence behaviour is found for the IRAS 18095+2704 and IRAS 20056+1834, two objects with strong indications of an edge--on disk. IRAS 20160+2734 is a lowly polarized object.

\subsubsection{IRAS 20343+2625}
IRAS 20343+2625 is unpolarized despite the relatively high errors. This result is in agreement with the work by Nook et al. (1990). 

\subsubsection{IRAS 21546+4721}
The first polarimetric data of IRAS 21546+4721 show very low DoP between 0.35\% and 0.82\% and PA between 30\degree\ and 50\degree. However, a possible flat wavelength dependence cannot be ruled out due to the high uncertainties. IRAS 21546+4721 is likely a very lowly polarized object.

\subsubsection{IRAS 22023+5249}
IRAS 22023+5249 is a PPN or a young nebula surrounding a central star of 24000K$\pm$1000K (Sarkar et al. 2012). 
$\rm{H_2}$ images of this source have revealed an apparently axisymmetric structure aligned towards PA$\sim$130\degree (Volk, Hrivnak \&
Kwok 2004).

Its intrinsic polarization resembles the profile of the Serkwoski law and in conjunction with the constant PA, we claim an interstellar origin. The reddening E(B-V)=0.62$\pm$0.06~mag (Vickers et al. 2015) implies ISP=2.38\%, almost 2-3 times higher than our value. According to this analysis, we may underestimate the ISP in the direction of this object. The PA is $\sim$55\degree, almost perpendicular to the major axis of the nebula. We, thus, argue that IRAS 22023+5249 is a lowly polarized object with an intrinsic DoP between 1\% and 1.6\%.

\subsubsection{IRAS 22223+4327}

IRAS 22223+4327 is a variable bright post-AGB star with a period of 90~d, surrounded by a C-rich reflected nebula (Hrivnak et al. 2013). {\it HST} images display an ellipsoidal nebula oriented along PA=25\degree\ (Si\'odmiak et al. 2008). CO emission has been detected in this object, likely originate from a compact central region, such as a torus or disk,  and there is a molecular outflow expanding with velocities of 11--22\kms\ (Omont et al. 1993, Bujarrabal et al. 2001).

Polarimetric observations of IRAS 22223+4327 in the {\it J} band have revealed a bipolar nebula embedded in an extended spherically symmetric envelope (Gledhill et al. 2001). The extended optical envelope shows a centrosymmetric pattern resulting in the very low polarization degree. Its central region, on other hand, shows an alignment of polarization vectors towards PA=40\degree. This is attributed to the presence of ellipsoidal dust grains aligned to PA=40\degree\ (Wolf, Voshchinnikov \& Henning 2002).

Bieging et al. (2006) reported a DoP of 1.66\% and a PA of 22.3\degree . Our measurements, before the correction of the ISM, agree with the aforementioned work. The ISP derived from the star field method,  namely 2.57\%, implies an unpolarized object. However, this is inconsistent with the very low reddening value of 0.15$\pm$0.02~mag (Vickers et al. 2015), which yields ISP=0.77$\pm$0.08\%. This ISP is found to be equal to or higher than the DoP$_{\rm{MAS}}$. We conclude that IRAS 22223+4327 is an unpolarized source.

\subsubsection{IRAS 22223+5556}
The first polarimetric observations of IRAS 22223+5556, a variable RV Tauri star (Szczerba et al. 2007), demostrate a moderately polarized object with DoP between  5\% and 7\% and PA between  40\degree\ and 50\degree. The intrinsic polarization of IRAS 22223+5556 can be fitted with the Serkwoski law, which means that we may have underestimated the ISP (1.35\%). However, the high DoP of this object cannot be explained by the interstellar medium. IRAS 22223+5556 is evidently a moderately polarized object.

\subsubsection{IRAS 22272+5435}
The morphology of IRAS 22272+5435 has been intensively studied in various wavelengths (Ueta et al. 2000, 2001; Gledhill et al. 2001). A seemingly multi--polar envelope has been revealed as well as a dusty inclined toroidal shell. CO emission has also been detected (Bujarrabal et al. 2001; Hrivnak \& Bieging 2005; Fong et al. 2006; Nakashima et al. 2012). In accordance with these works, the molecular envelope has a spherically symmetric structure, extended up to 20~arcsec, whereas there is no evidence for high-velocity molecular outflows.

IRAS 22272+5435 is a lowly polarized object. In particular, its DoP decreases gradually towards the red part of the spectrum  (from $\sim$1.5\% in the {\it U} band to zero in the {\it I} band) suggesting an intrinsic origin. The PA is found to be nearly constant at 55-60\degree. Our DoP measurements, before the correction of ISM, agree with those from Trammell et al. (1994), Parthasarathy et al. (2005) and Bieging et al. (2006). The ISP in the direction of this object is calculated as 1.13\%, very close to the value reported by Trammell et al. (1994) of 1.67\%. IRAS 22272+5435 is apparently a lowly polarized object.

\subsubsection{IRAS 22327--1731} 
We present the first optical broad-band polarimetric observations of this binary star. Our polarimetric data show a nearly flat wavelength-dependent polarization of $\sim$0.4\%.

\subsubsection{BD +39\degree 4926}
The first polarimetric data of BD +39\degree 4926 reveal an unpolarized source.

\section{Discussion}  
Here, we discuss the multi-band ({\it U, B, V, R} and {\it I}) polarimetric survey of 53 post-AGB stars. 24 out of 53 (45\% of the sample) were presented for the first time, whereas the rest 29 (55\%) have been observed and studied before by other authors.

The ISP degree of our programme stars was calculated by means of the field star method, and for this we used the catalogue from Heiles (2000). For all the objects, the ISP was estimated within a circle of angular radius of 5\degree, except from six cases (IRAS 06338+5333, IRAS 07134+1005, IRAS 07430+1115, IRAS 09371+1212, IRAS 11472+0800 and IRAS 17436+5003), where a circle of angular radius of 10\degree\ was selected due to an insufficient sample of field stars within the 5\degree\ angular radius. Besides, for those objects with available reddening values in the literature, we also calculated the ISP using the empirical formula from Fosalba et al. (2002). It should be noted that this empirical formula, which can be uncertain due to the high dispersion of the data, is not applicable for reddening values higher than 1~mag. In most of the cases (72\% of our sample), the ISP degree was found to be distributed around PA=90\degree\ ($\pm$15\degree); see Fig. \ref{fig_IP}. From the remaining 28\% (or 15 stars), only three sources exhibit DoP higher than 1\%. We thus conclude that our measurements of the ISP are reliable within their uncertainties. 

For those cases in which the stellar polarization, after correcting for the ISM and bias noise (MAS), still resemble the Serkwoski's law profile (with a constant PA), we considered a likely interstellar origin for the residual DoP. These cases correspond to objects with very low DoP.

\begin{figure}
\begin{center}
  \includegraphics[scale=0.30]{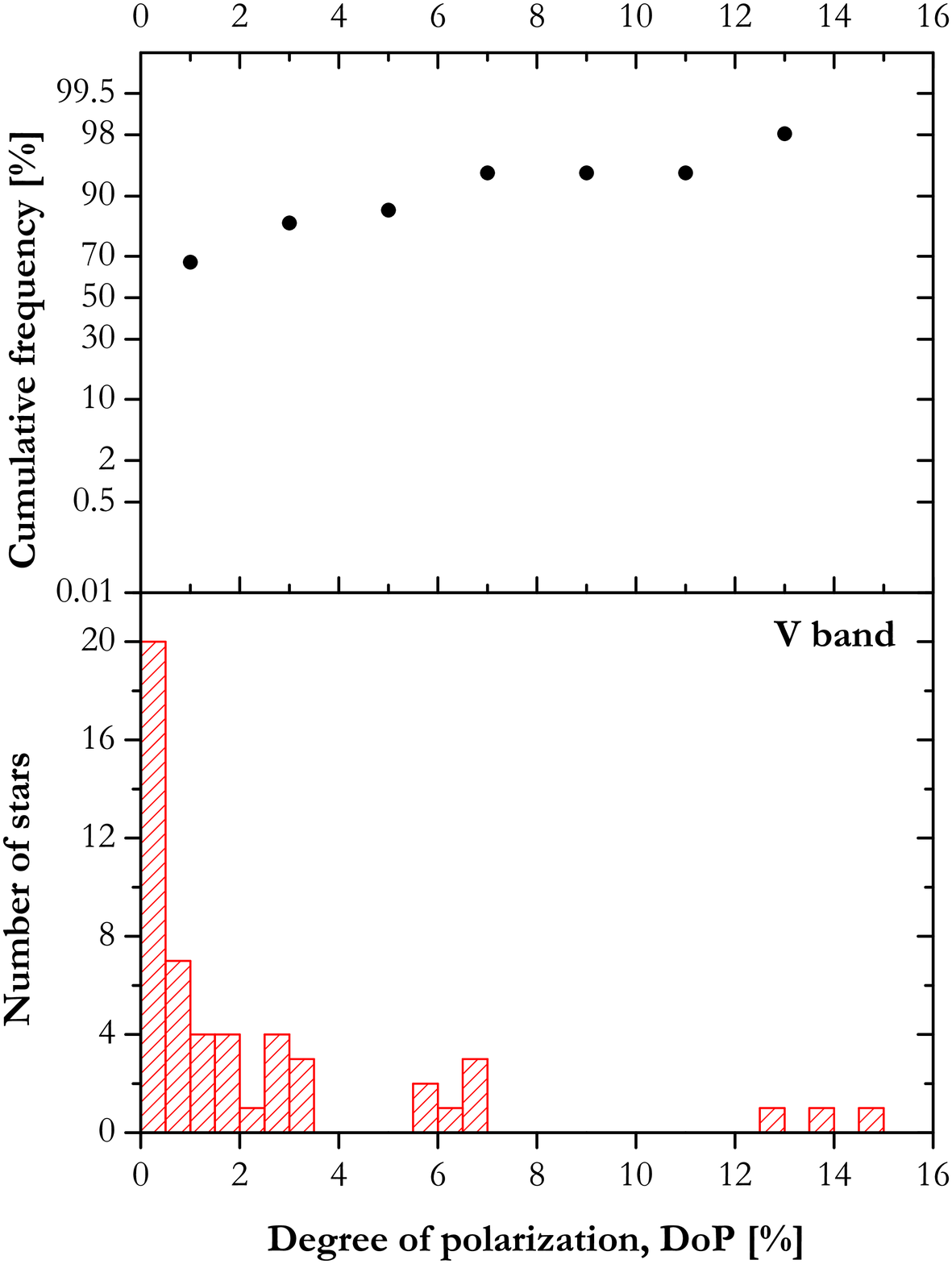}
  \includegraphics[scale=0.05]{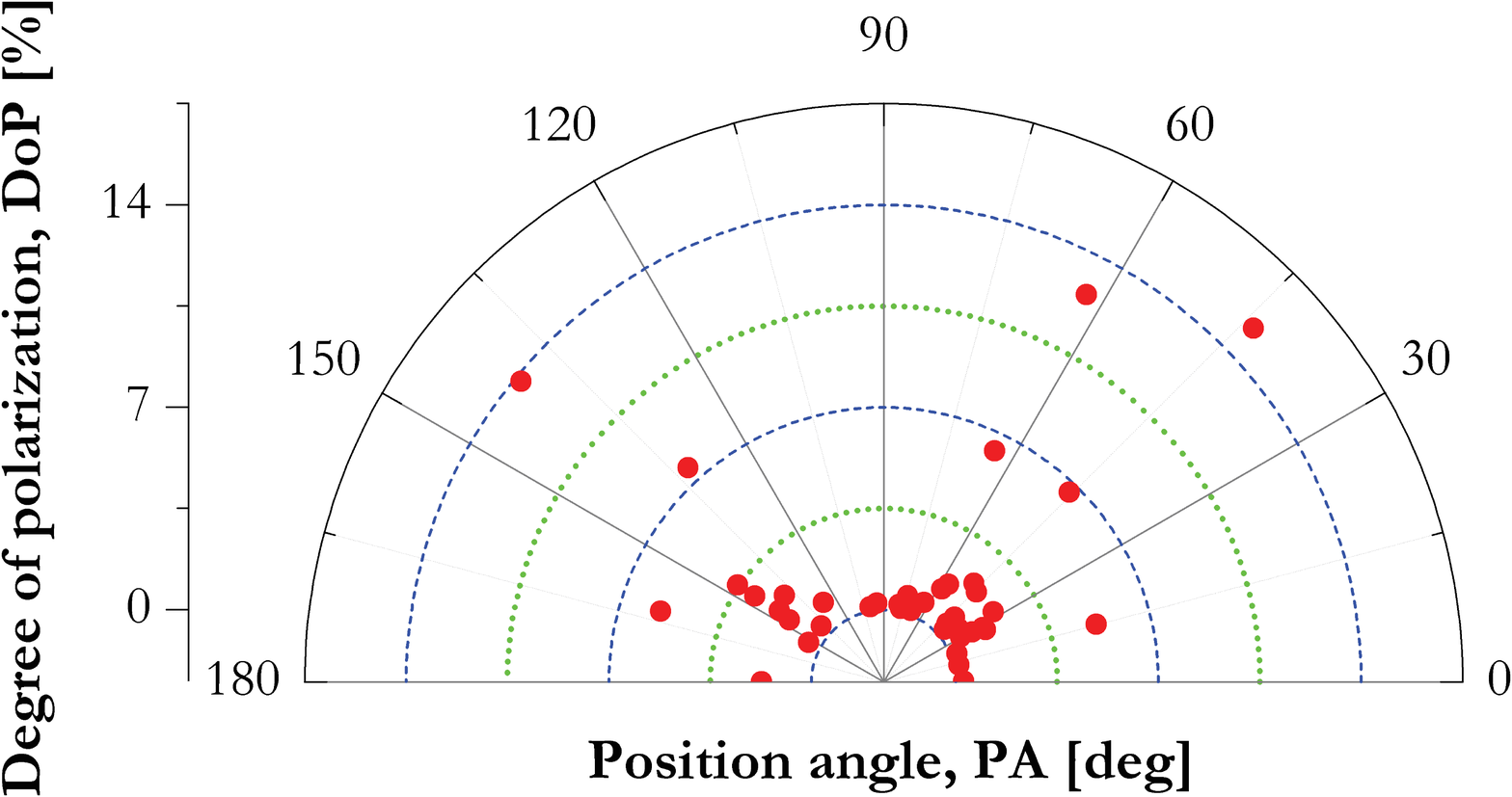}	
 \caption{Upper panel: Distribution of DoP in the {\it V} band toward the programme stars. Three discernible augmentations of the DoP are found at 1.66, 6.25 and 13.86\%. The size of DoP bins is 1\%. Lower panel: Polar diagram depicting the position of the polarization vectors in the {\it V} band for all the stars in our sample. The blue and green semicircles correspond to different DoP values, which differ by 3.5\%.}
\label{DOP_v}
\end{center}
\end{figure}

After precluding the unpolarized objects from the initial sample, a statistical analysis of the remaining data reveals the multimodal nature of the DoP distribution in all five photometric bands. In particular, in the upper panels of Fig. \ref{DOP_v}, we show the cumulative frequency and the histogram  of the DoP in the {\it V} band, and they clearly show that the statistical mass is concentrated in three distinctive regions. The first one includes the majority of the examined DoP values and concerns only weakly polarized sources [containing both very lowly (DoP$<$1\%) and lowly polarized sources (DoP$<$4\%)]. The distribution of this first region decreases exponentially with the DoP, and exhibits a strong accumulation close to DoP=0.7\%. The mean DoP of this region is estimated as low as 1.04$\pm$0.15\%. The other two regions are likely represented through a Gaussian profile. In particular, a clear mode appears at 6.3$\pm$0.4\% for the second region (the group of moderately polarized objects; hereafter group M1), while another mode appears at 13.9$\pm$0.9\% for the third region (the group of highly polarized objects; hereafter group M2). Any conclusion about the third region is doubtful due to the limited number of sources belonging to this group (three stars). The same behavior is found in the rest of the bands as well (see Fig.~5).

\begin{figure*}
\includegraphics[width=17cm]{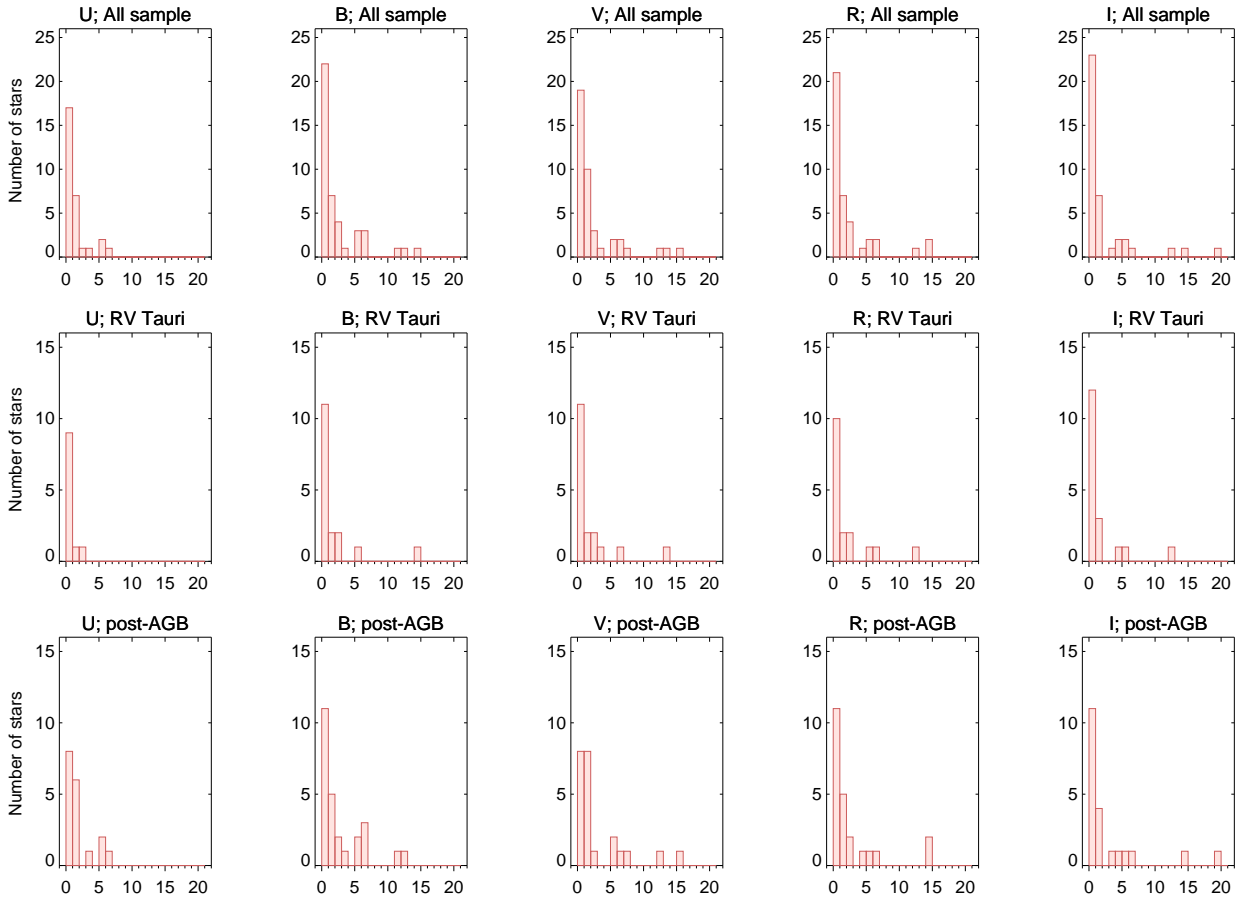}
 \caption{Histograms of DoP in each of the {\it U, B, V, R} and {\it I} broad-bands. From top to bottom, the upper panels correspond to the total sample, the middle panels to the RV Tauri stars and the lower panels to the post-AGB stars.  The size of the DoP bins are of 1\%.} 
 \label{histos_all}
\end{figure*}

\begin{figure}
\includegraphics[width=8cm]{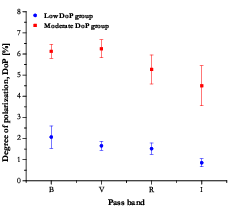}
 \caption{Mean DoP as a function of wavelength for the M1 (red dots) and M3 group (blue squares). This sort of
wavelength-dependennt polarization resembles the Rayleigh-type scattering spectrum.} 
 \label{fig_M1_M3}
\end{figure}

Although the sample of the moderately polarized group is not as large as the weakly polarized group (five and 33 objects, respectively), it is reliable enough to infer that the mode appearing at 6.3\% of DoP in the {\it V} band decreases to lower mode values at longer wavelength (see Fig.~6). This means that the moderately polarized group M1 is dominated by a Rayleigh-like scattering process, which is typical for non-symmetrical envelopes composed principally of small dust grains. Moreover, this wavelength-dependent polarization of the moderately polarized group M1 provides a strong evidence of an intrinsic origin. At shorter wavelengths, the contribution of the polar regions to the total polarized flux is higher than the contribution from the equatorial disk, because the optically thick disk blocks the scattered radiation, whereas at longer wavelengths, the disk is no longer optically thick and the scattered light can escape, resulting in a higher contribution to the total polarization. 

Note that only a few measurements had been obtained for the moderately polarized objects in the U band and, reasonably, they were omitted from our statistic analysis.

Taking into consideration the impact of the equipment in the validity of our measurements, a significant fraction of the weakly polarized objects might fall into the unpolarized category. To examine this possibility, we proceed to a 3-$\rm{\sigma}$ statistics -- with $\rm{\sigma}$ representing both the instrumental error and the error of the measurement itself -- attempting to isolate the objects that still remain polarized even after subtracting the possible false polarization induced by instrumental and/or observational effects. The re-analysis of our data shows that 70\% of the weakly polarized sources (23 out of 33) can potentially drift to the group of unpolarized, leaving only 10 objects unaffected. Given that this remaining subgroup of 10 weakly polarized objects (hereafter group M3), are expected to genuinely represent the properties of the low polarized stars. It is of interest to compare it  with moderate-polarized group M1.

In Fig.~\ref{fig_M1_M3}, we plot the mean values of DoP for the (moderate) M1 and (low) M3 groups. For the former group, the polarization degree descends progressively from  shorter  to longer wavelengths -- from  6.1$\pm$0.3\% and  6.3$\pm$0.4\% in the {\it B} and {\it V} bands to 5.3$\pm$0.7\% and 4.5$\pm$1.0\% in the {\it R} and {\it I} bands, respectively. For the latter group, we find a similar, but weaker, tendency -- decrease from 2.1$\pm$0.5\% and 1.66$\pm$0.20\% in the {\it B} and {\it V} bands, to 1.52$\pm$0.27\% and 0.86$\pm$0.19\% in the {\it R} and {\it I} bands, respectively.

To interpret the different strength of polarization between the M1 and M3 groups along with their wavelength-dependent behavior, we searched for any possible similarities in the fundamental physical parameters, such as effective and dust temperatures, among their stellar associates. We find that the {\it B -- V} index of the lowly polarized group gives an average color index of 1.43$\pm$0.11~mag with a standard deviation (SD) of 0.35~mag, which corresponds to a late K or early M spectra type star 
with an average temperature of 3850~K. Regarding the M1 group, we find an average {\it B -- V} color index of 1.82$\pm$0.35~mag with a SD of 0.77~mag, which corresponds to a M spectral type star with an average temperature of $\sim$3000~K. This implies that the moderately polarized objects tend to be redder compared to the lowly polarized objects.

Although the estimation of the dust temperature is beyond of the scope of this work, we may extract useful information by examining the distribution of the IR colour indices of the two groups. We find that both the {\it J -- H} and {\it H -- K} colour indices have no significant differences for the two examined groups. Regarding the first index, mean value of 0.64$\pm$0.05 and 0.72$\pm$0.08~mag were derived for both lowly and moderately polarized group, respectively, showing a common SD of 0.16~mag. As a result, the {\it J -- H} index can be considered almost the same for both groups within errors. As for the second index, mean value of 0.62$\pm$0.12 and 0.48$\pm$0.06~mag were estimated for the lowly and moderately polarized group, respectively, indicating a slight deviation between the groups. These values suggest a color temperature close to $\sim$2500 K for both groups, further implying that their stars share common dust temperatures.

\begin{figure*}
\includegraphics[width=14cm]{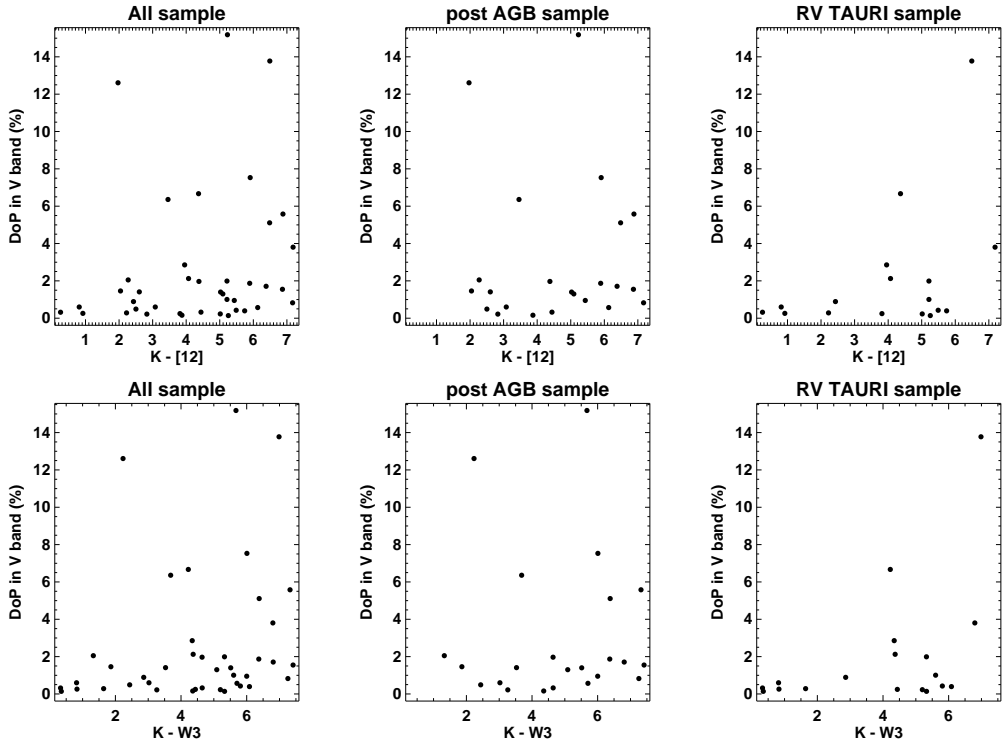}
 \caption{DoP in the {\it V} band as a function of the {\it K} -- [12] colour (upper panels) and {\it K -- W3} colour (bottom panels) for the whole sample, as well as the post-AGB and RV Tauri groups.} 
 \label{fig_DOP_K}
\end{figure*}

A statistical correlation between the DoP and the {\it K} -- [12] colour index has been reported for carbon and AGB stars (e.g. Bieging et al. 2006; Lopez \& Hiriart 2011a). Lopez \& Hiriart (2011a) show that DoP increases with {\it K} -- [12] colour index, which is associated with the mass-loss rate of the star. Our DoPs measurements as a function of  {\it K} -- [12] and {\it K -- W3} colour indices, (where {\it W3} is the WISE band at 11.6~$\micron$; Wright et al. 2010), are presented in Fig.~\ref{fig_DOP_K}. No clear correlation is found between the DoP and these two colour indices, probably due to the complex evolution of the post-AGB stars. In particular, the {\it K} -- [12] colour index has been found to change significantly during the evolution of these objects covering a range of values from 2 to 8~mag (see Fig.~12 in van Hoof, Oudmaijer \& Waters 1997). At the beginning of the post-AGB phase, the central star is not hot enough to ionize the circumstellar gas and, as it evolves, the {\it K} -- [12] colour index decreases. When the star becomes hot enough to ionize the circumstellar gas, {\it K} -- [12] increases. Hence, the {\it K} --[12] colour decreases with time up to a certain value and then starts increasing again. On the other hand, the mass-loss rate, which is a crucial parameter that affects the evolution of post-AGB stars, decreases with time (e.g. van Hoof et al. 1997). Therefore, the {\it K} -- [12] (or/and {\it K -- W3}) colour index may not be a good indicator to describe the mass-loss rate for post-AGB stars. In addition, the molecular hydrogen emission lines, which some post-AGB show in the {\it K} band, can be one of the causes that makes the DoP vs. {\it K} -- [12] relation dubious. Unlike post-AGB stars, carbon and AGB stars do not show any H$_2$ emission lines, making the DoP vs. K-[12] relation perceptible. Regarding the M1 and M3 groups, we found substantial different {\it K -- W3} colour indices. In particular, the average {\it K -- W3} colour indices of the M1 and M3 groups are calculated as 5.52$\pm$0.68~mag and 4.78$\pm$0.38~mag, respectively. This indicates a probable increase of polarization with the mass-loss rate. Yet, these two numbers can be considered the same within the errors.

\begin{figure*}
\includegraphics[width=14cm]{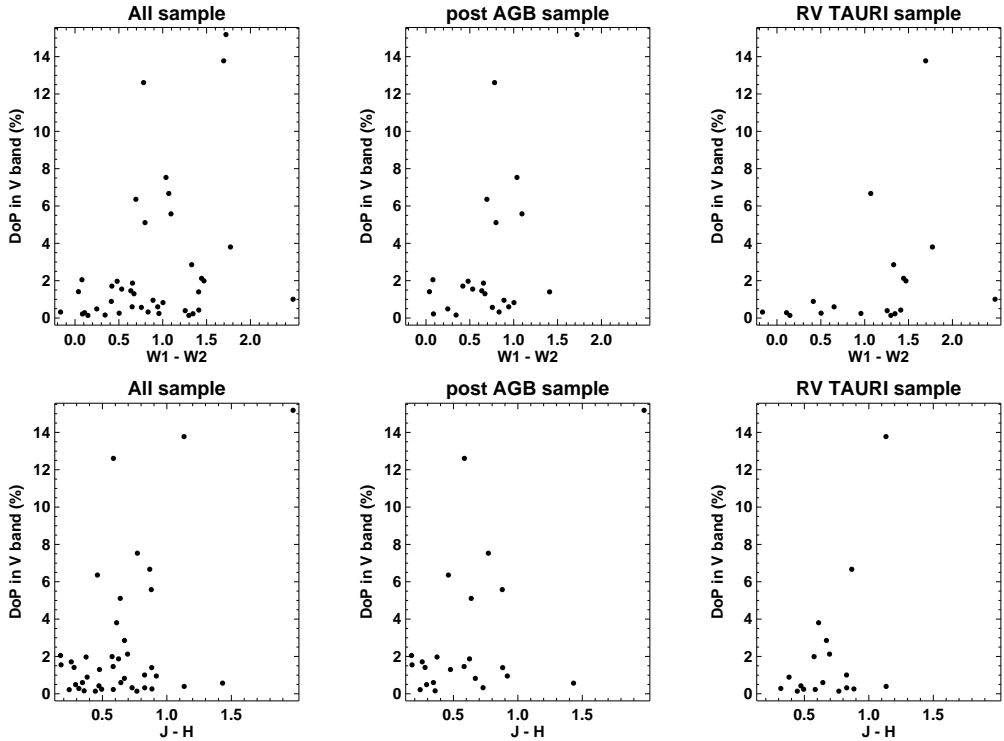}
 \caption{DoP in the {\it V} band as a function of the WISE {\it W1 -- W2} colour (upper panels) and 2MASS {\it J -- H} colour (bottom panels) for the whole sample, as well as the post-AGB and RV Tauri groups.} 
 \label{fig_DOP_W}
\end{figure*}

Messineo et al. (2012) show that the GLIMPSE [3.6] -- [4.5]\footnote{The IRAC camera on board the {\it Spitzer Space Telescope} provides the magnitudes in the four GLIMPSE filters centred at 3.6, 4.5, 5.8, and 8.0~$\mu$m (Fazio et al. 2004).} of late-type stars increases with the amplitude of the pulsation, whereas it can also be used as an indicator of the mass-loss. The analogy between the GLIMPSE [3.4] -- [4.5] and the WISE {\it W1 -- W2} colour indices ({\it W1} and {\it W2} are the WISE bands at 3.35~$\micron$ and 4.6~$\micron$, respectively) allow us to use the latter as an indicator for the mass-loss rate. As Fig.~\ref{fig_DOP_W} demonstrates, a weak tendency of an increase in the {\it W1 -- W2} colour index with DoP is found. On average, however, the groups M1 and M3 are found to exhibit the same {\it W1 -- W2} colour index of 0.94$\pm$0.11 and 1.14$\pm$0.20~mag, respectively.

A significant difference between the two groups is found in the {\it W1 -- W3} colour index, though. In particular, the M1 group has an average {\it W1 -- W3} colour index of 4.945$\pm$0.70~mag, whereas the M3 group of lowly polarized object shows an average value of 3.94$\pm$0.39~mag. Overall, the moderately polarized objects (M1 group) are systematically redder than the lowly polarized group (M3 group).  In conclusion, the moderately polarized stars (M1) exhibit higher {\it K -- W3} and {\it W1 -- W3} than the lowly polarized stars. This may be associated  with the mass-loss rate (Bieging et al. 2006; Lopez \& Hiriart 2011a) and/or the central star. In particular, the M1 group exhibits a higher {\it B-V} colour index than the M3 group, which mean that colder stars have larger amount of dust resulting in high polarization.

Concerning the circumstellar envelope, two-thirds of our sample appear to exhibit a nearly spherical structure (DoP$<$1\%). This number is significantly higher than the percentage of spherical PNe. Morphological classifications of PNe have shown that the majority of PNe ($\sim$75\%) have an aspherical morphology, and only $\sim$25\% are spherically symmetric (Manchado et al. 1997, Manchado 2004; Parker et al. 2006). Similar percentages have also been found in the IPHAS\footnote{ Isaac Newton Telescope Photometric $\rm{H\alpha}$ Survey of the Northern Galactic plan (Drew et al. 2005).} catalogue of 159 new PNe in the Northern Galactic plane (Sabin et al. 2014). The causes that may be responsible to this discrepancy are: (i) the dilution of polarized flux from the CSE resulting in a low net aperture DoP, due to the direct observation of the bright unpolarized star, (ii) the number of unpolarized objects is biased because of mixing with objects with very low DoP ($\sim$1\%), which may have formed a somewhat aspherical envelope (e.g. elliptical structure) and (iii) a possible over-subtraction of the ISP. 

By scrutinizing the objects with DoP$<$1\%, we found five likely very lowly polarized stars with DoP very close to 1\%, rather than unpolarized (e.g. IRAS 07008+1050, IRAS 21546+4721). This shortens the number of unpolarized stars to 18 or 34\%, which is very close to the percentage of round PNe from statistical studies.

The histograms of the DoP for each band and the two subgroups of post-AGB and RV Tauri stars are 
presented in Fig.~\ref{histos_all}. From 21 RV Tauri stars in our sample, only three show a significant level of polarization (DoP$>$3\%). The distribution of PA of the polarization vectors of our sample in the {\it V} band is presented in Fig.~4. All the objects with a DoP$>$3\% exhibit a PA between 0 and 60\degree\ and 120 and 180\degree. This provides a further evidence for the intrinsic origin of polarization, taking into account that the PA of interstellar polarization is distributed around 90\degree\ (see Fig.~\ref{DOP_v}). 

Nine objects in our sample are known binary systems (see Table~1), 15 have been classified as candidate binary system due to their SED characteristics (De Ruyter et al. 2006) and six are known single stars, whereas there is no available information for the rest 23 objects. Interestingly, five out of the nine known binary systems are found to be very lowly polarized or unpolarized objects. Even though binary systems are expected to be responsible for the deviation of CSE from spherical symmetry, this result implies that most of them likely have a spherically symmetric envelope or they are seen face-on. For the face-on disk or toroidal structure, the net aperture DoP is expected to be zero or even very low due to the cancellation of the perpendicular polarization vectors in centrosymmetric pattern. Moreover, the direct unpolarized light from the central star also dilutes the polarized light emanating from the CSE, resulting in a lower DoP. This can be why some objects have highly polarized polar outflows and very low aperture DoP (e.g. IRAS Z02229+6208, IRAS 04296+3429 and IRAS 17436+5003). 

Three of the known single stars are very lowly polarized or unpolarized, two are lowly polarized, and one (IRAS 04296+3429) is highly polarized. The latter has a multi-polar morphology (Sahai 1999; Ueta et al. 2000, 2005, 2007) and in conjunction with the wavelength-dependent polarization found in this object, we claim that it is very likely a binary system but further investigation is required to confirm this.

Three objects show DoP$>$8\% (highly polarized): IRAS 09371+1212, IRAS 19343+2926 and IRAS 20056+1834. Two are PPNe (IRAS 09371+1212, IRAS 19343+2926) and one is a variable RV Tauri star (IRAS 20056+1834). All of them have a bipolar or multi-polar envelope, likely with a circumstellar torus or disk. Multiple scattering photons in the optically thin polar outflows/envelopes of these objects can explain the high DoPs through the decrease of photons that escape from the dusty equatorial region (Wood et al. 1996a,b; Hoffman et al. 2003).

Regarding the wavelength dependence of polarization, we found 10 cases where the DoP increases towards shorter wavelengths, in contrary to only three cases (IRAS 09371+1212, IRAS 11472--0800 and IRAS 19343+2926) where the DoP increases towards longer wavelengths and just one post-AGB (IRAS 18564--0814) with a more complex profile (see Fig.~A1). All other sources exhibit a nearly flat wavelength-dependence within their errors.

\section{Conclusions}
 Optical, multi-band polarization measurements of 53 post-AGB stars were presented. 24 of them were observed and presented for the first time. DoP was found to vary mainly from 0\% up to 8\%, whereas three objects were found to be highly polarized with DoP up to 16\%. Overall, 66\% of our sample was found to be polarized due to scattering in the surrounding aspherical dusty CSEs. This number is close to the number of aspherical PNe from previous statistical studies. The very high DoP found in some bipolar or multi-polar outflows is very likely due to the multiple scattering of photons in the optical thin polar outflows.

A detailed statistical analysis of all available data revealed the multimodal nature of the DoP distribution of the examined post-AGB stars in all five photometric bands. Two distinctive groups of low (DoP$\sim$2\%) and moderate (DoP$\sim$6\%) polarization are clearly present at all available bands, with their mode shifting toward lower DoP values as the wavelength increases. The scattering spectrum of both these two groups was found to have a similar Rayleigh-like form, possibly reflecting common dust properties of the envelopes that surround the post-AGB stars. The moderately polarized group seems to contain on average cooler central stars (or higher {\it B-V}) with respect to the lowly polarized group; however this does not hold for the dust grain temperatures, which appear almost the same for these two groups. Subsequently, the condensation process in the moderately polarized group may occur closer to the central star, lending strong support to their higher DoP values as the result of the formation of denser dusty envelopes.
Moreover, the moderately polarized stars appear to have systematically higher {\it K--W1} and {\it W1--W3} colour indices compared to the lowly polarized stars. This may indicate a possible relation between DoP and mass-loss rate of the stars. 

\section*{Acknowledgments}
We thank the anonymous referee for her/his thorough review. We highly appreciate the comments and suggestions, which significantly contributed to improving the quality of the publication. We also thank to L. Sabin for her comments that helped to improve the manuscript and his content. This work is based upon observations carried out at the Observatorio Astron\'{o}mico Nacional on the Sierra San Pedro M\'{a}rtir (OAN-SPM), Baja California, M\'{e}xico. 
We also thank the daytime and night support staff at the OAN-SPM for facilitating and helping obtain our observations. S.A. is supported by the Brasilian CAPES post-doctoral fellowship \lq\lq Young Talents Attraction" - Science Without Borders (grant \mbox{A035/2013}). J.R.V. and D.H. acknowledge the CONACyT financial grant numbers 240441 and 180817, respectively. N.N. gratefully acknowledges financial support under the \lq\lq MAWFC" project. Project MAWFC is implemented under the \lq\lq ARISTEIA II" action of the Operational Programme \lq\lq Education and Lifelong Learning". The project is co-funded by the European Social Fund (ESF) and National Resources. 
D.P. acknowledges support from Brazilian agency FAPESP (grant \mbox{2013/16801-2})

\bibliographystyle{mnras}

\newpage

\appendix

\section{Wavelength dependence of the linear degree of polarization and position angle}
The degree of polarization (upper panels), and the position angle of the polarization vector (lower panels) as 
a function of wavelength for the {\it U, B, V, R} and {\it I} bands are given for all the members of our sample.
 
\begin{figure*}
\begin{center}
\vspace{-0.25cm}
\includegraphics[width=8cm]{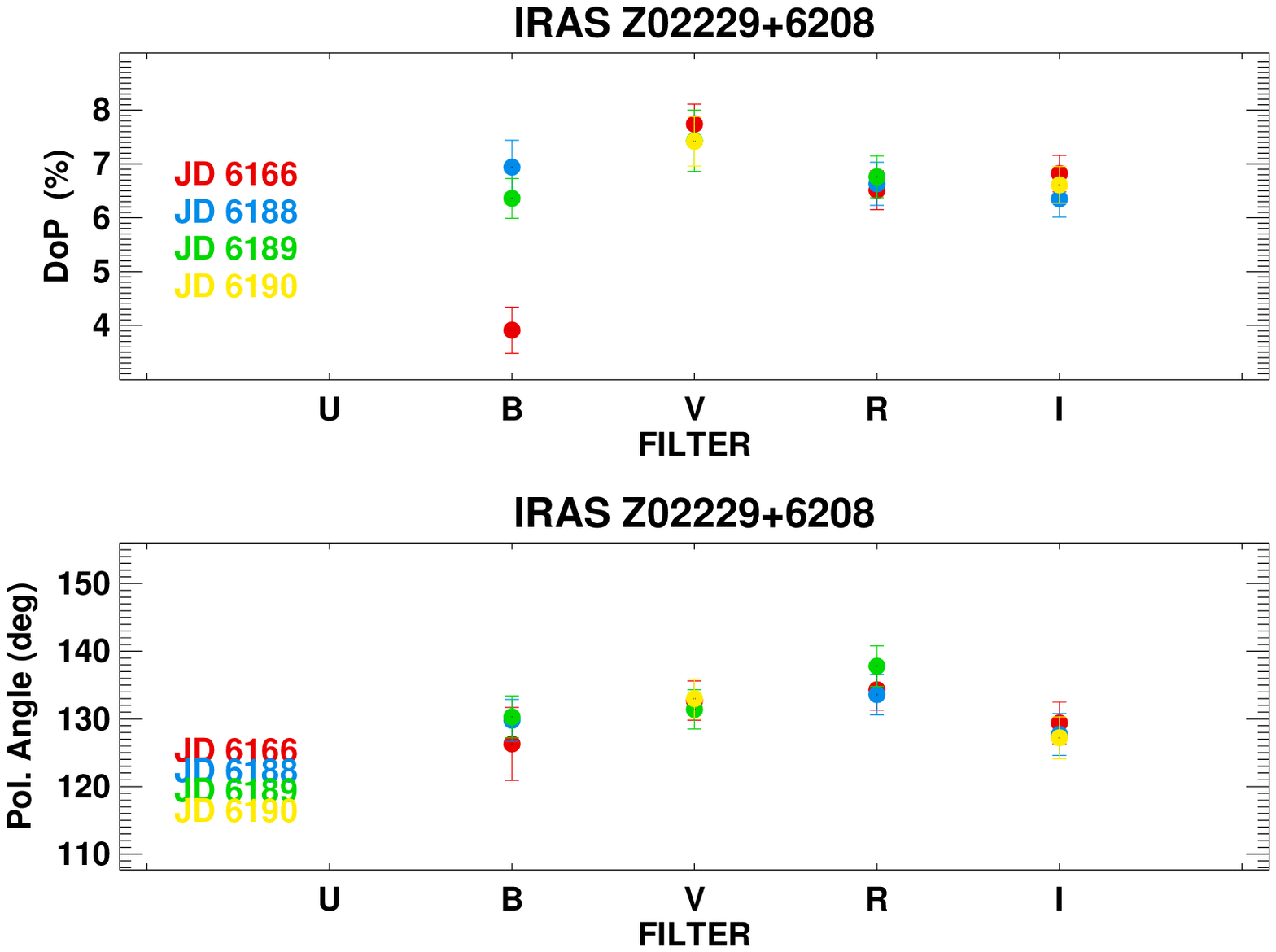}\includegraphics[width=8cm]{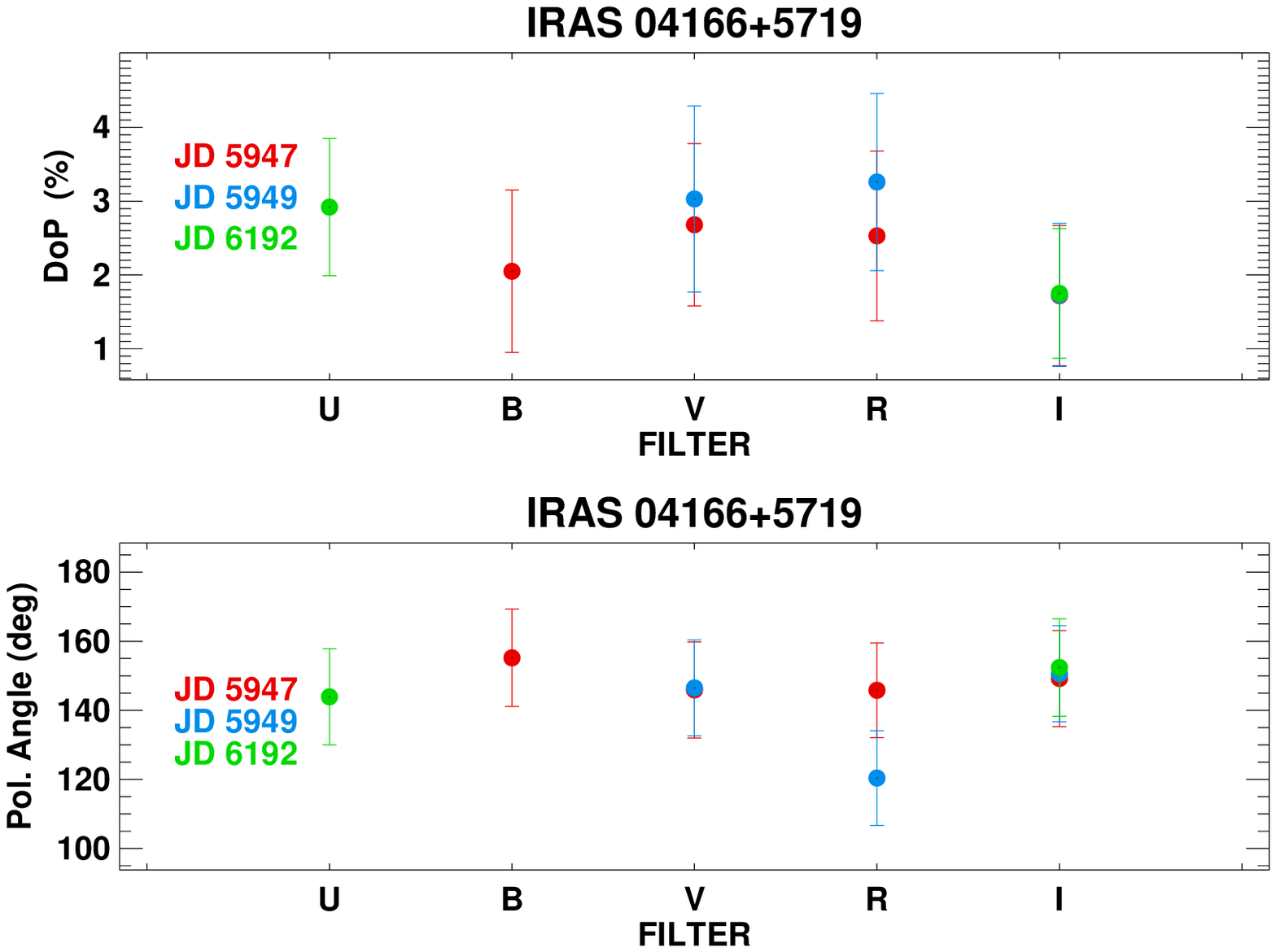}
\includegraphics[width=8cm]{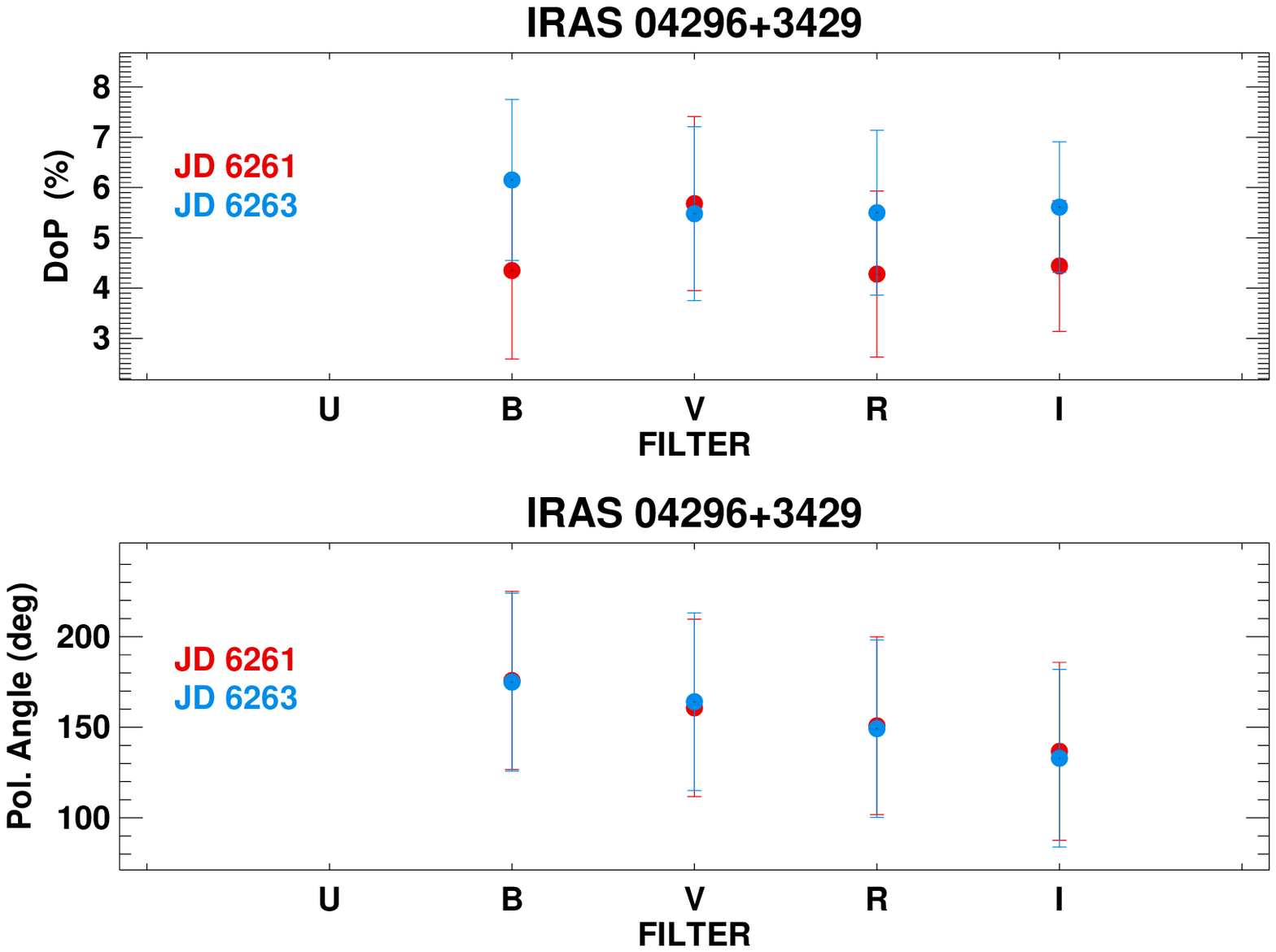}\includegraphics[width=8cm]{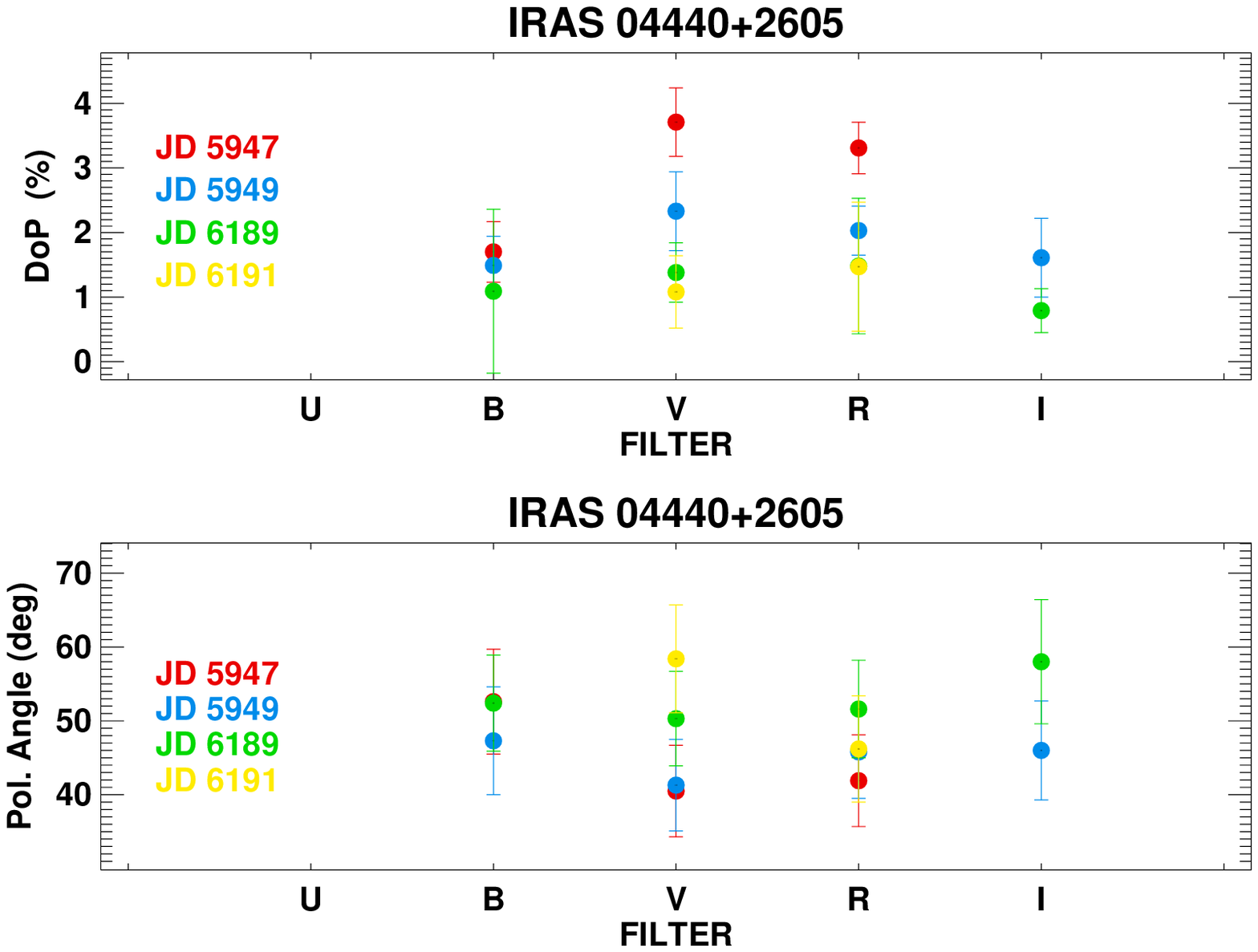}
\includegraphics[width=8cm]{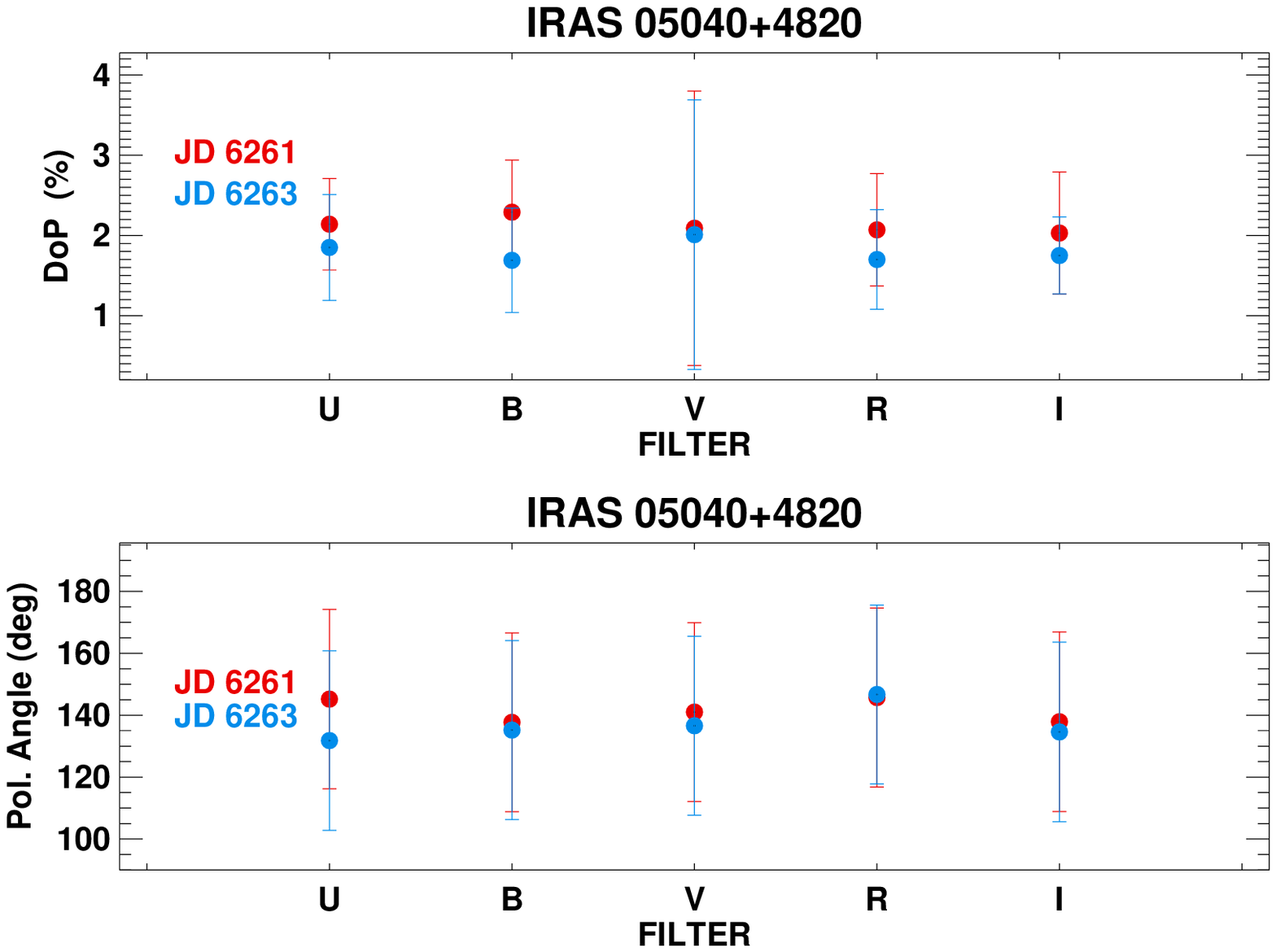}\includegraphics[width=8cm]{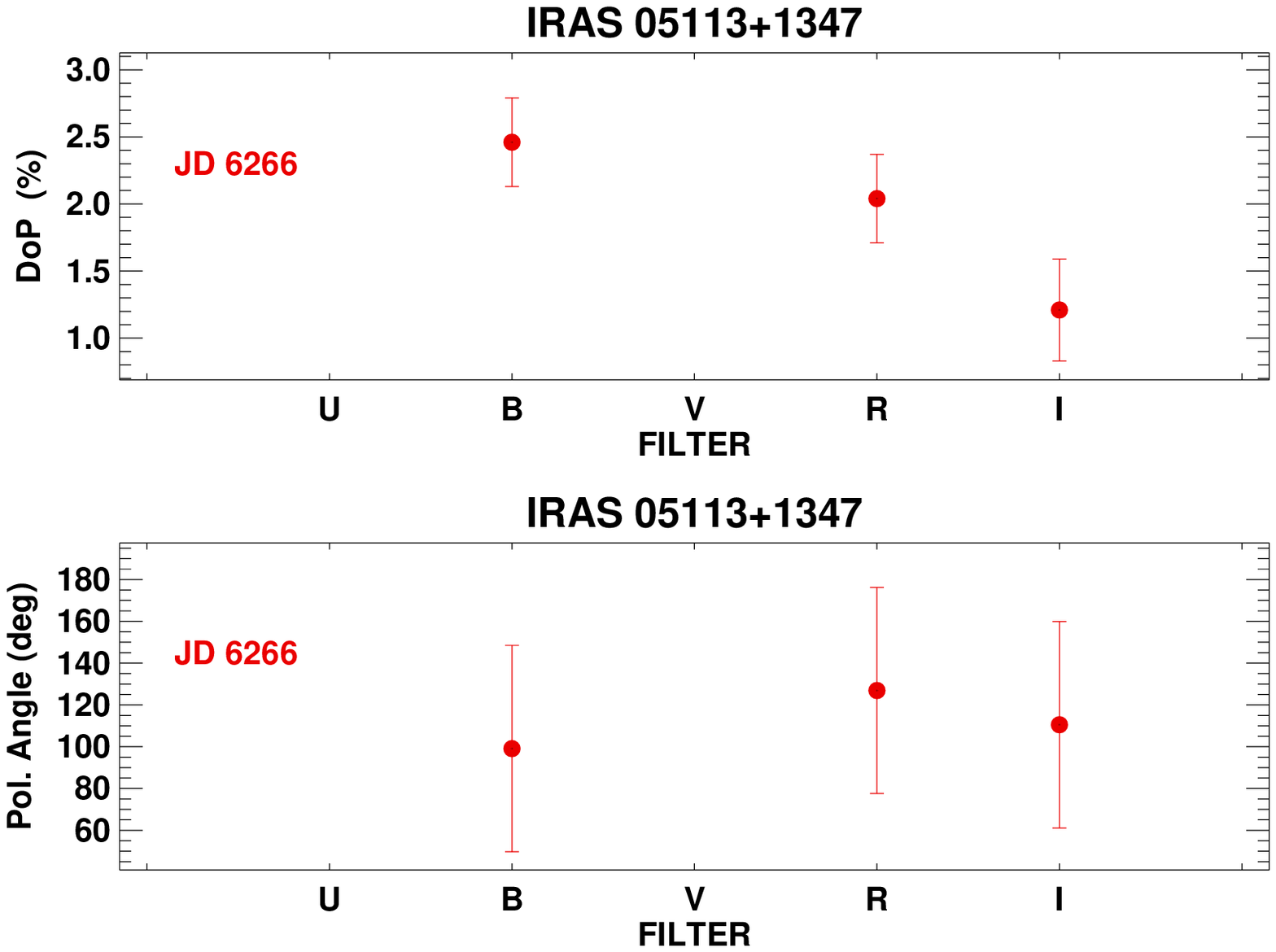}
\includegraphics[width=8cm]{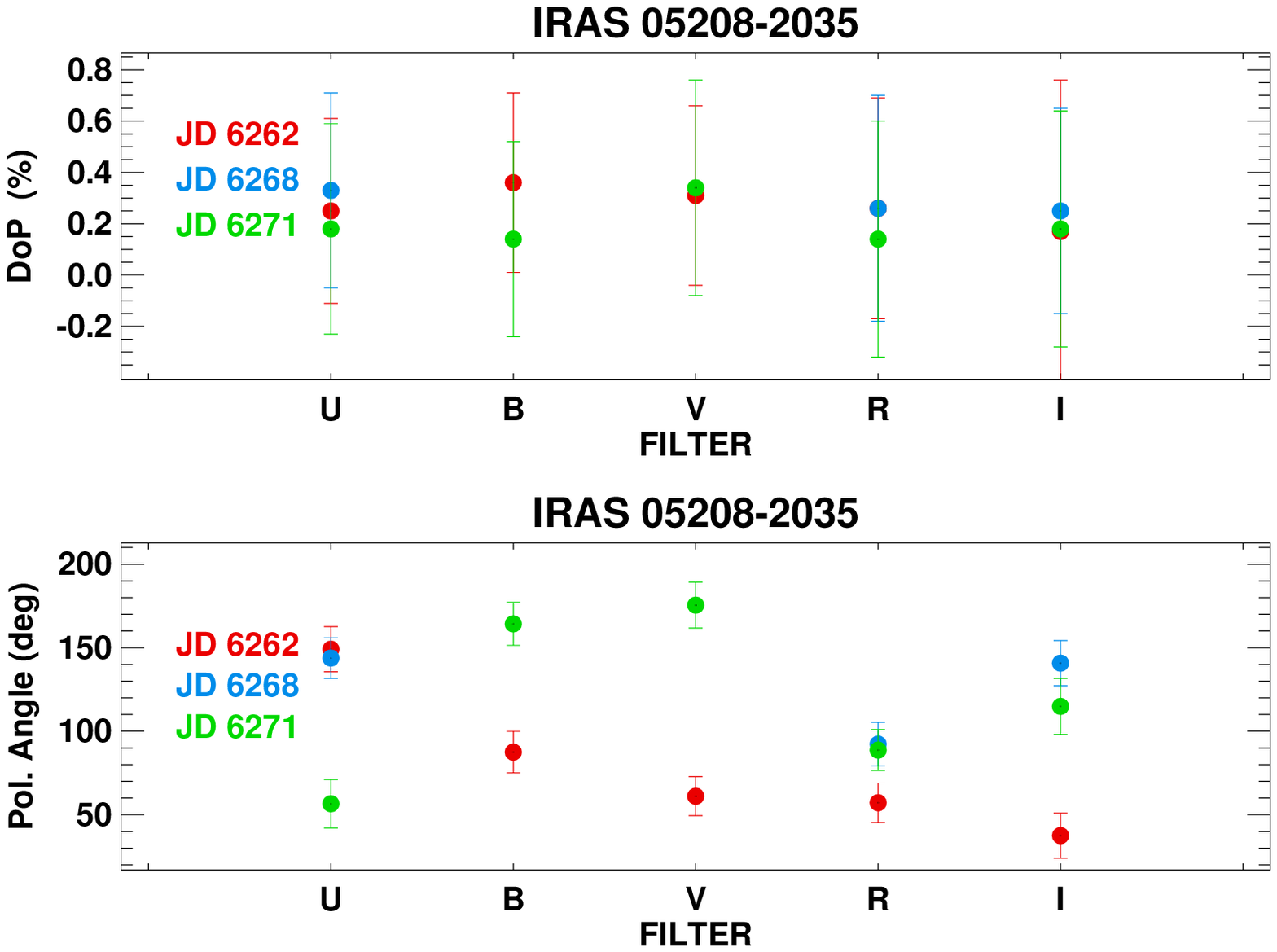}\includegraphics[width=8cm]{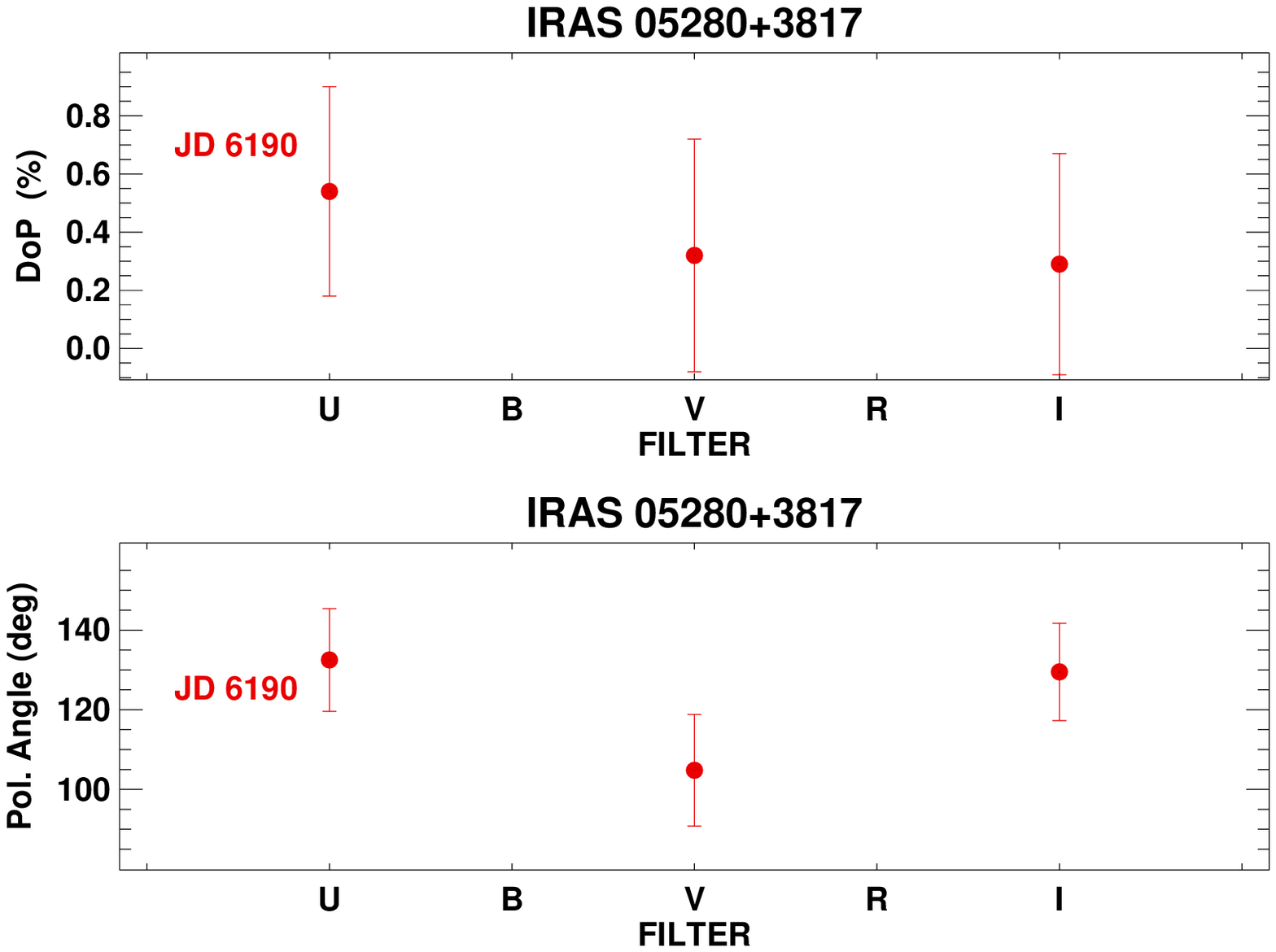}
\caption{Polarimetric observations of the programme stars. Different colours correspond to 
different observing nights. (A color version of this figure is available in the online journal). }
 \label{fig3}
\end{center}
\end{figure*}

\begin{figure*}
\addtocounter{figure}{-1}
\begin{center}
\vspace{-0.25cm}
\includegraphics[width=8cm]{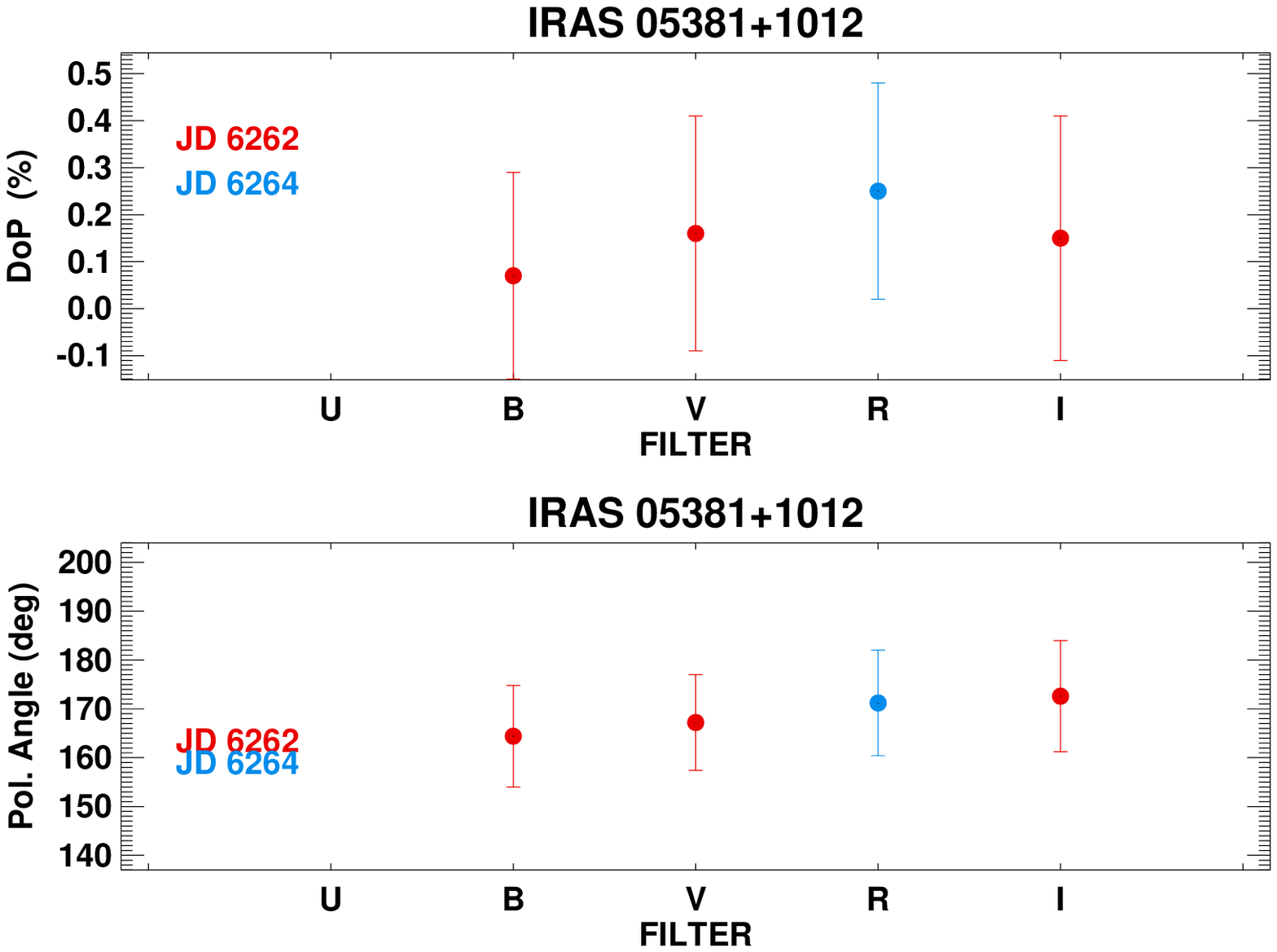}\includegraphics[width=8cm]{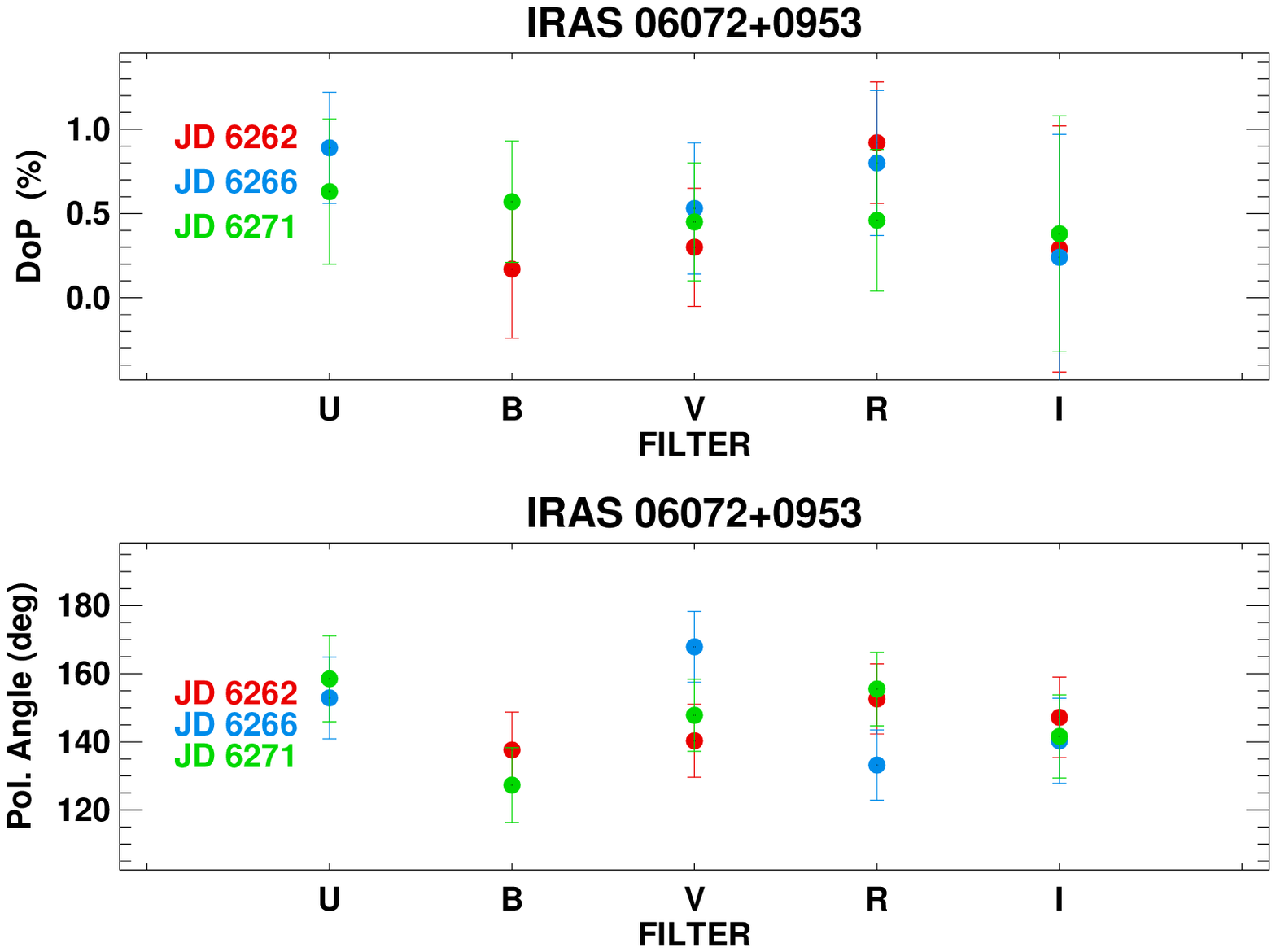}
\includegraphics[width=8cm]{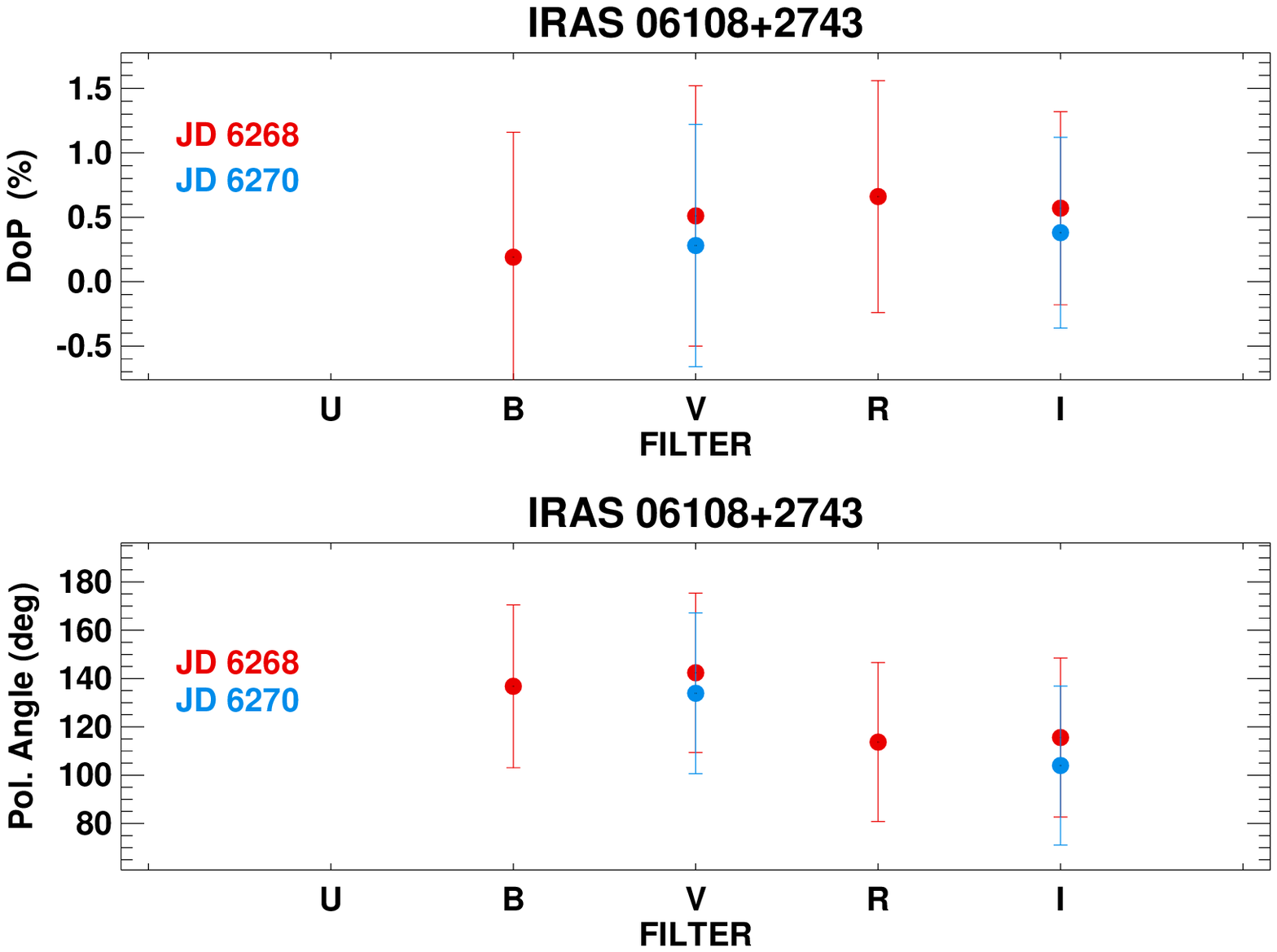}\includegraphics[width=8cm]{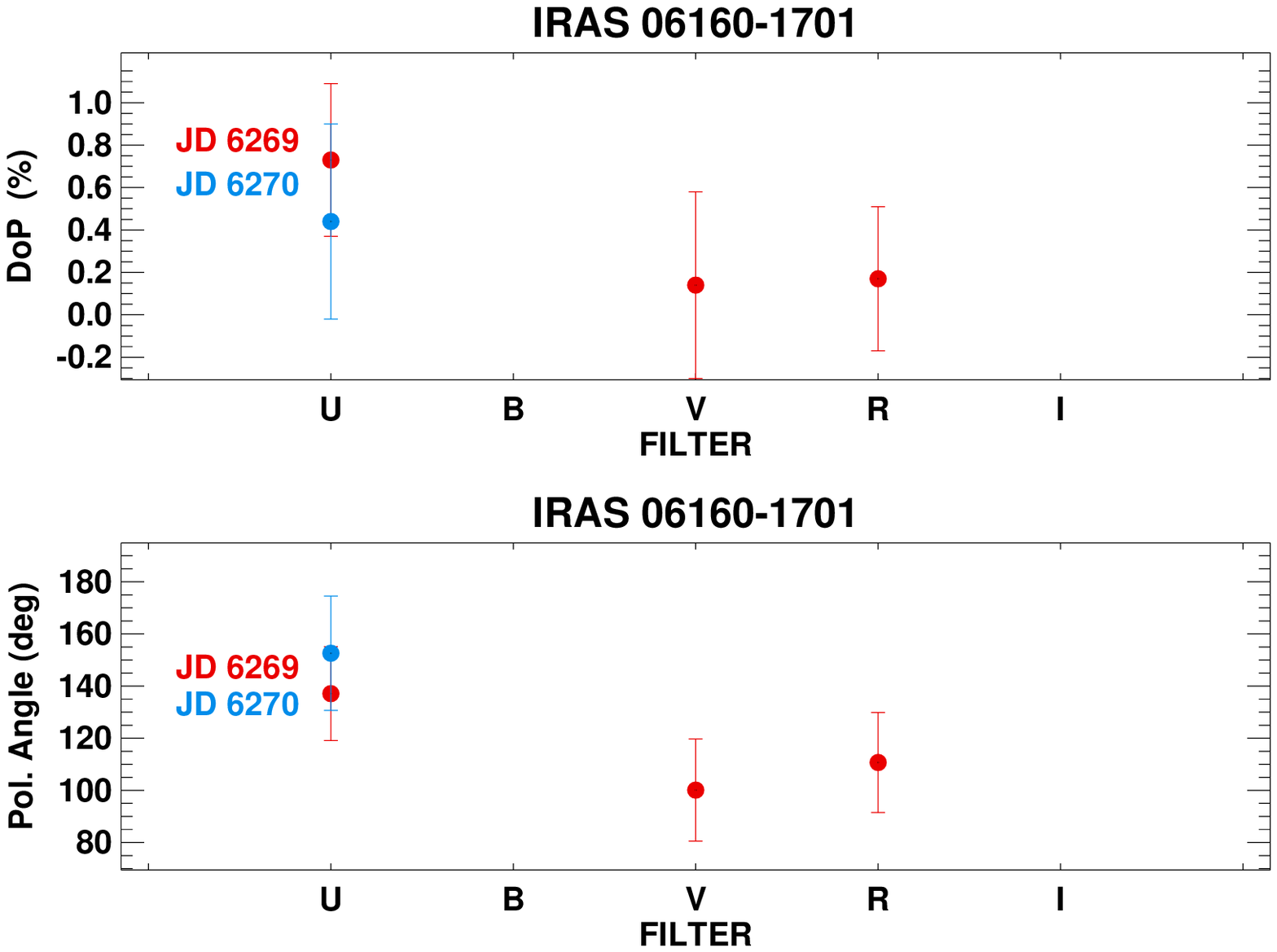}
\includegraphics[width=8cm]{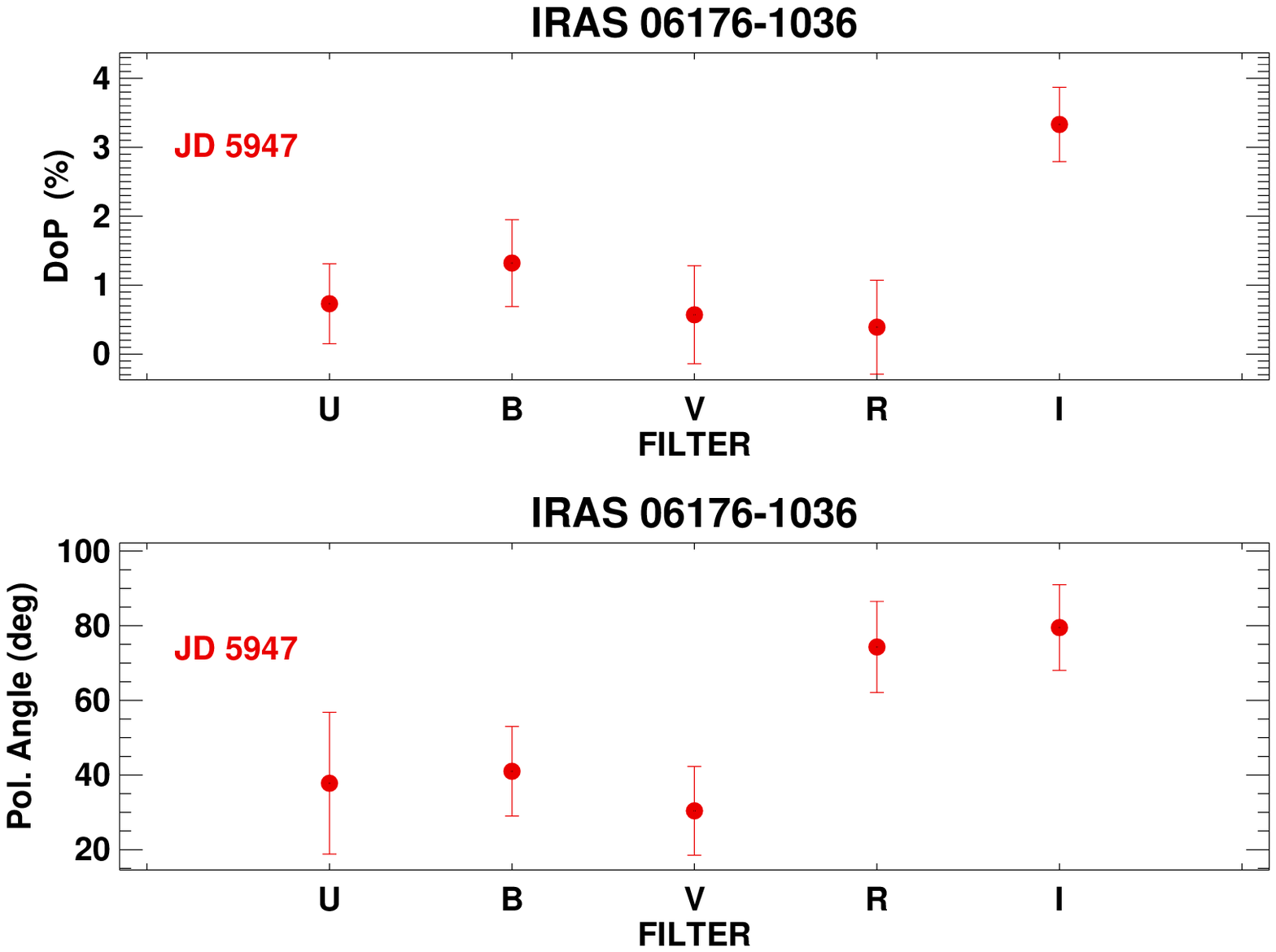}\includegraphics[width=8cm]{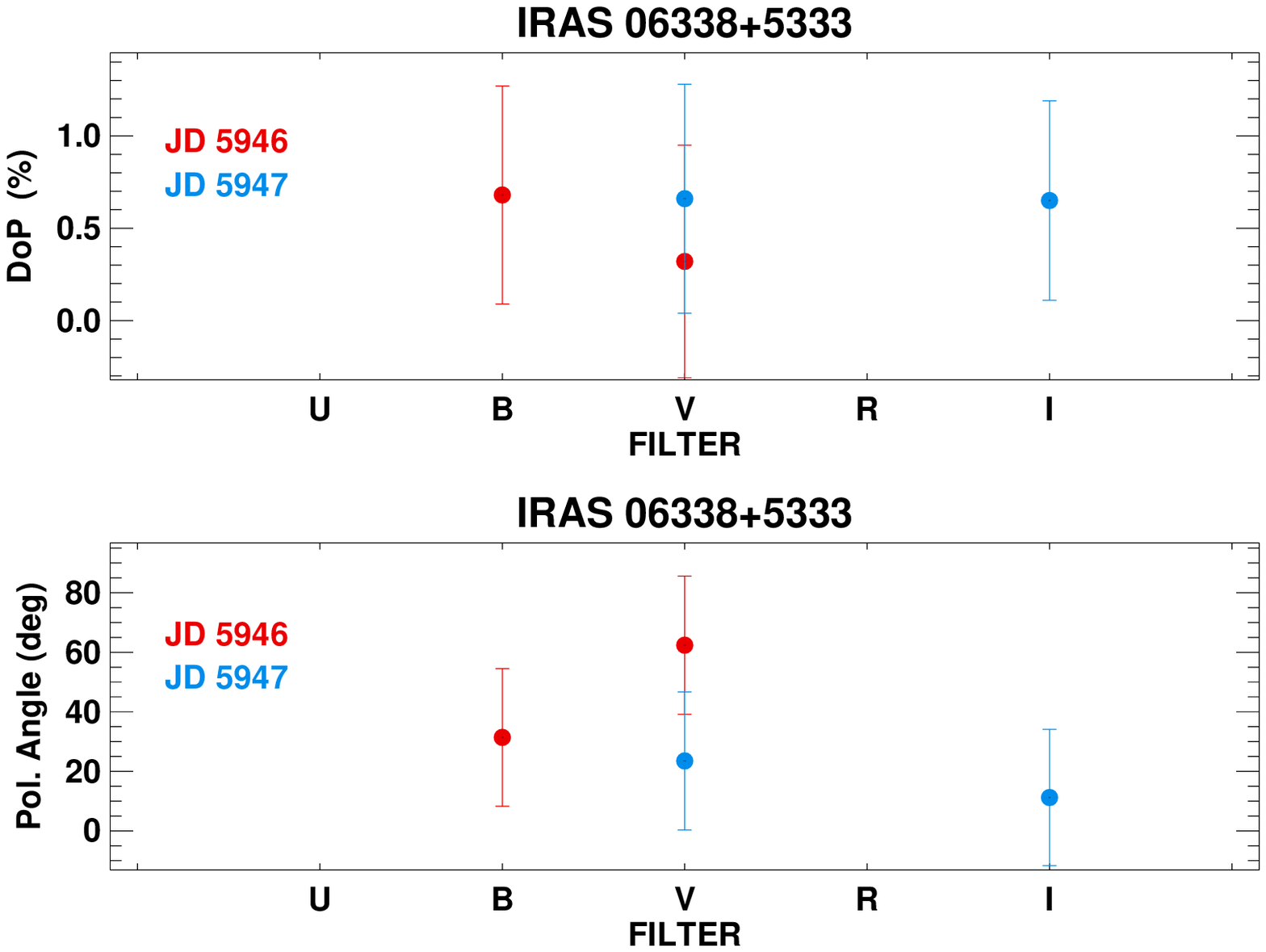}
\includegraphics[width=8cm]{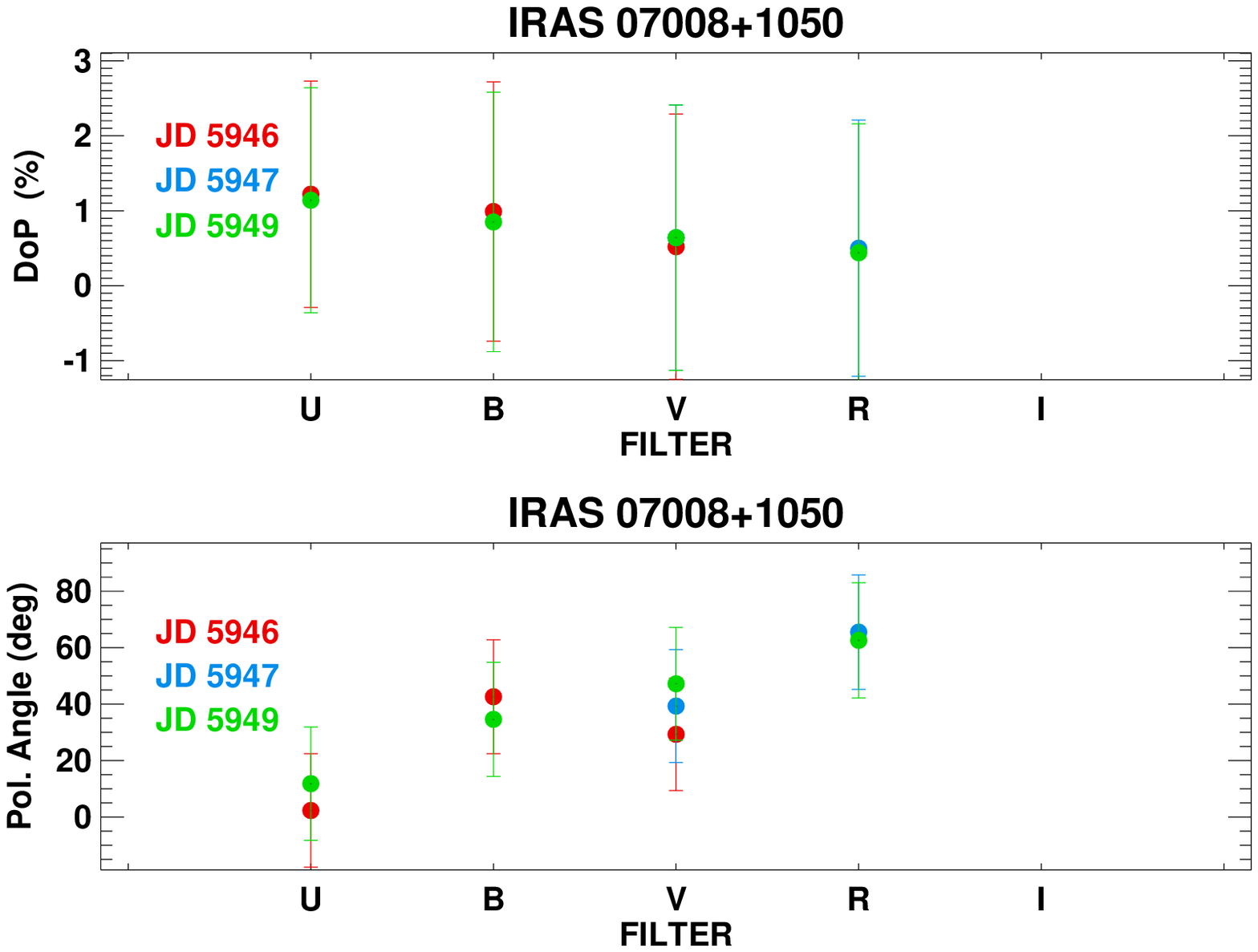}\includegraphics[width=8cm]{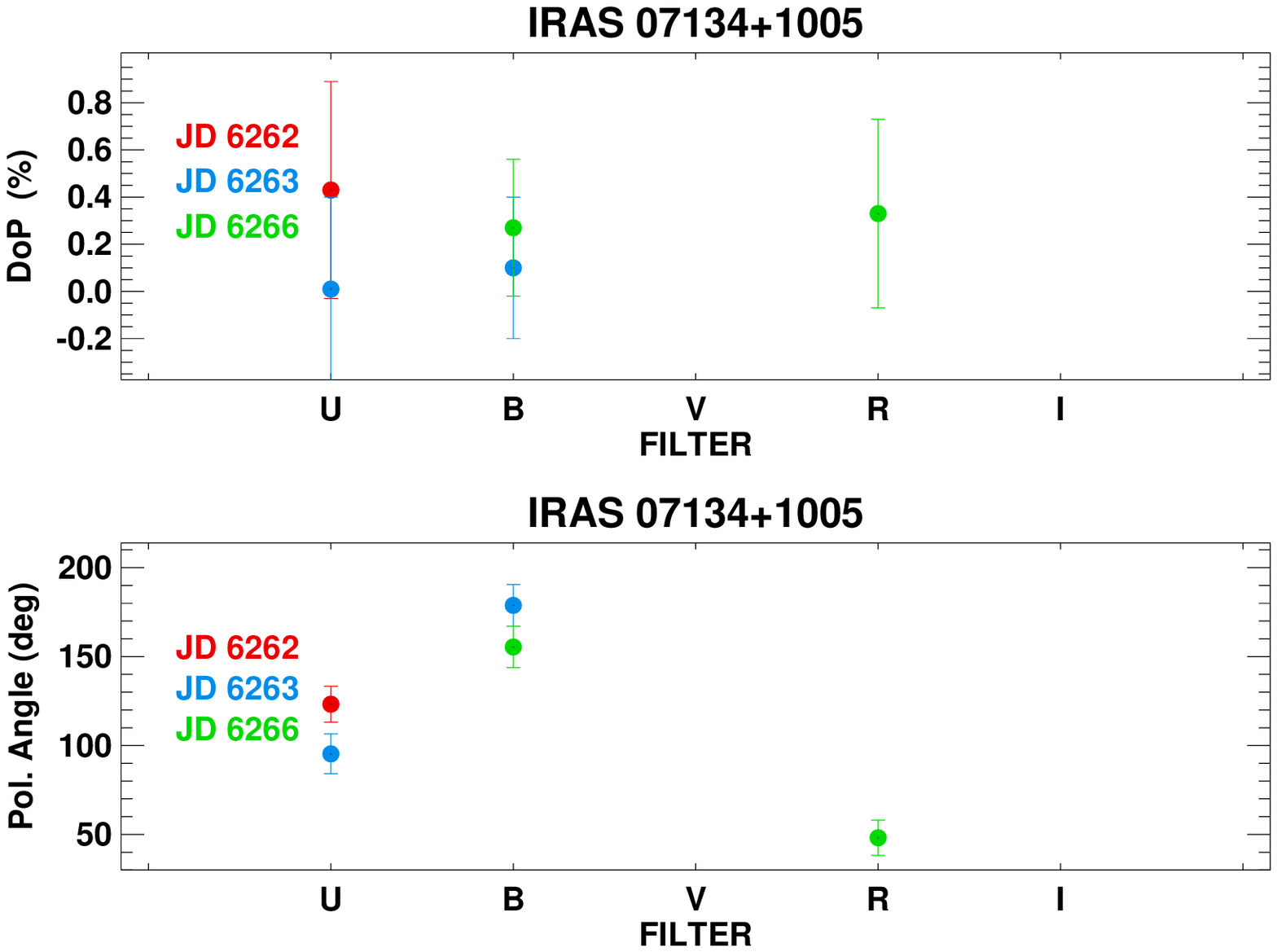}
\caption{continued}
 \label{fig4}
\end{center}
\end{figure*}

\begin{figure*}
\addtocounter{figure}{-1}
\begin{center}
\vspace{-0.25cm}
\includegraphics[width=8cm]{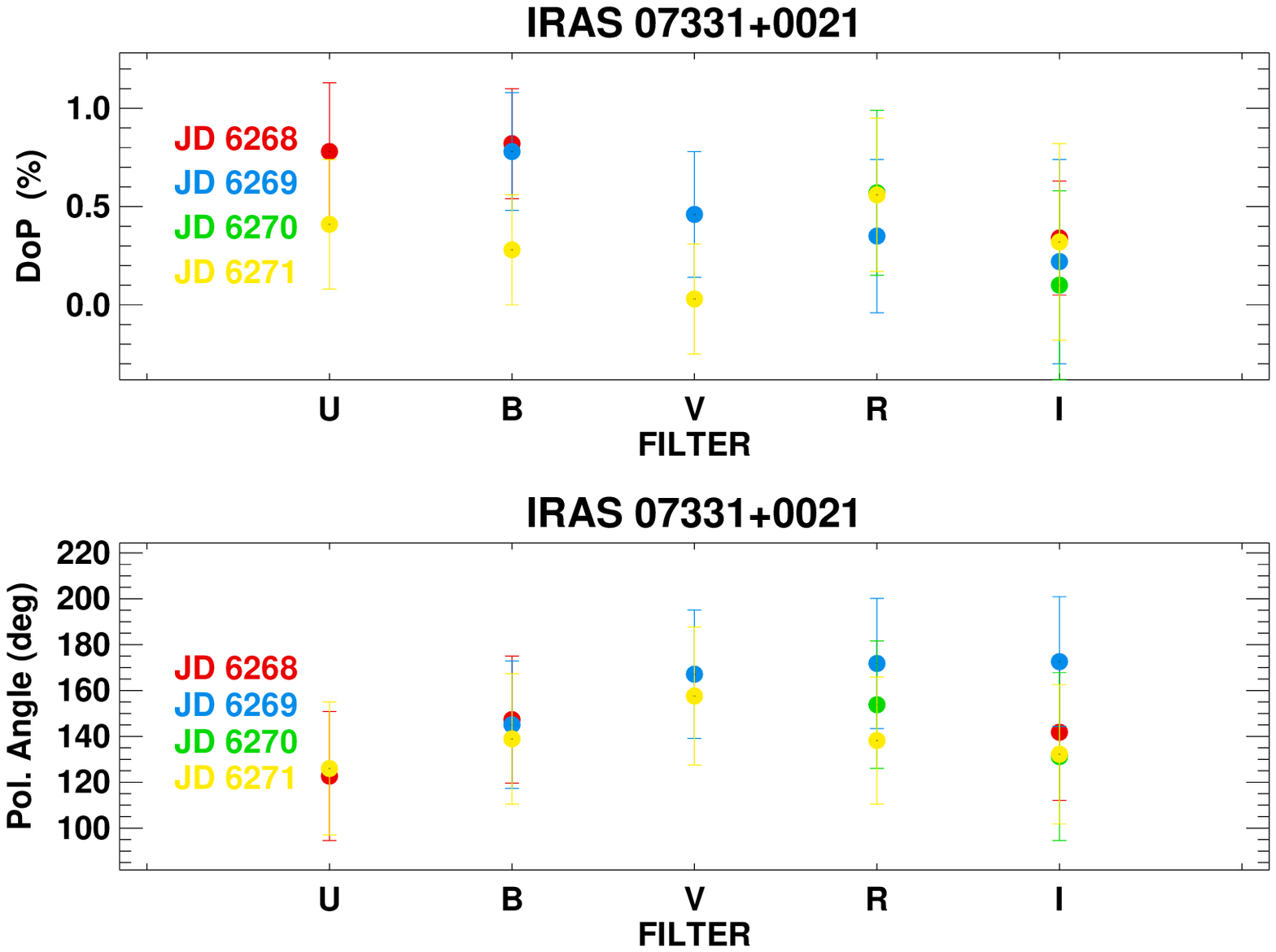}\includegraphics[width=8cm]{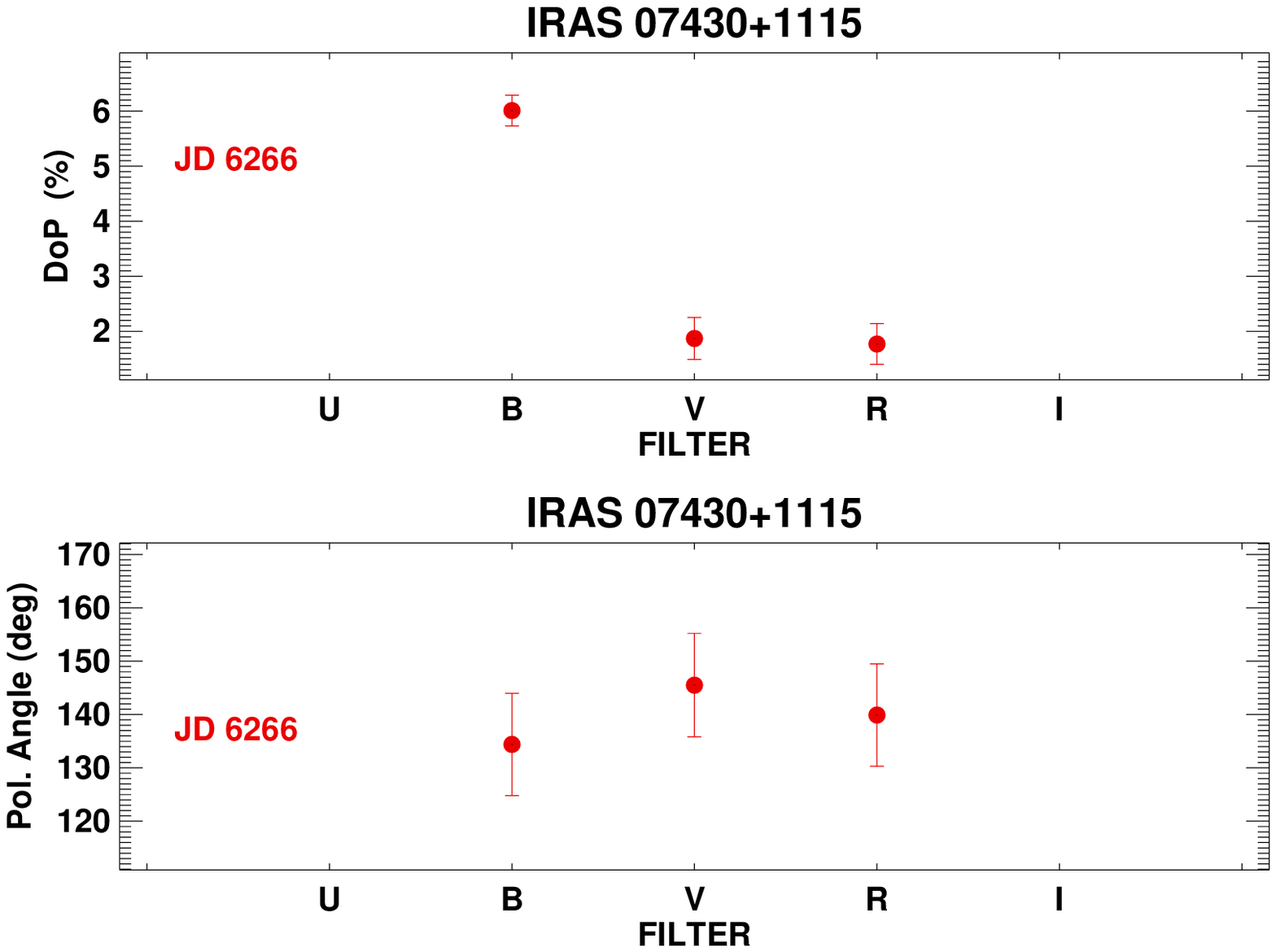}
\includegraphics[width=8cm]{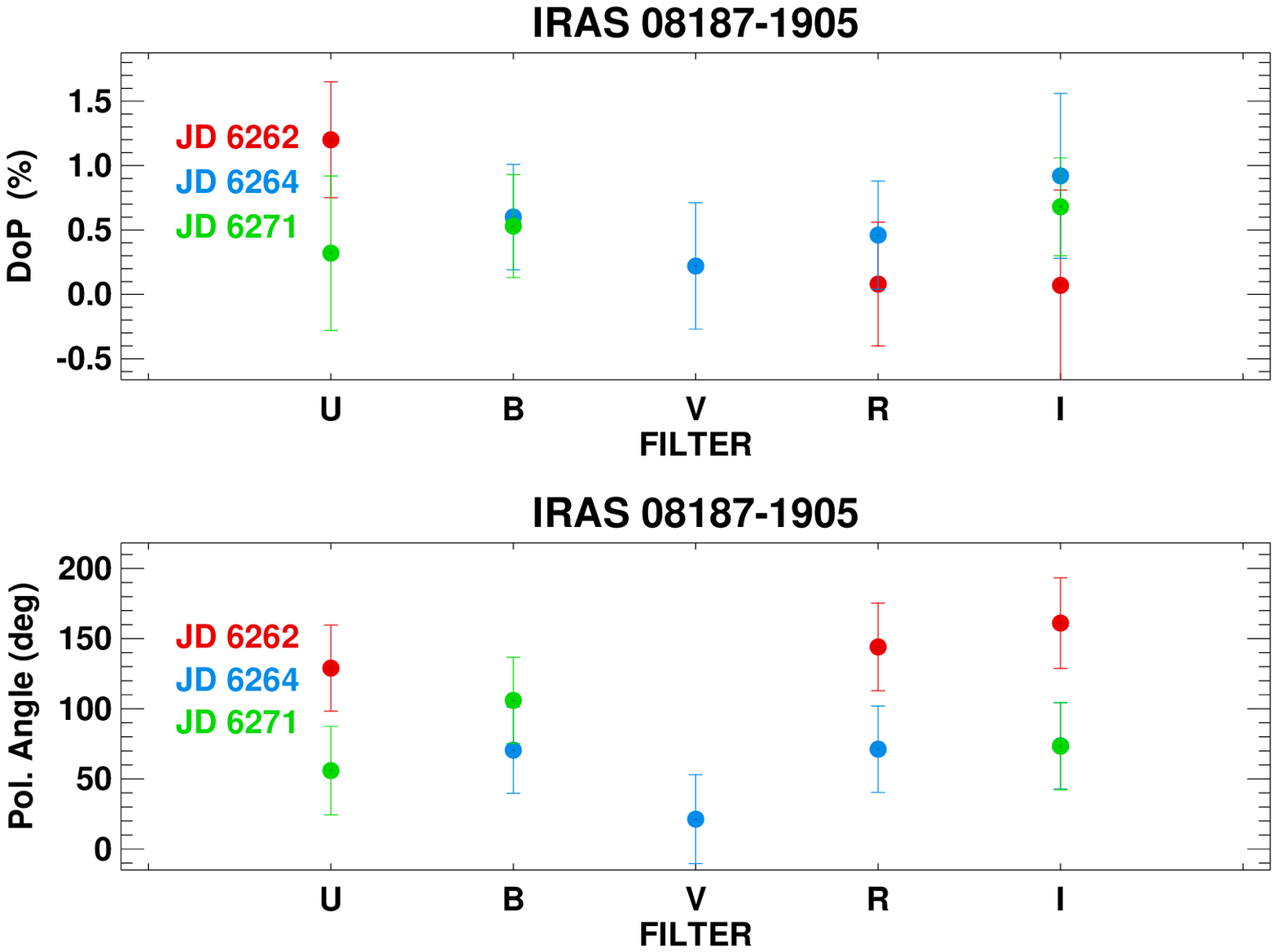}\includegraphics[width=8cm]{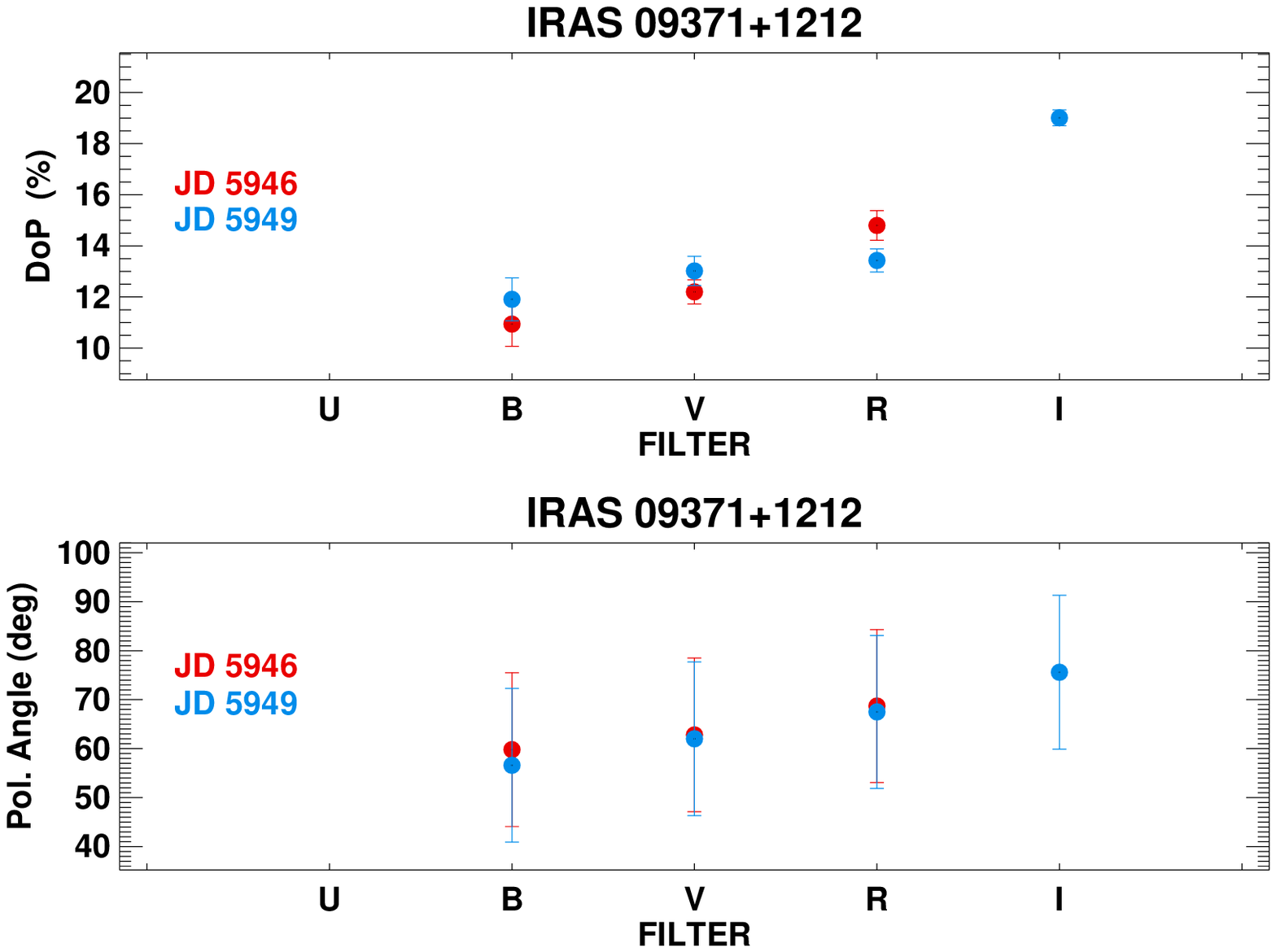}
\includegraphics[width=8cm]{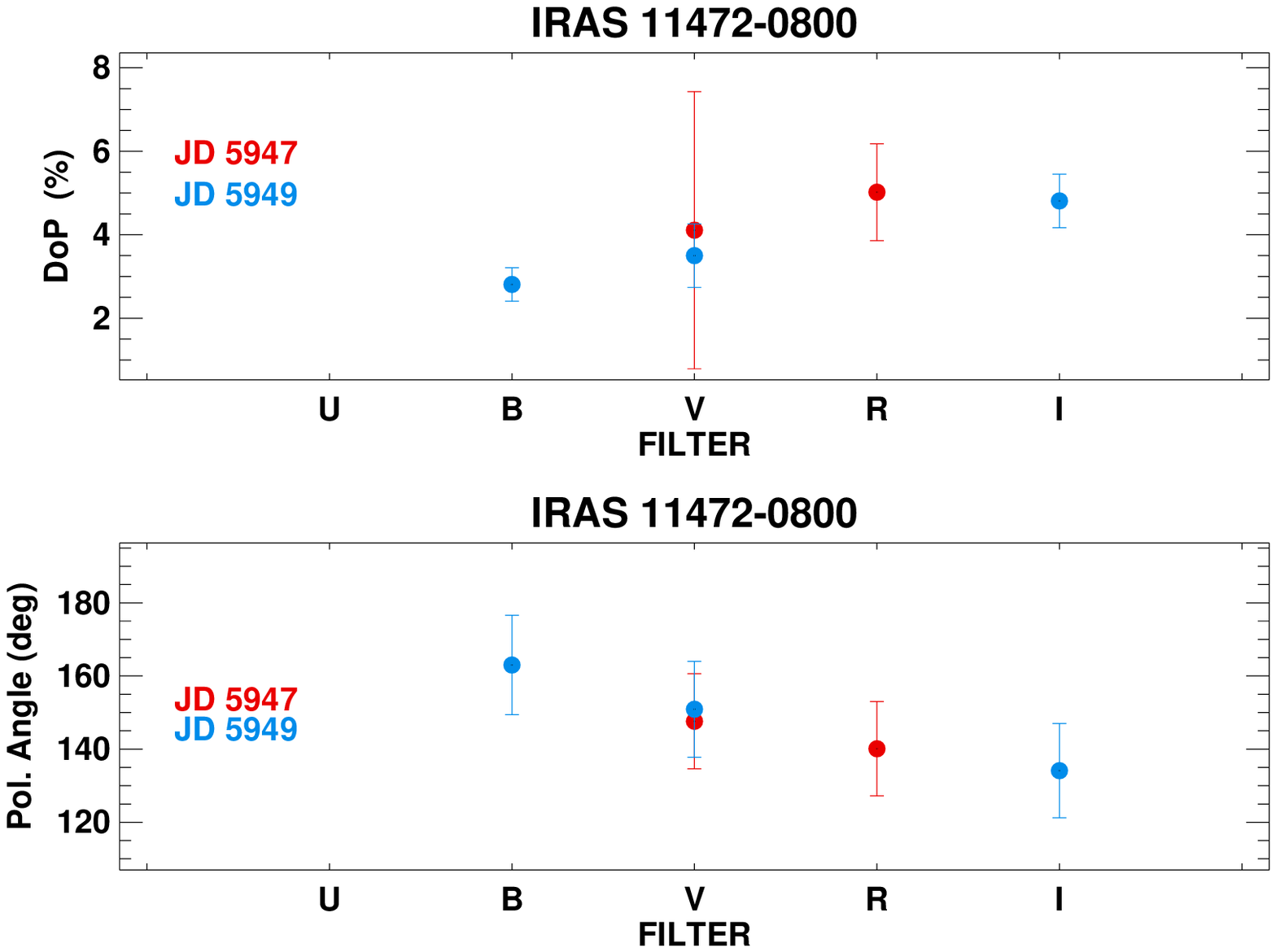}\includegraphics[width=8cm]{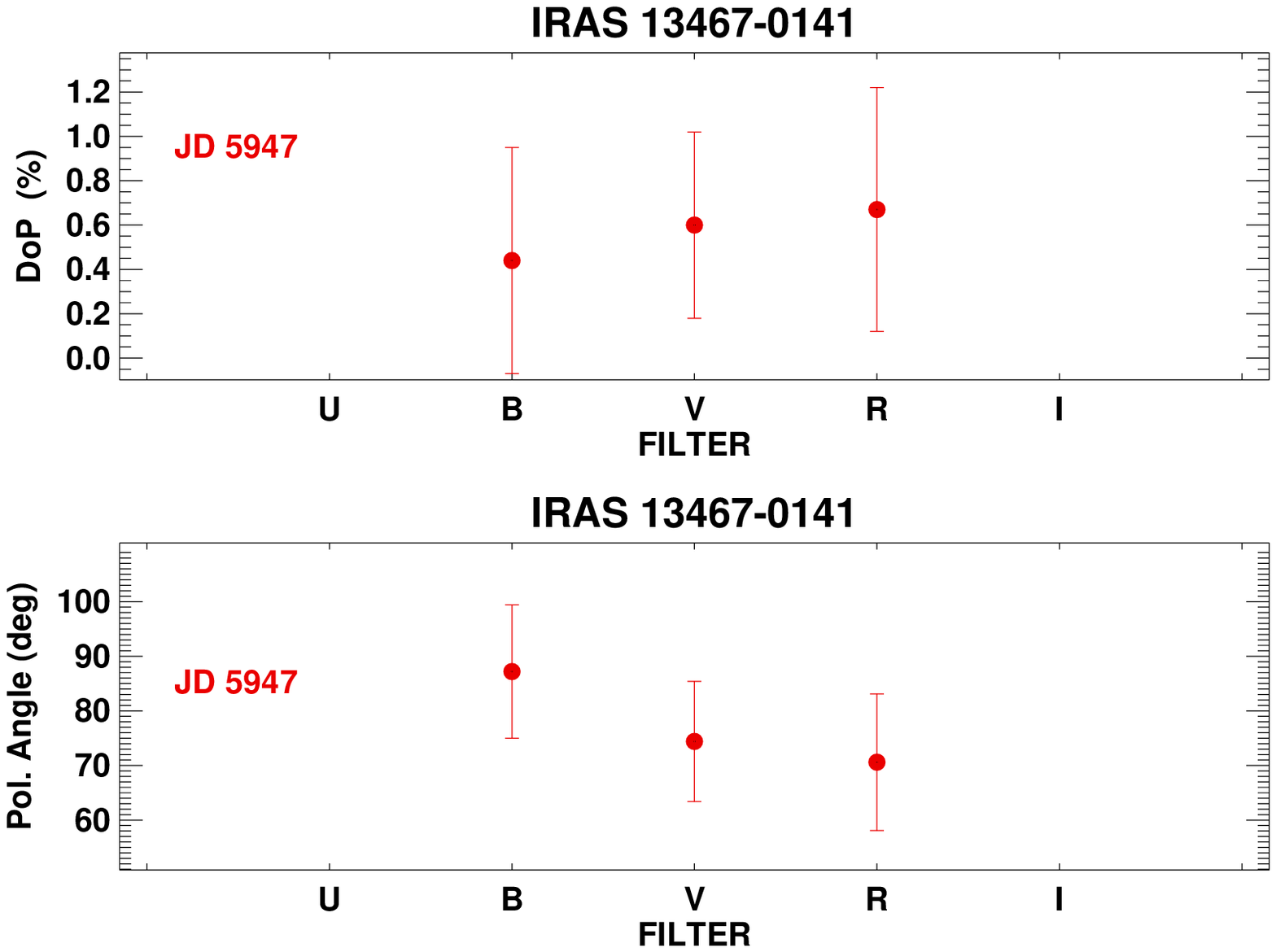}
\includegraphics[width=8cm]{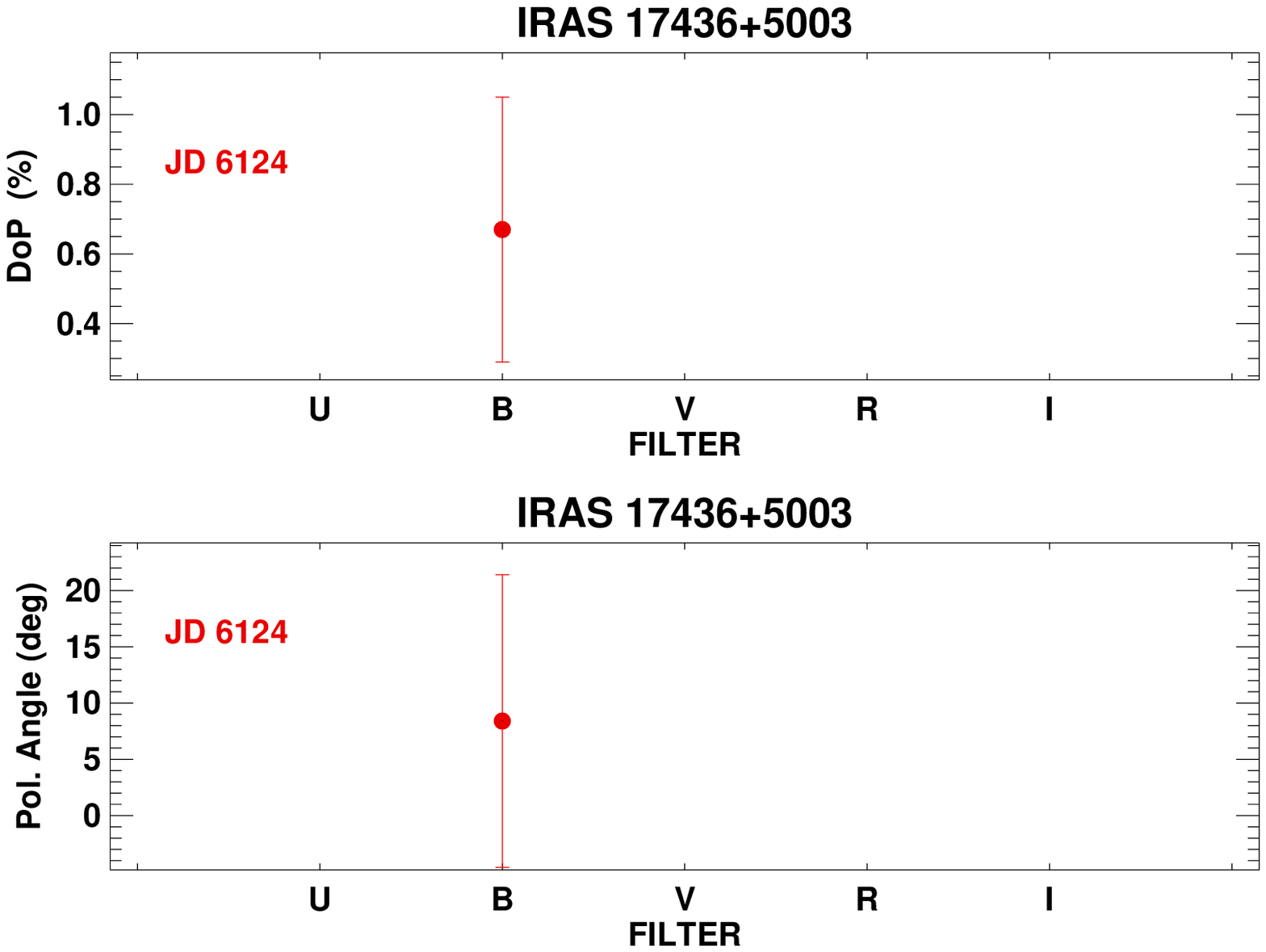}\includegraphics[width=8cm]{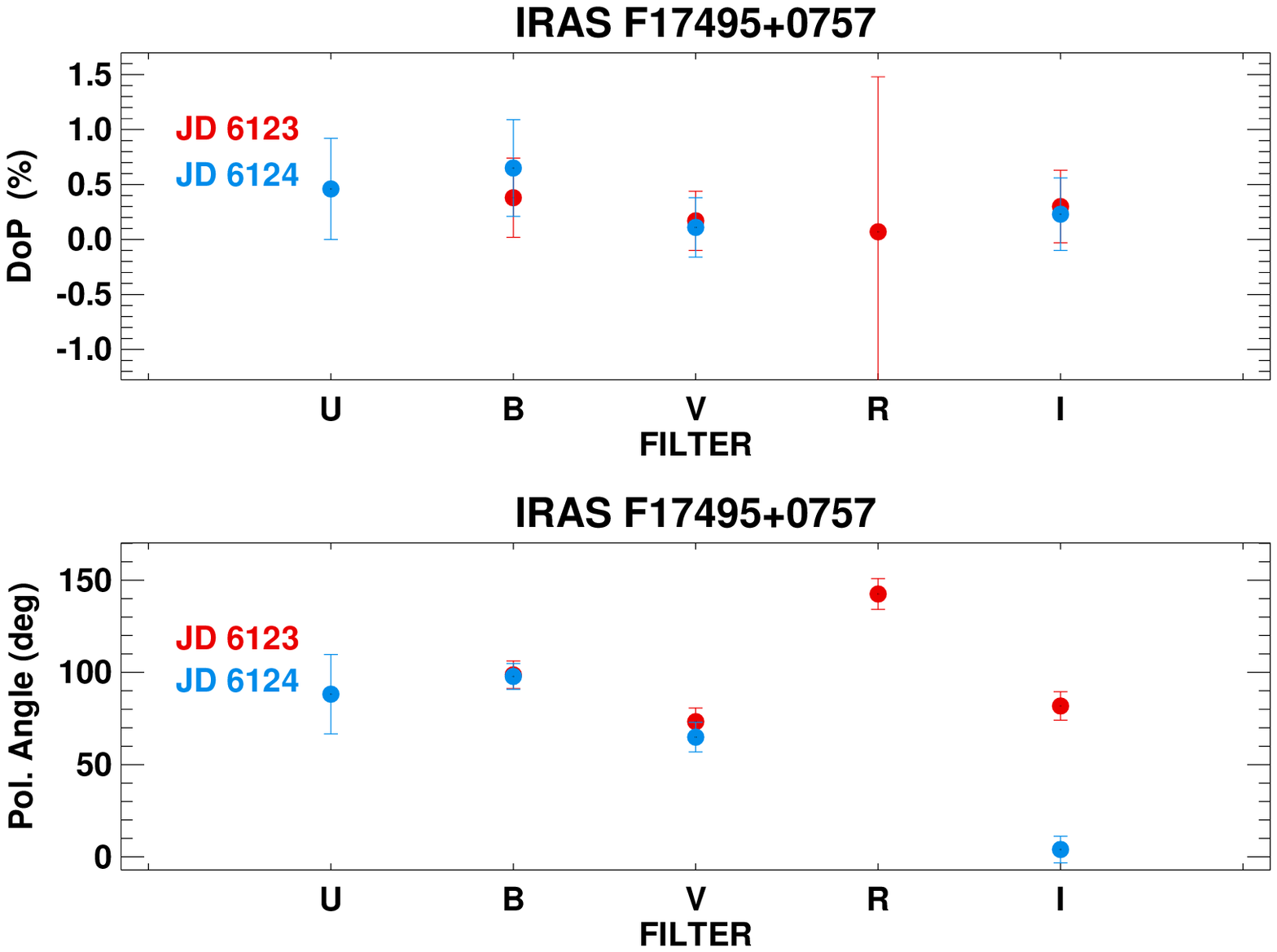}
\caption{continued}
\label{fig5}
\end{center}
\end{figure*}

\begin{figure*}
\addtocounter{figure}{-1}
\begin{center}
\vspace{-0.25cm}
\includegraphics[width=8cm]{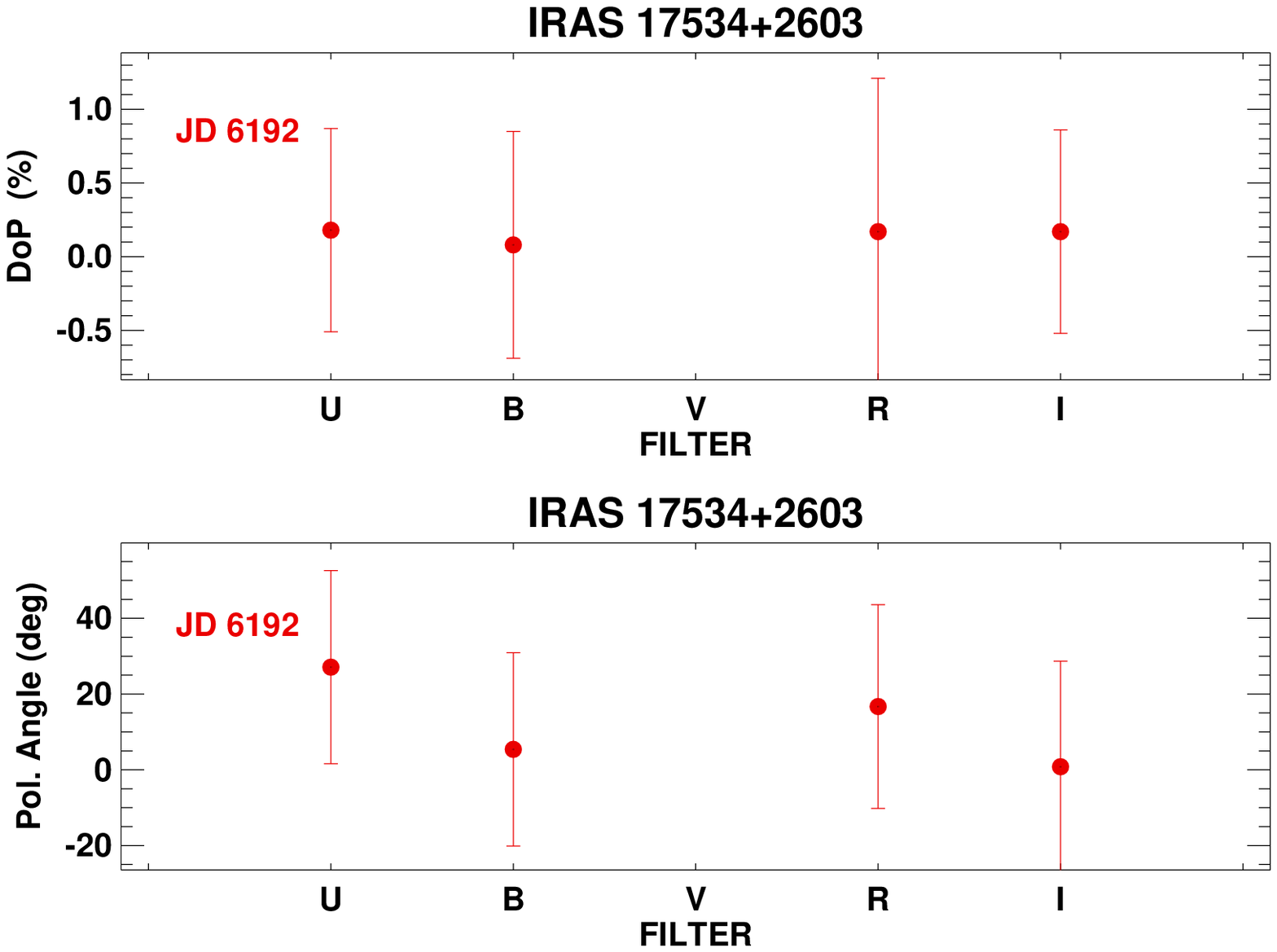}\includegraphics[width=8cm]{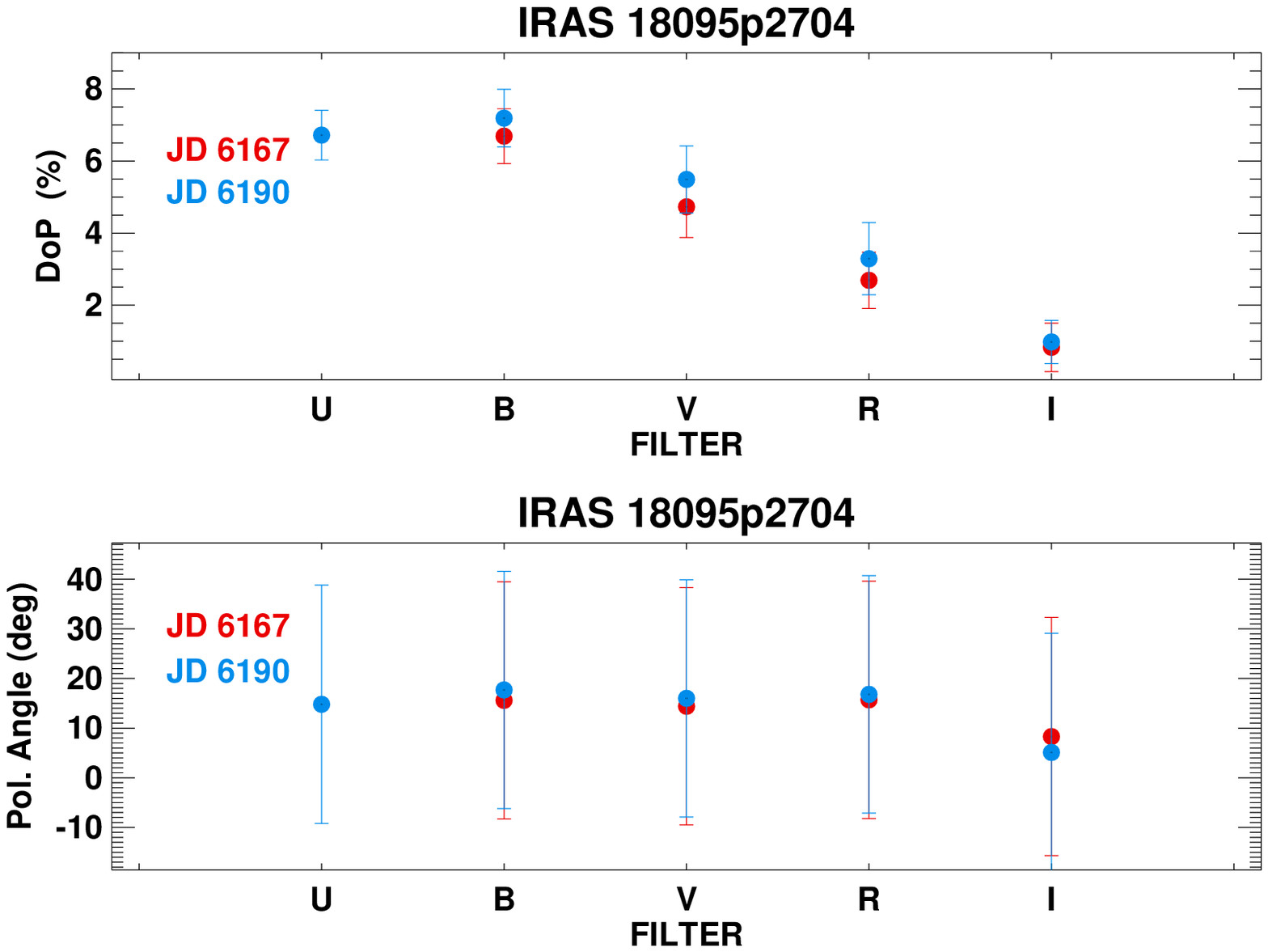}
\includegraphics[width=8cm]{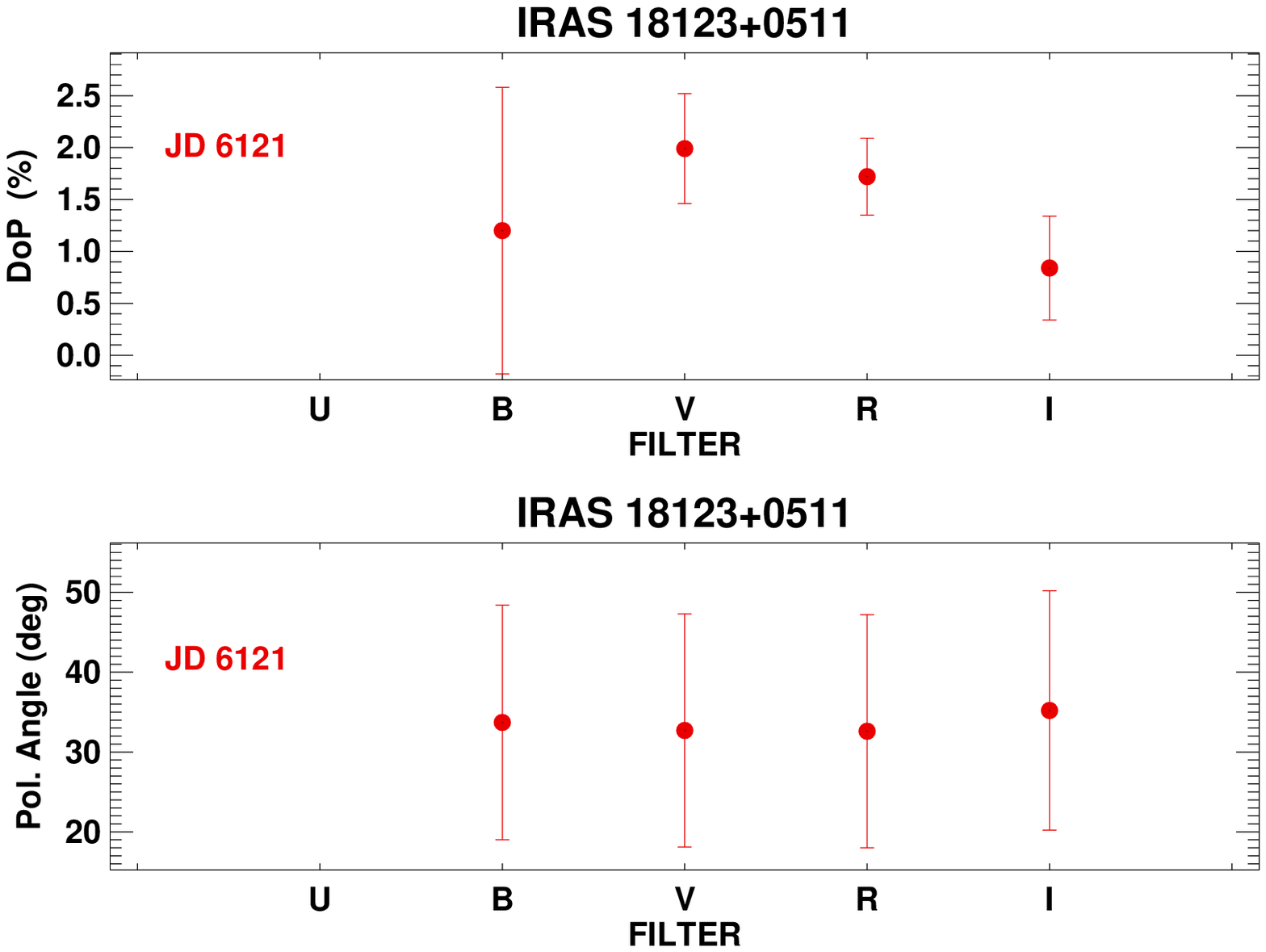}\includegraphics[width=8cm]{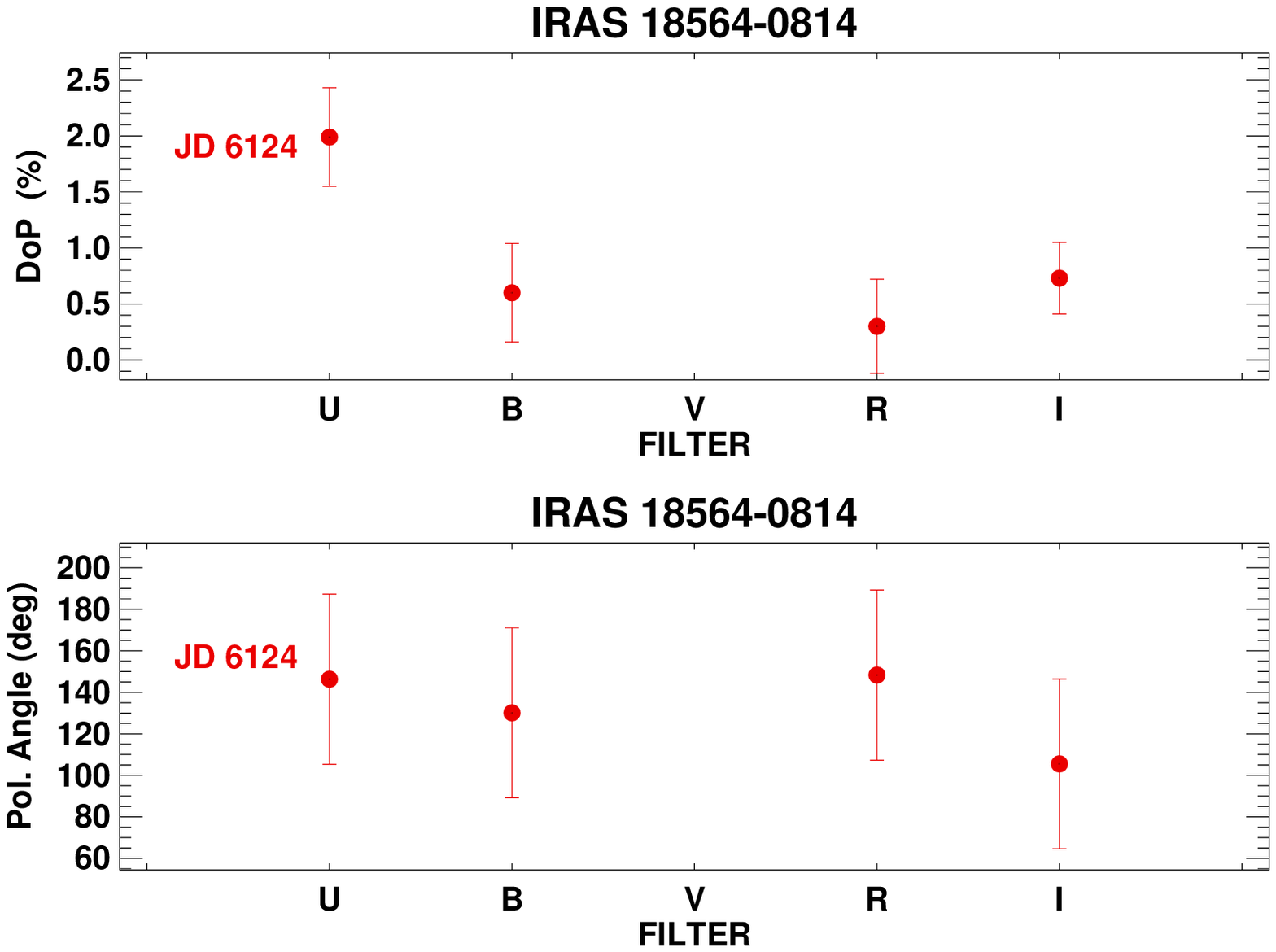}
\includegraphics[width=8cm]{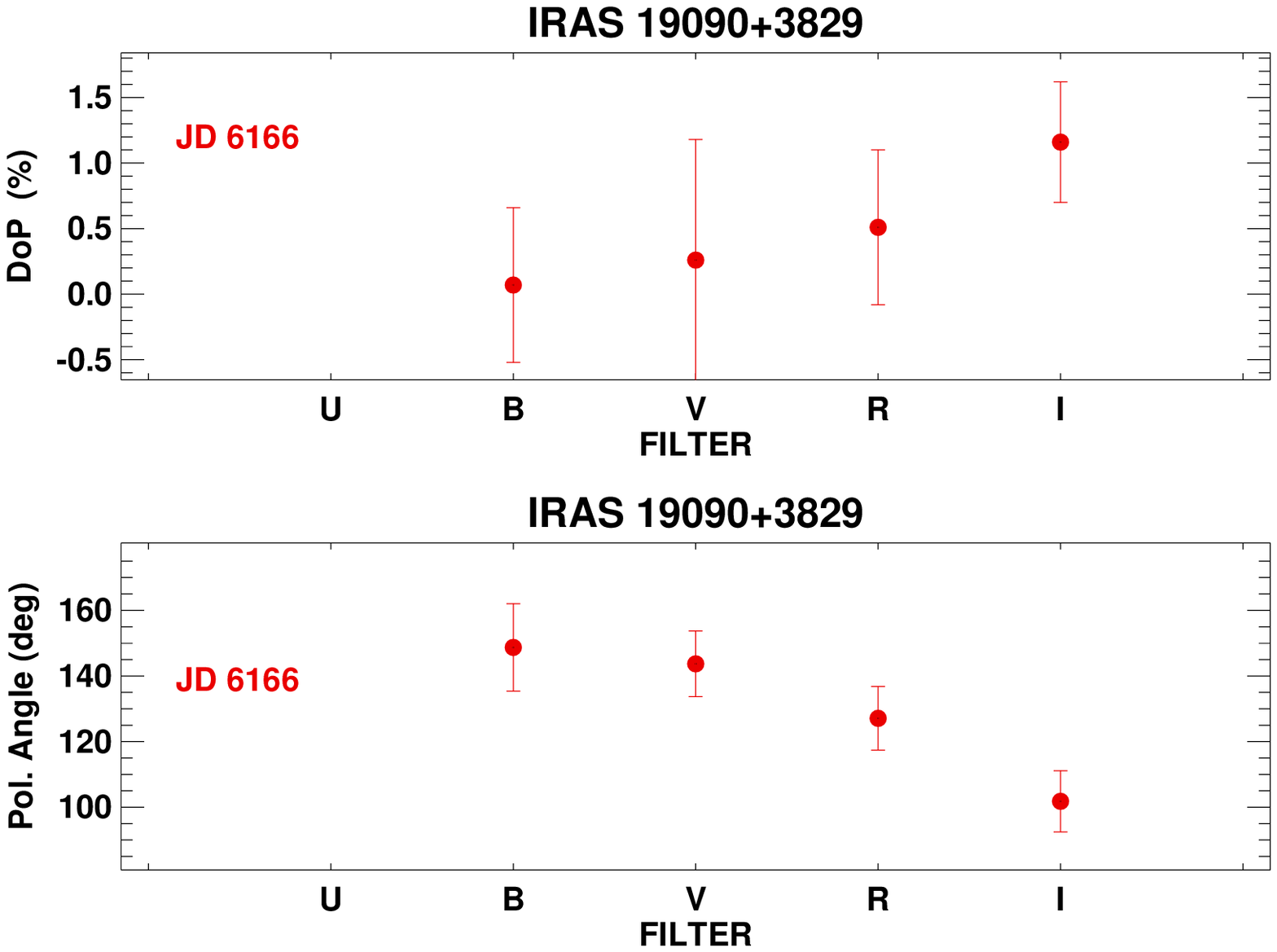}\includegraphics[width=8cm]{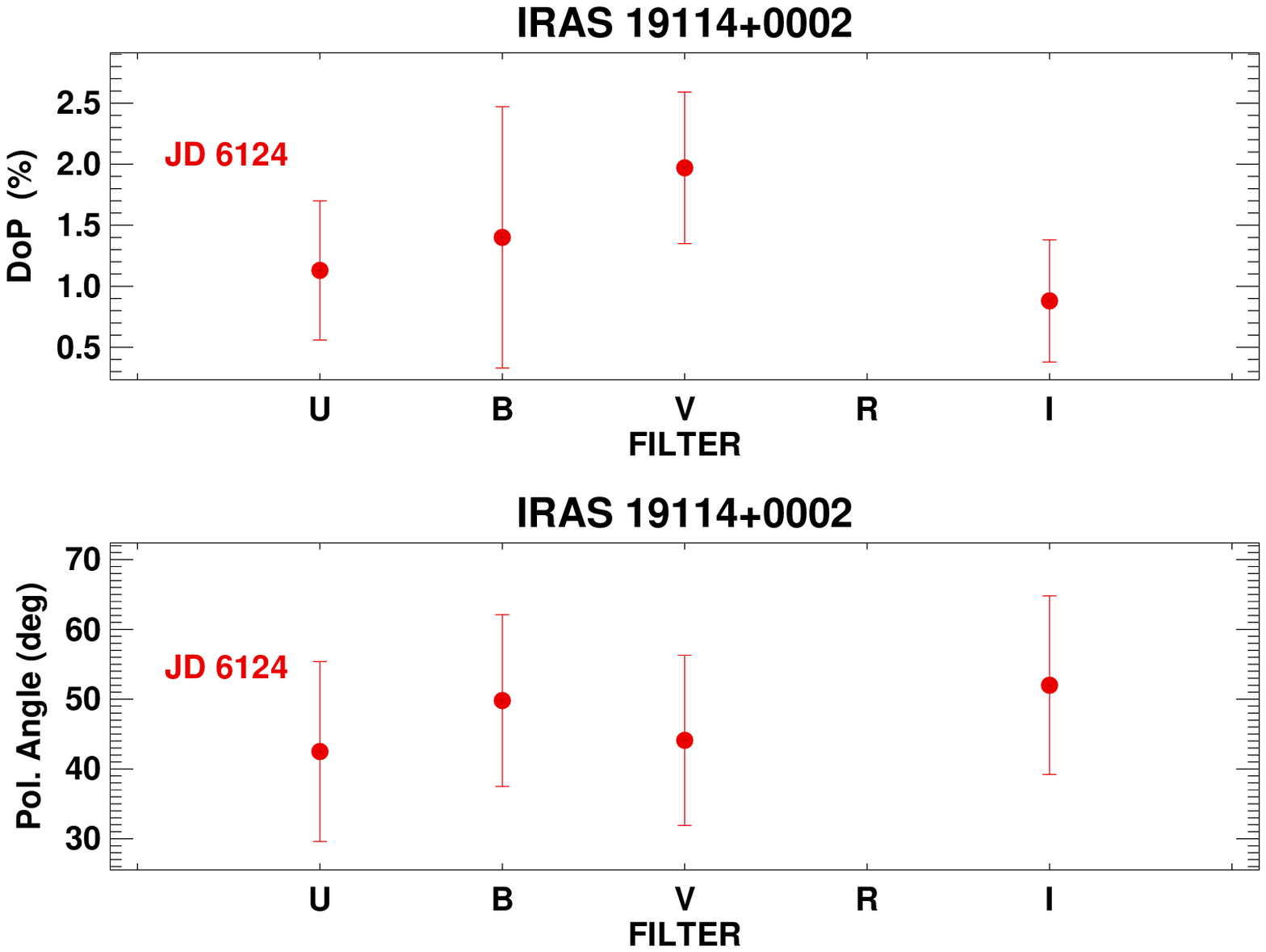}
\includegraphics[width=8cm]{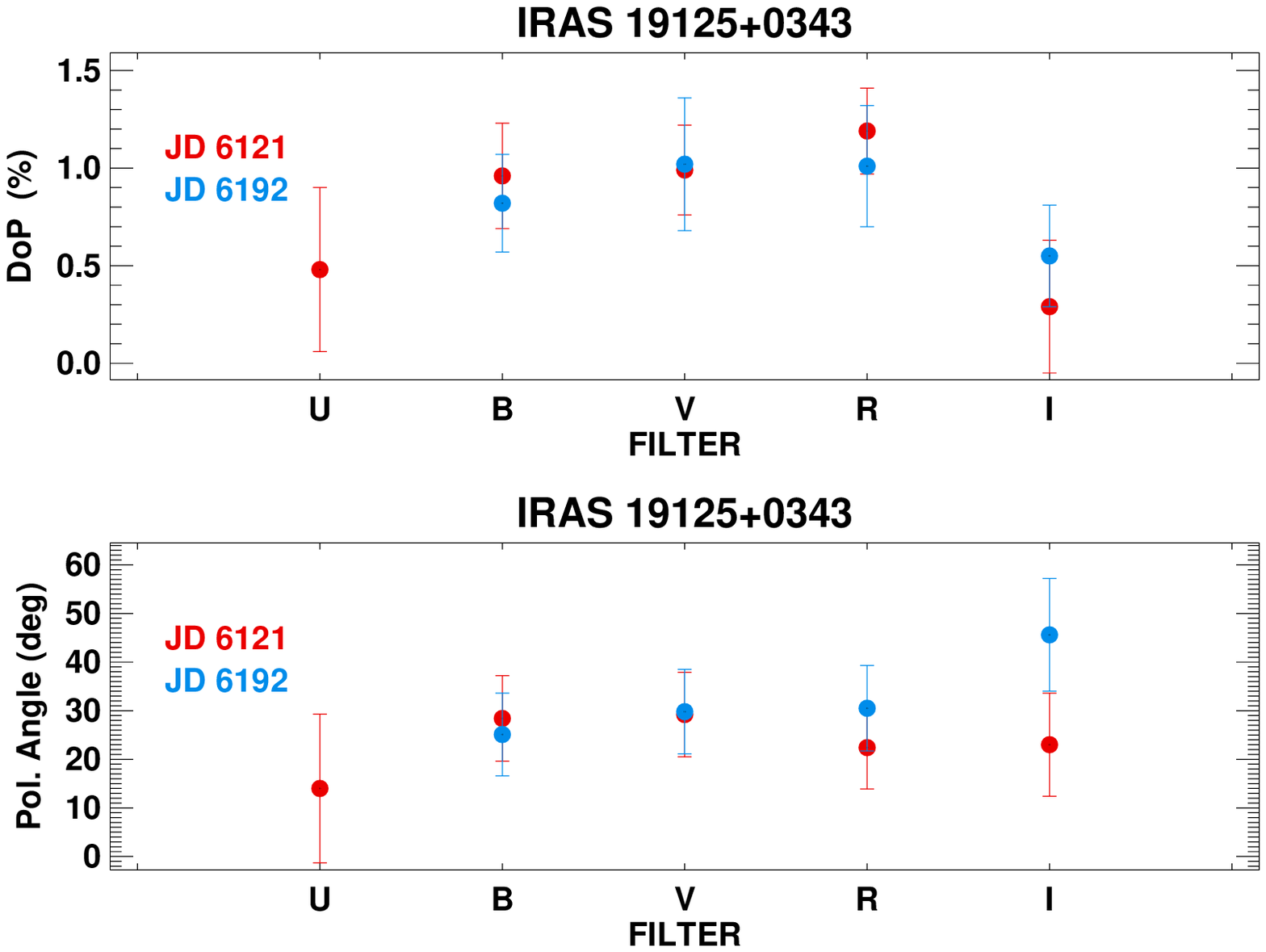}\includegraphics[width=8cm]{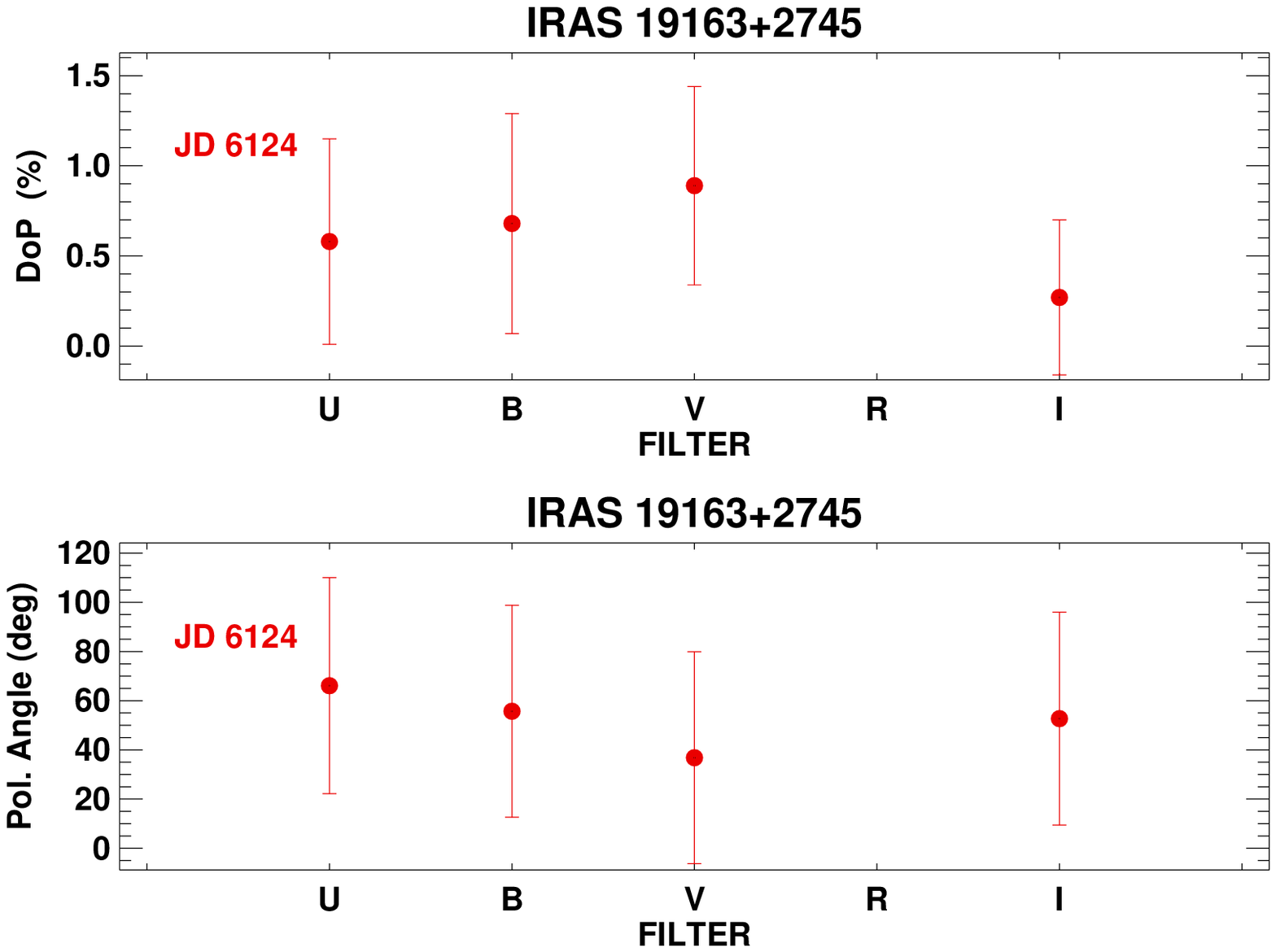}
\caption{continued}
\label{fig6}
\end{center}
\end{figure*}

\begin{figure*}
\addtocounter{figure}{-1}
\begin{center}
\vspace{-0.25cm}
\includegraphics[width=8cm]{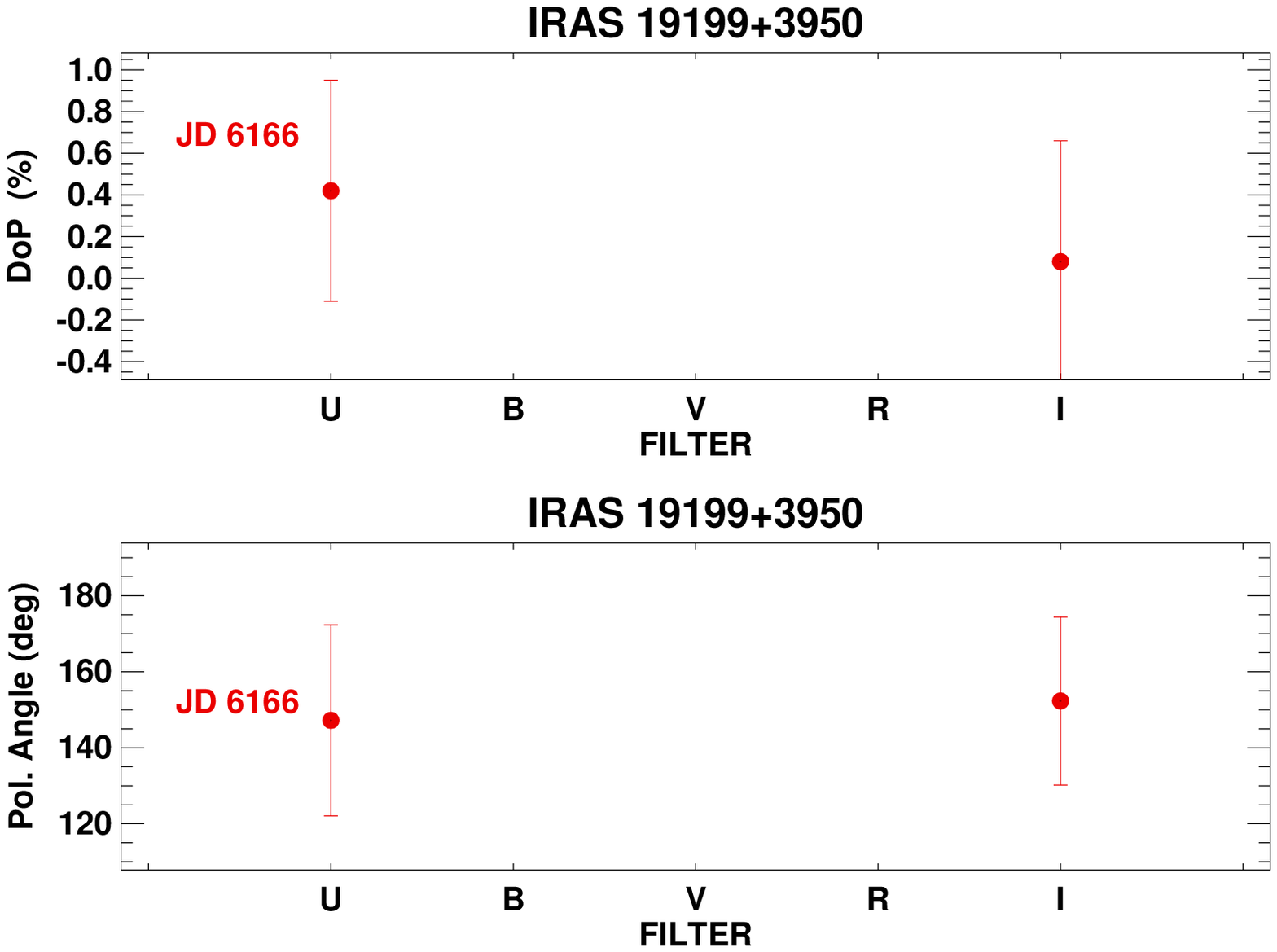}\includegraphics[width=8cm]{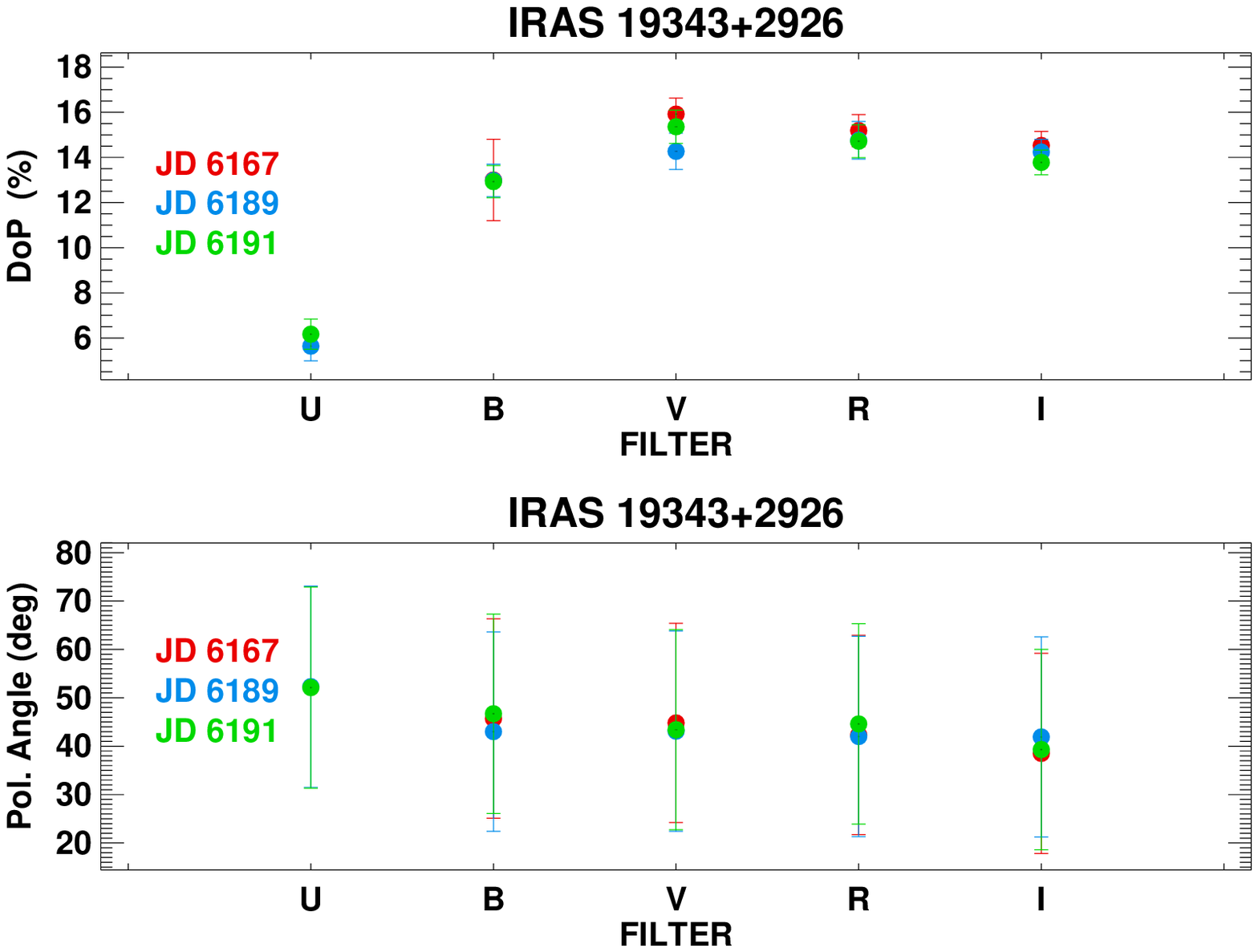}
\includegraphics[width=8cm]{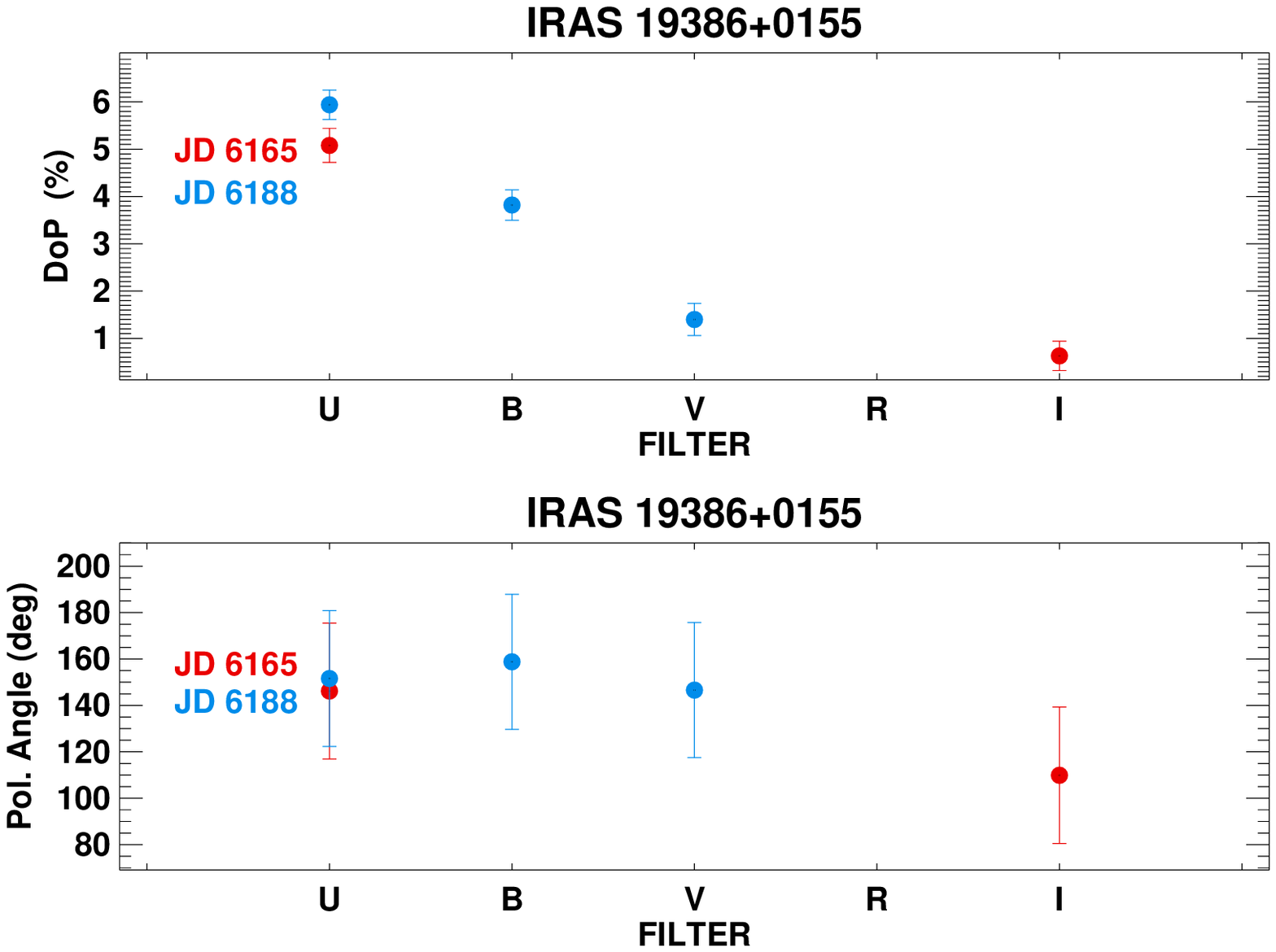}\includegraphics[width=8cm]{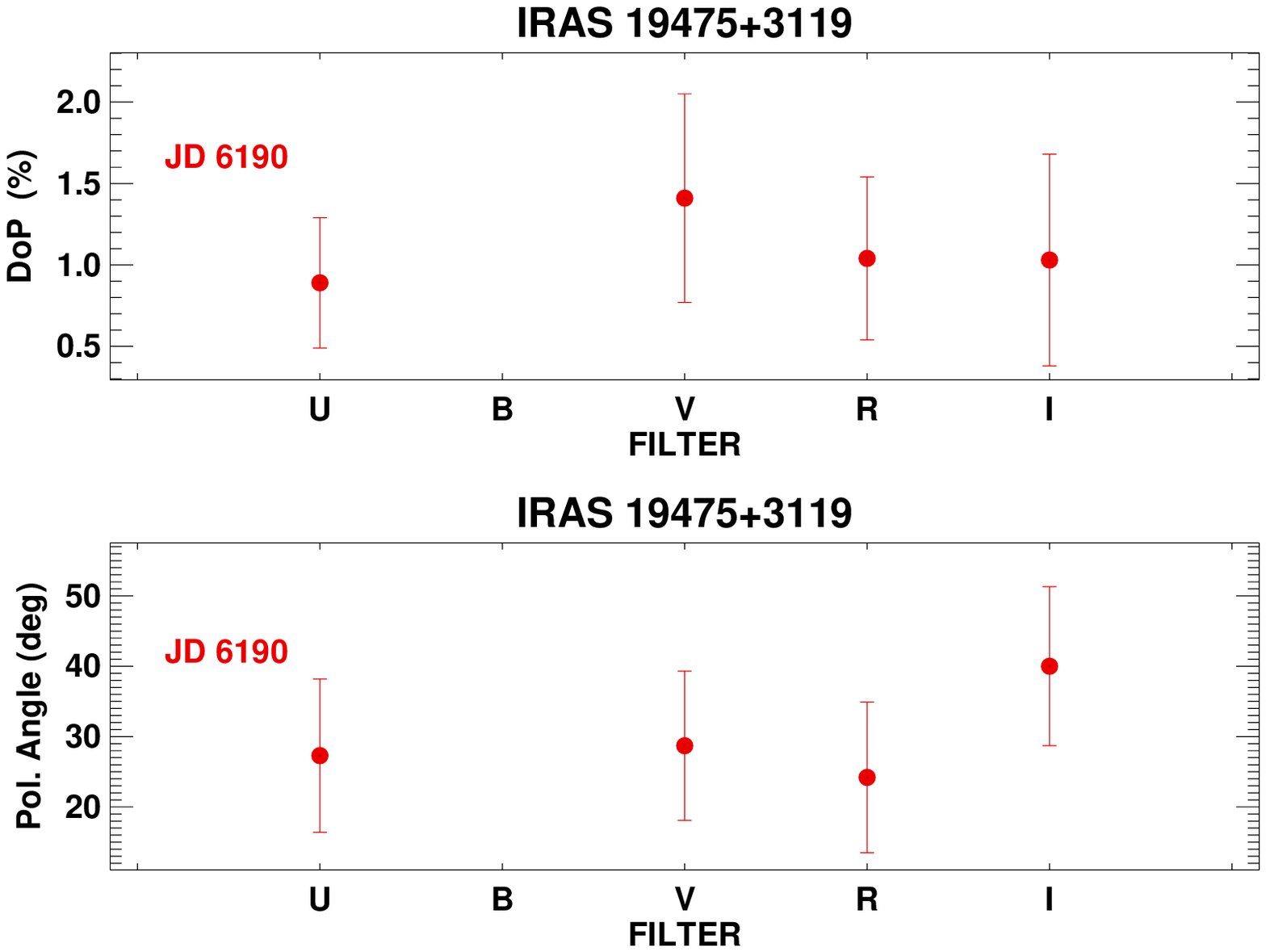}
\includegraphics[width=8cm]{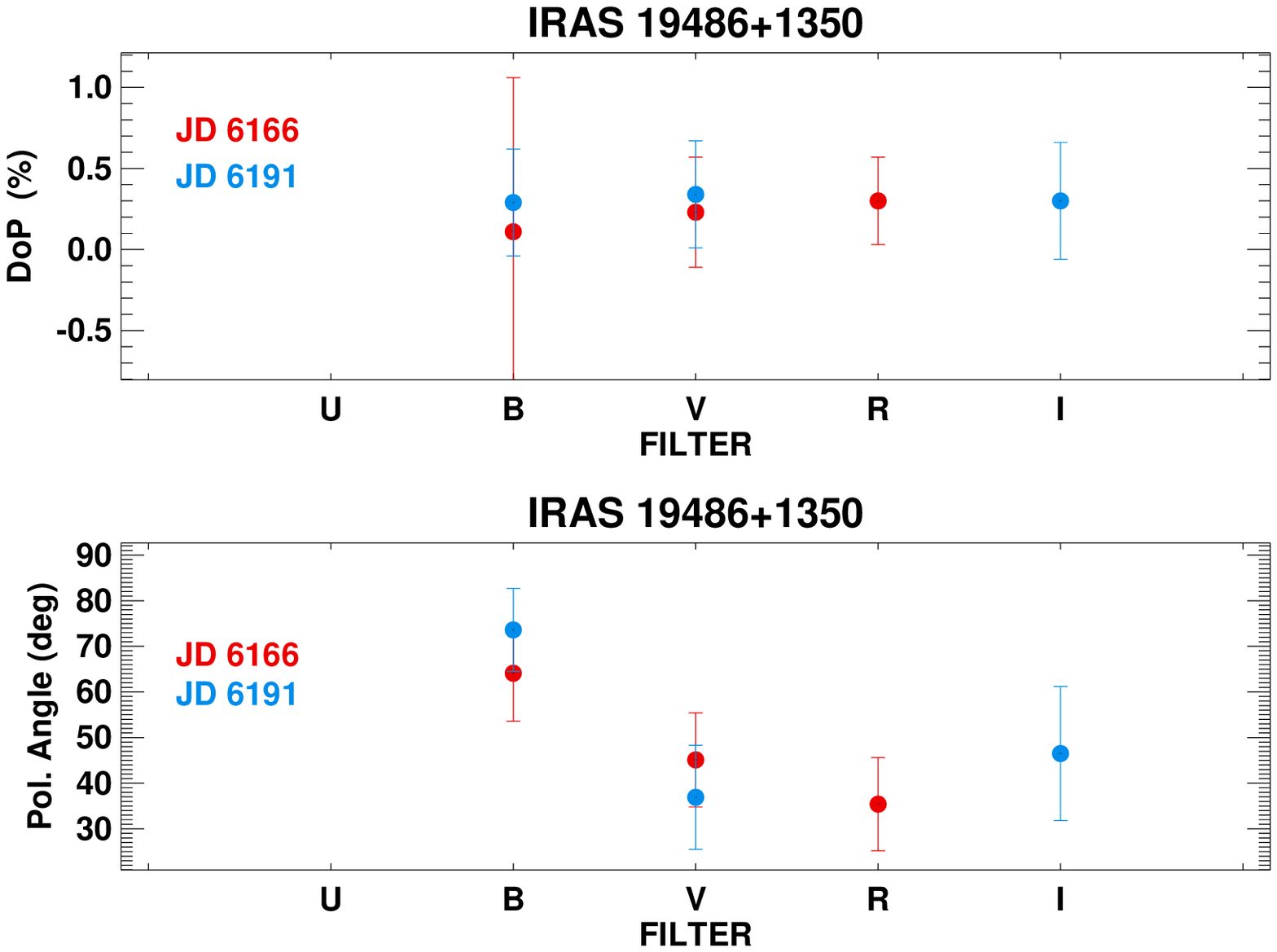}\includegraphics[width=8cm]{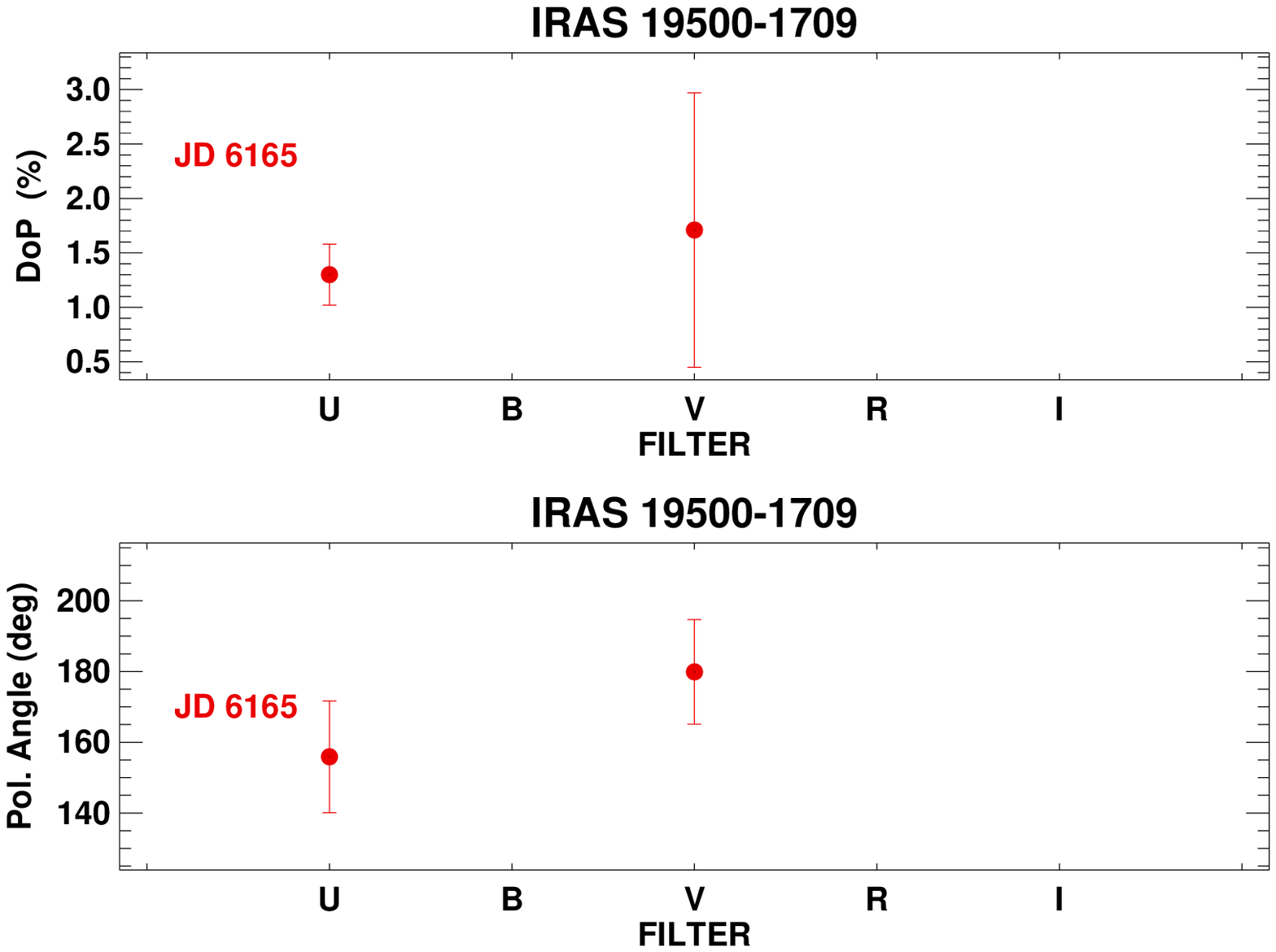}
\includegraphics[width=8cm]{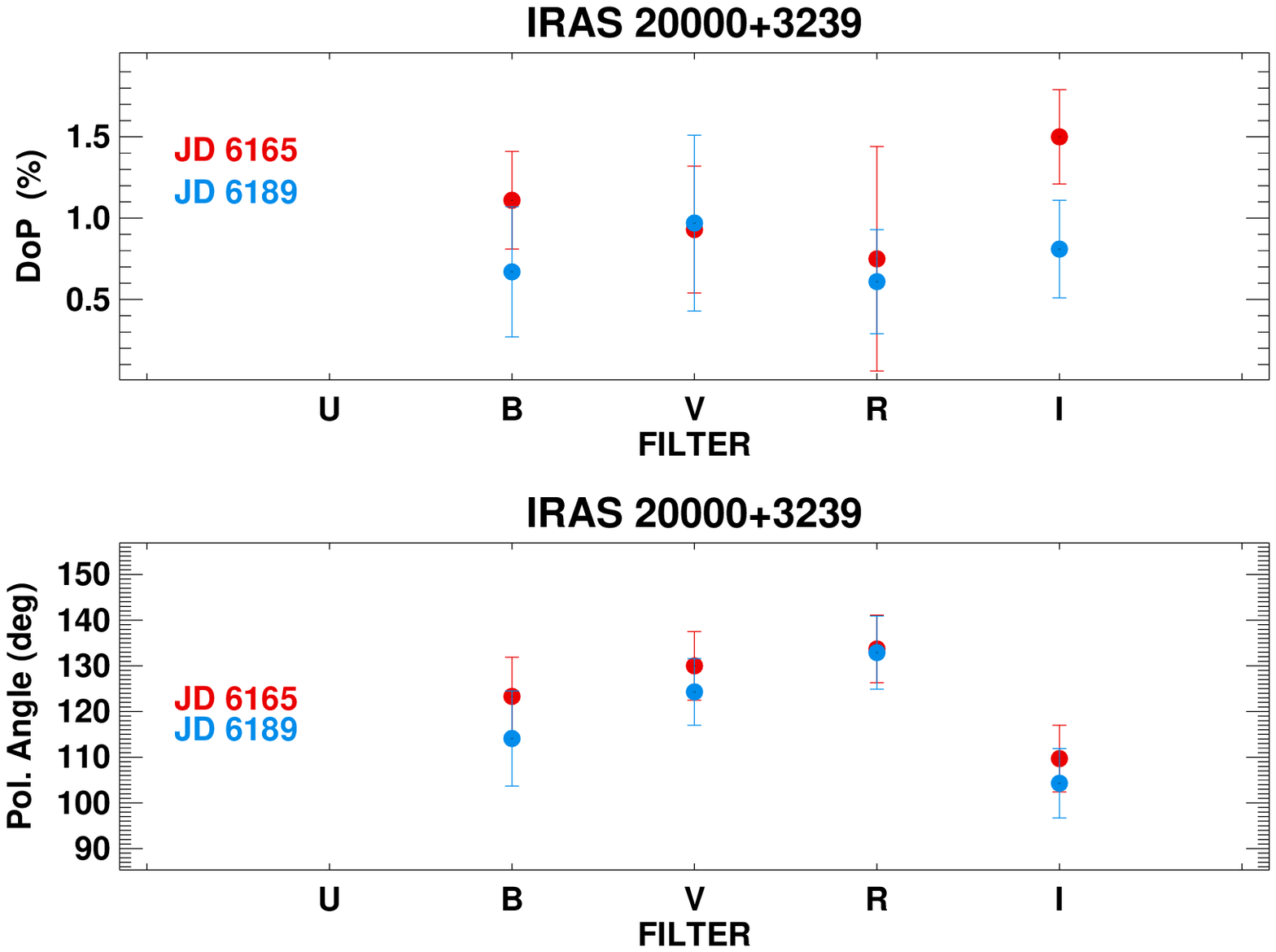}\includegraphics[width=8cm]{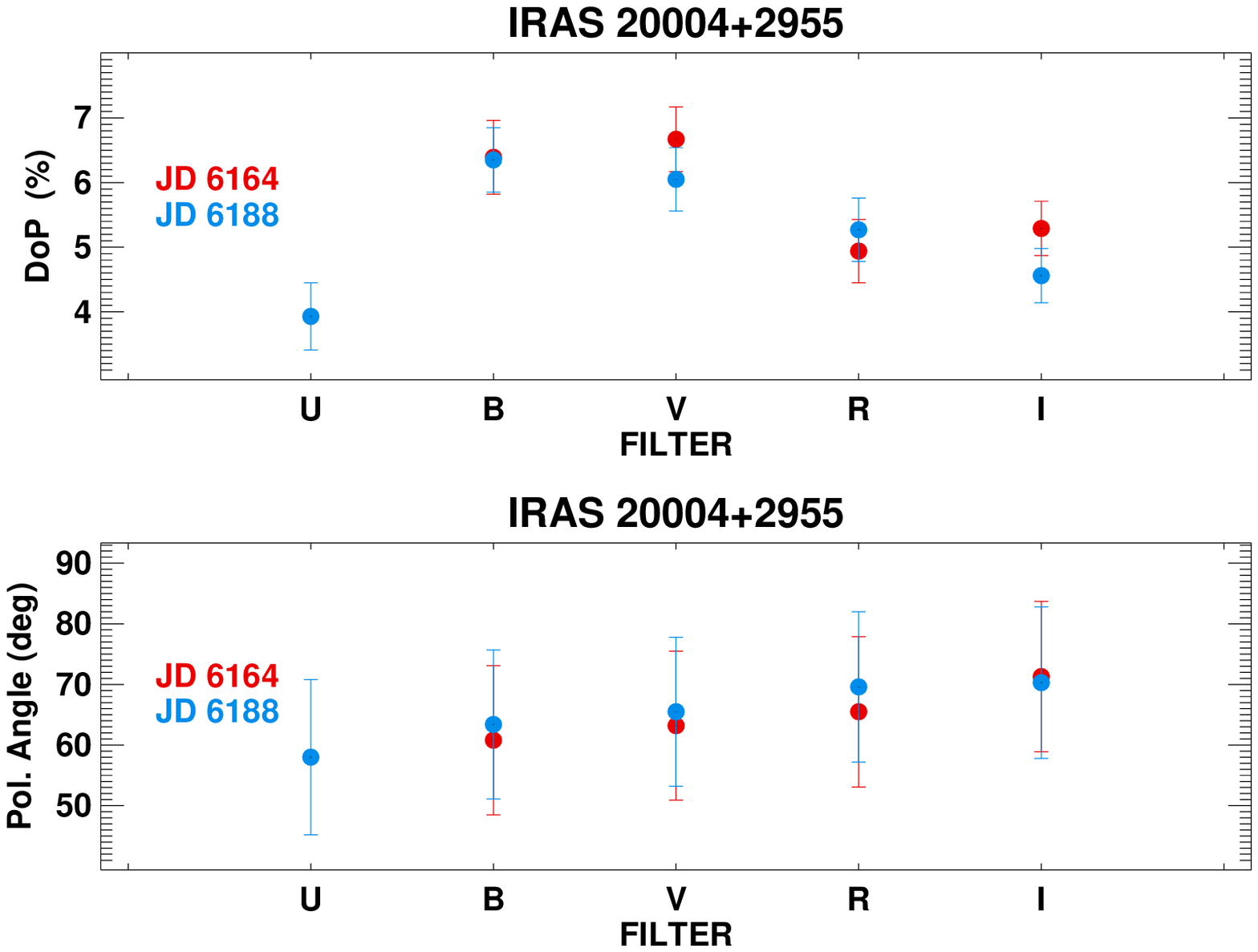}
\caption{continued}
 \label{fig7}
	\end{center}
\end{figure*}

\begin{figure*}
\addtocounter{figure}{-1}
\begin{center}
\vspace{-0.25cm}
\includegraphics[width=8cm]{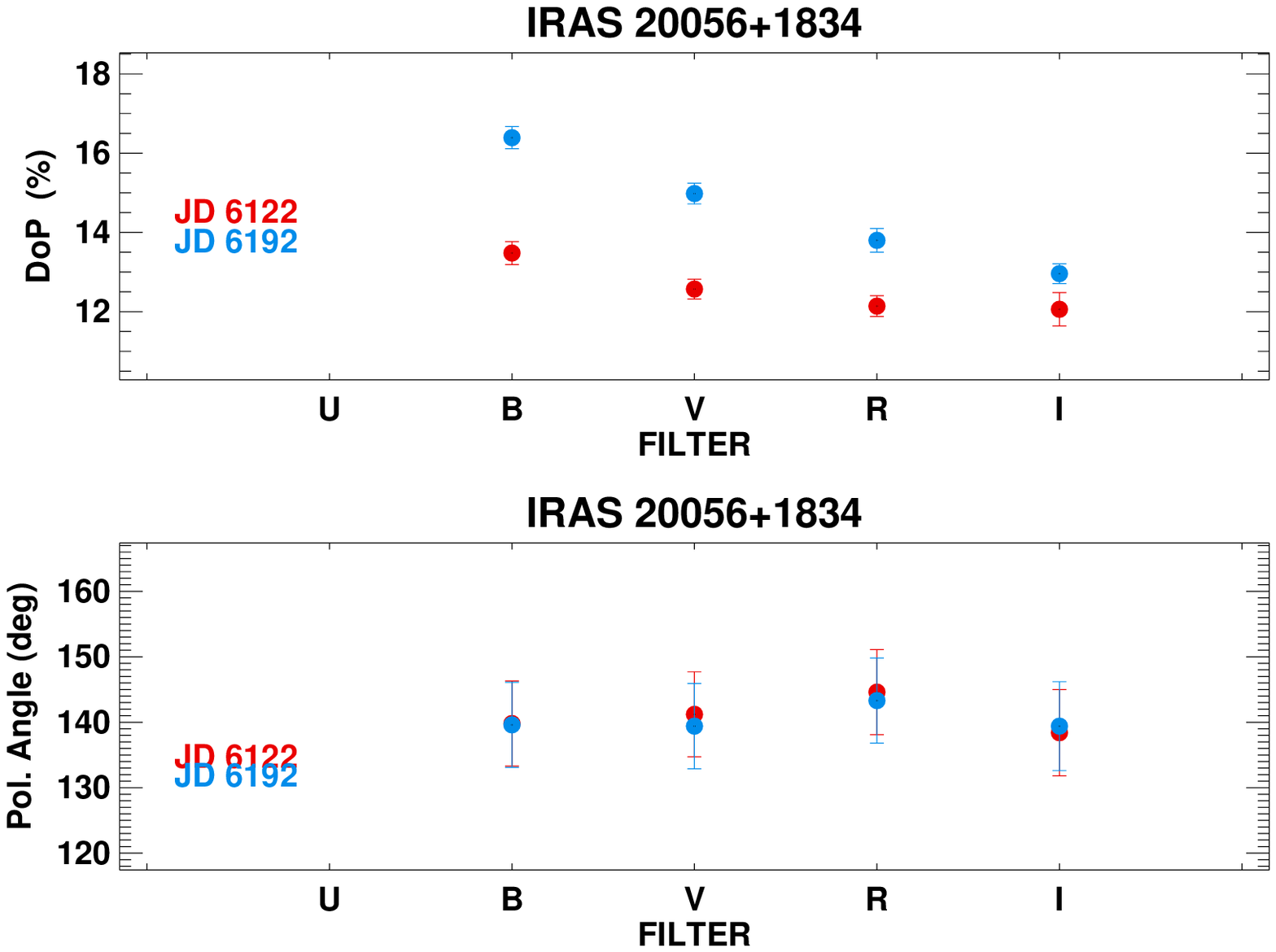}\includegraphics[width=8cm]{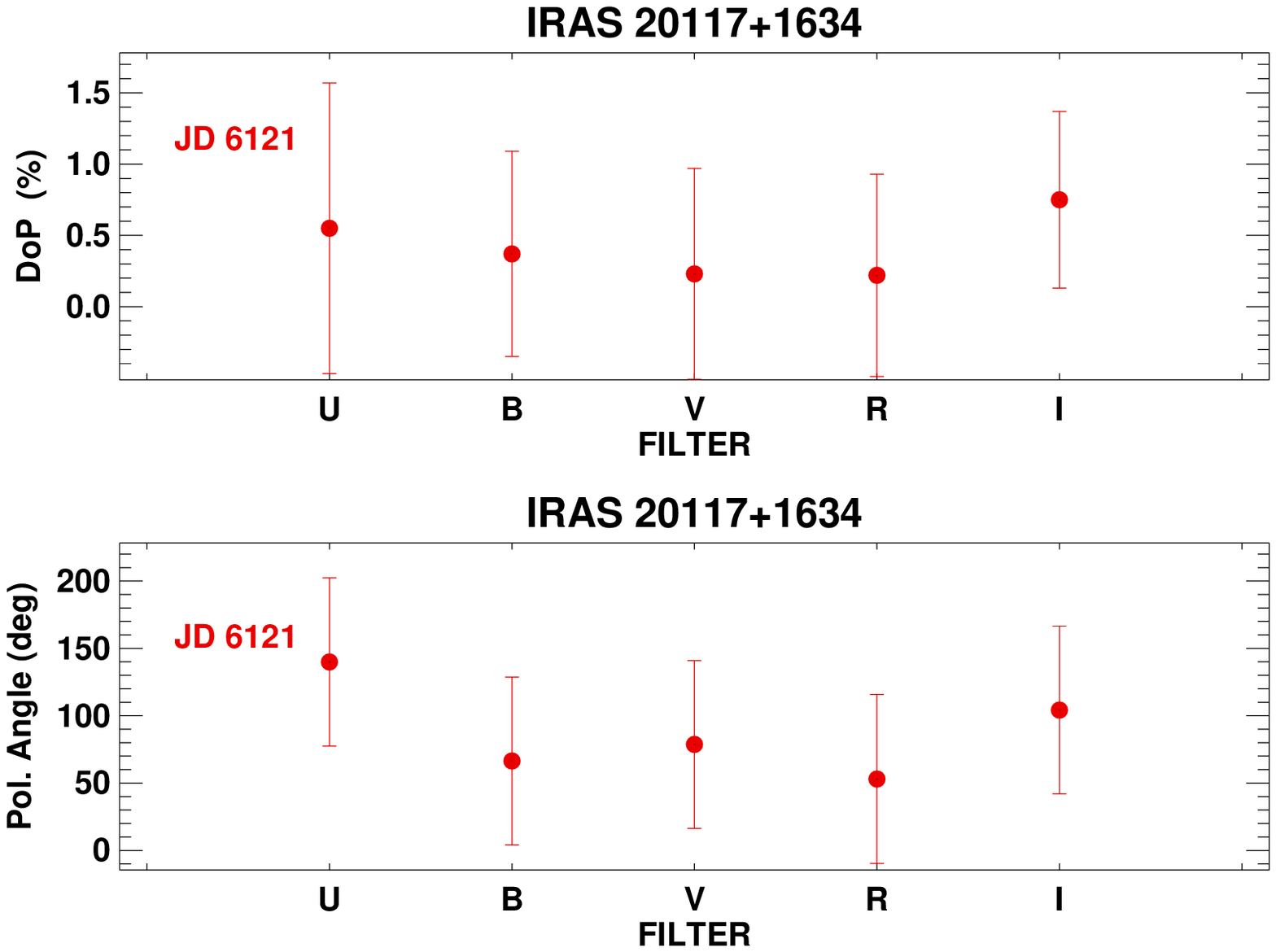}
\includegraphics[width=8cm]{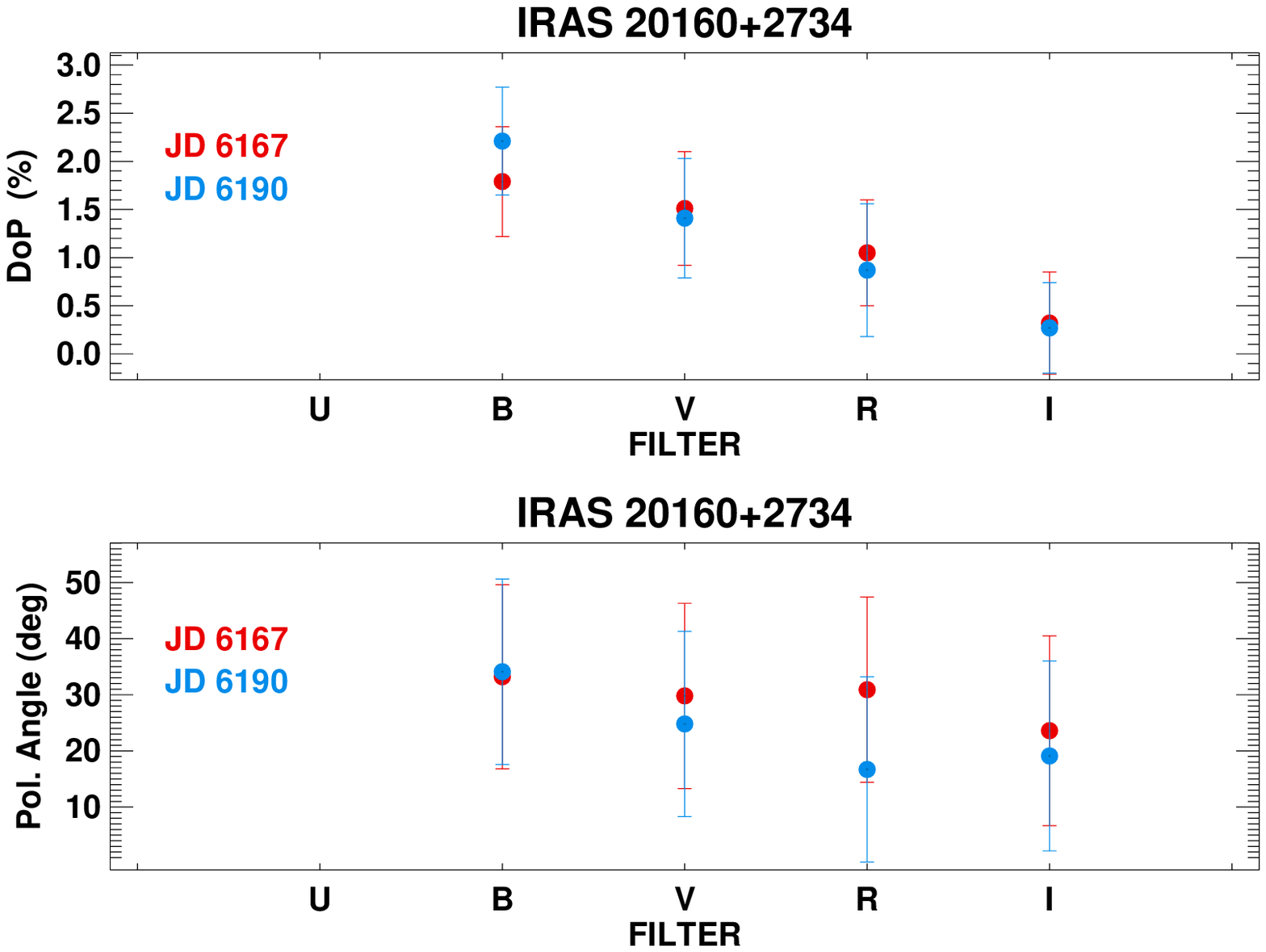}\includegraphics[width=8cm]{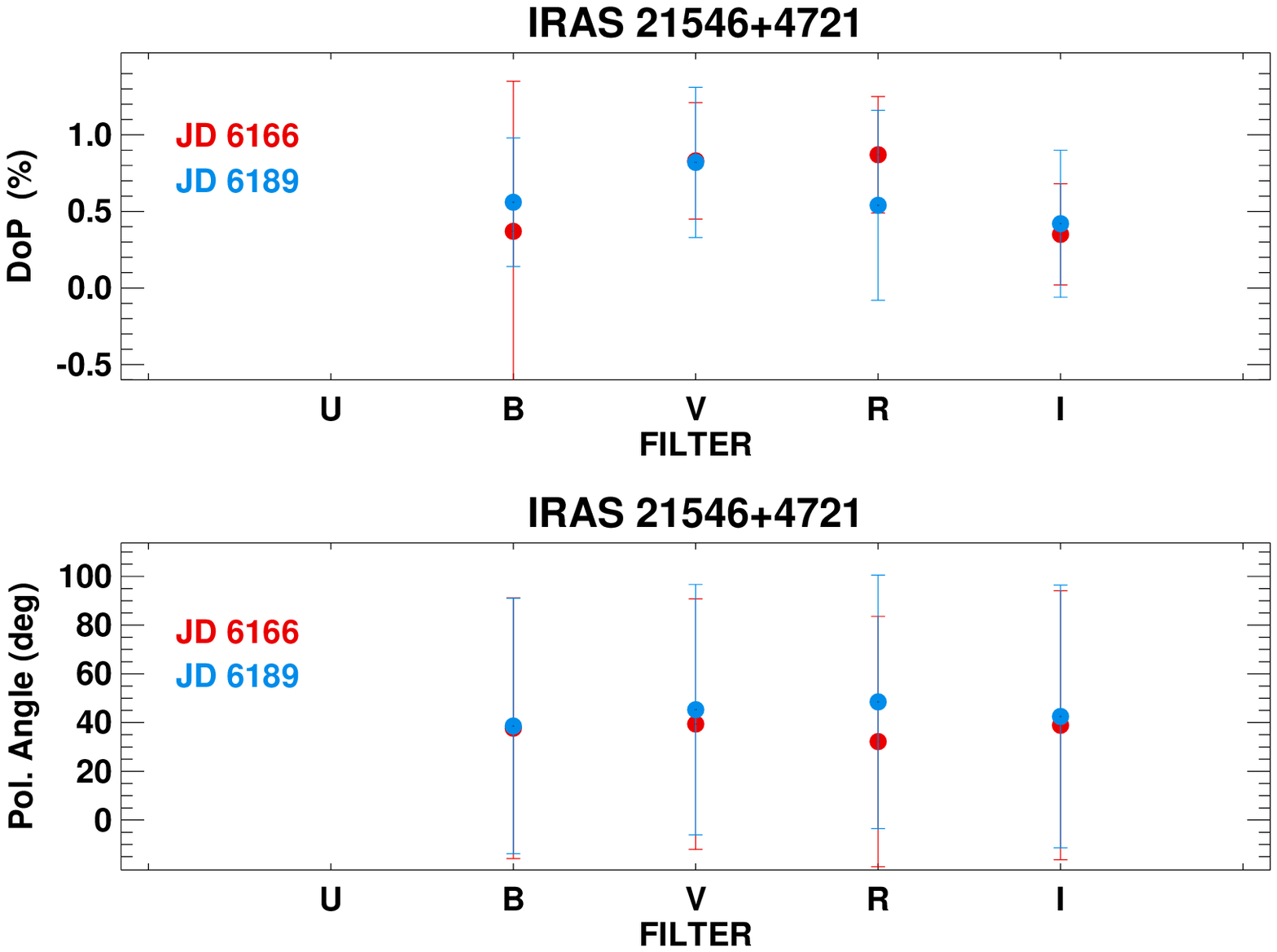}
\includegraphics[width=8cm]{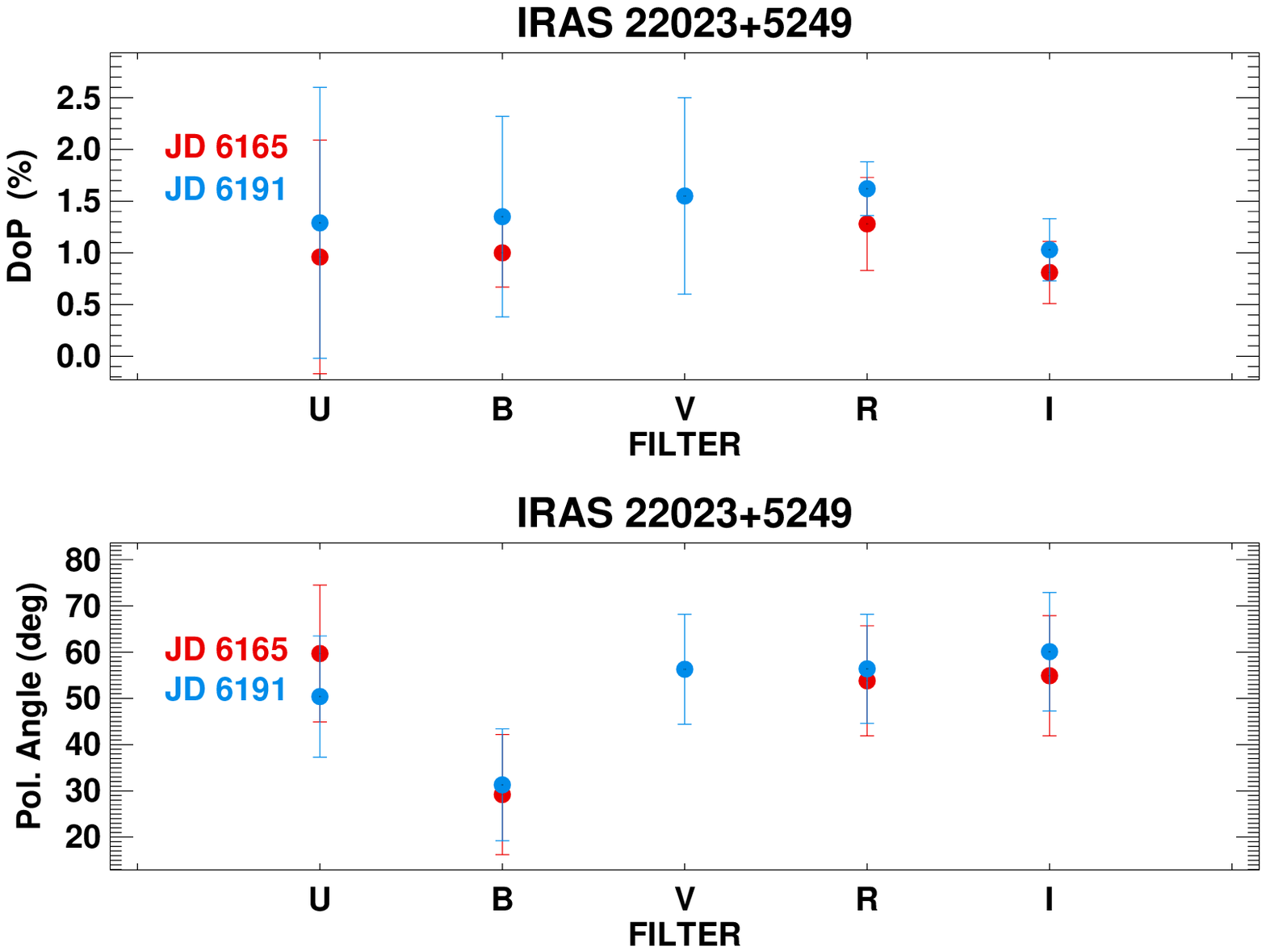}\includegraphics[width=8cm]{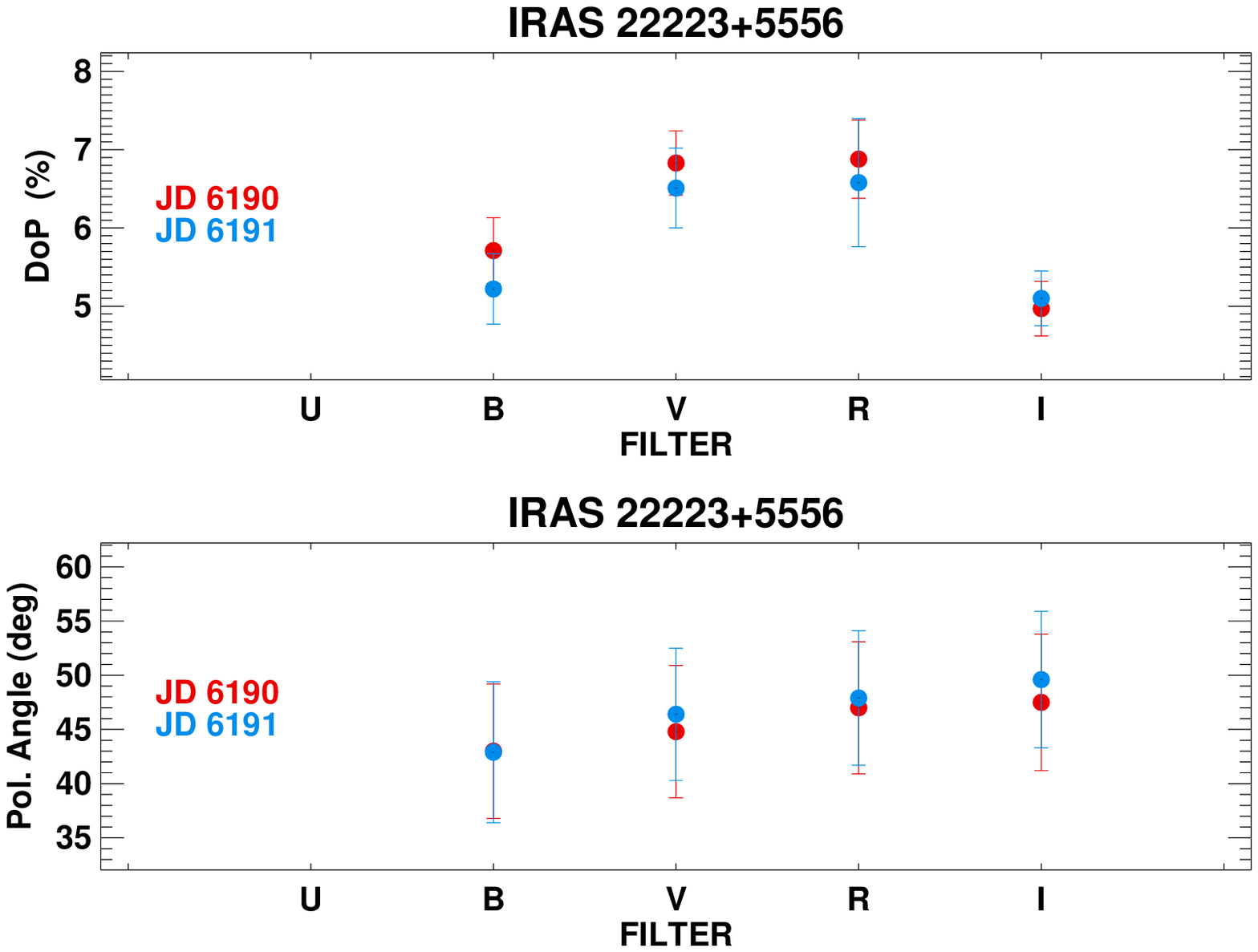}
\includegraphics[width=8cm]{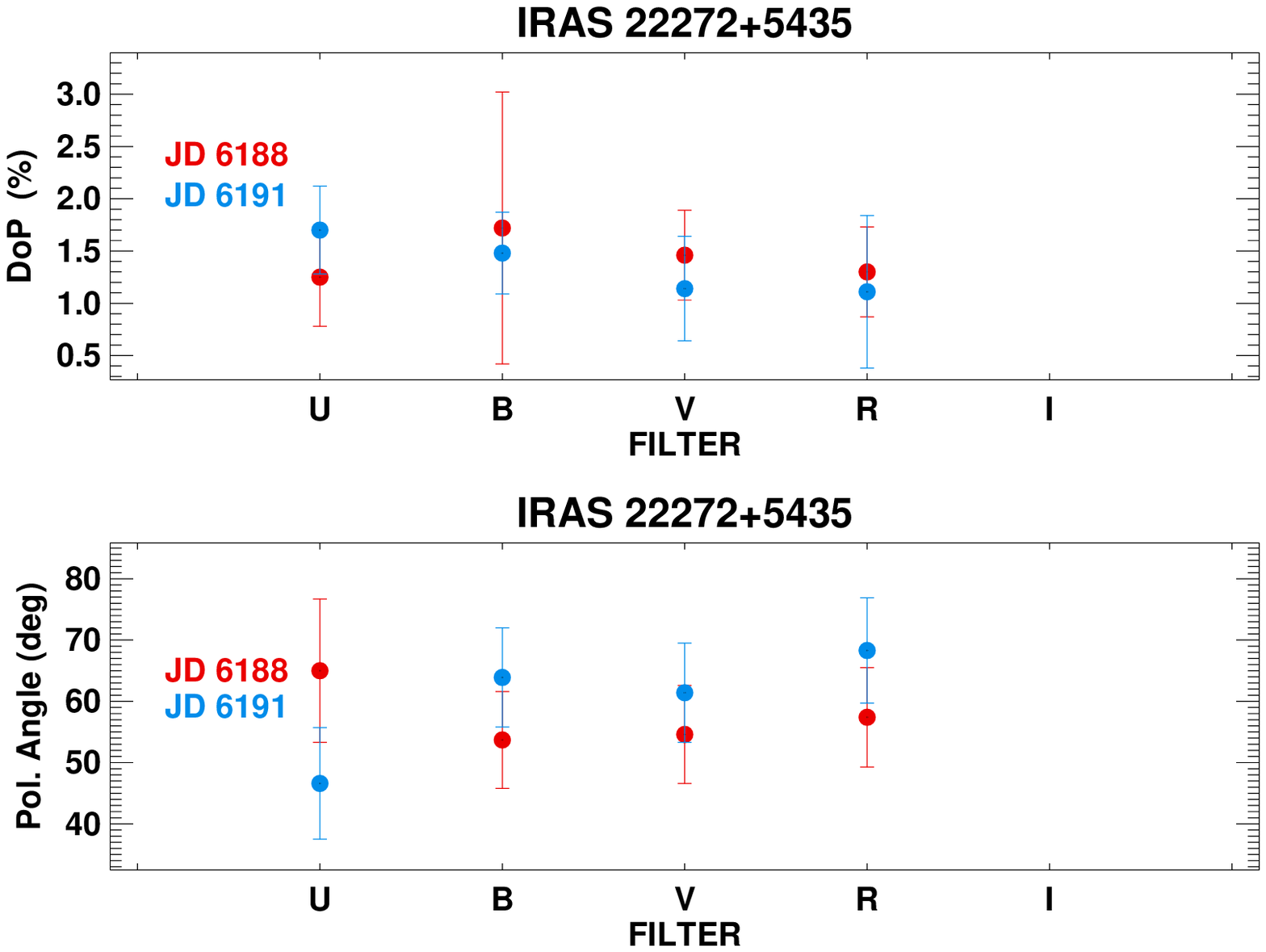}\includegraphics[width=8cm]{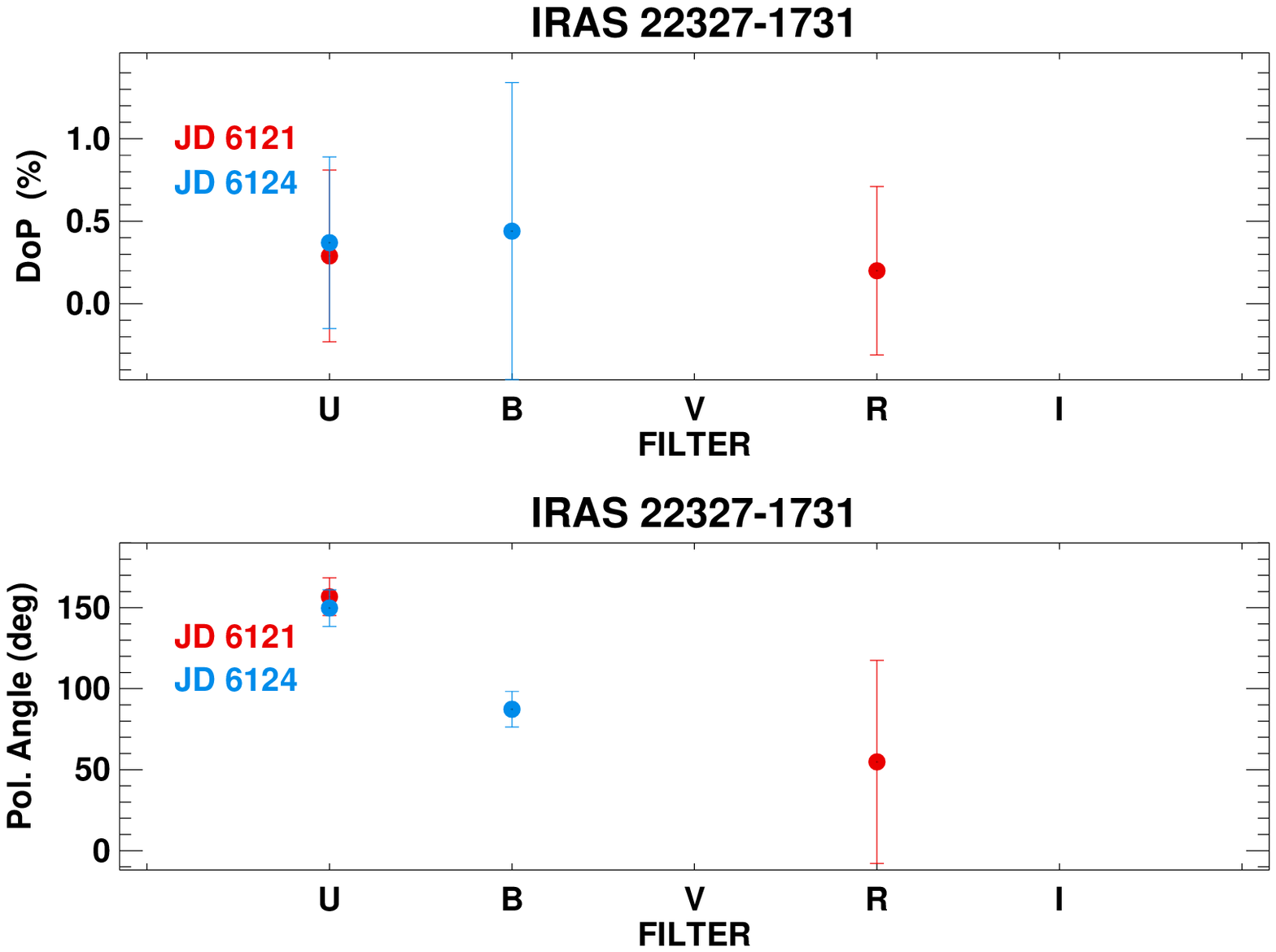}
\caption{continued}
\label{fig8}
\end{center}
\end{figure*}


\newpage
\section{Degree of polarization measurements}
In this section we list the intrinsic DoP, the PA of the polarization vectors after the correction of the interstellar medium and bias-noise, as well as the DoP and PA of the interstellar medium in the direction 
of each object using the field star method.

\begin{table*}
\caption{Intrinsic degree of polarization, position angle of the polarization vectors, total degree of polarization and interstellar polarization of all the stars in our sample. Column (1) gives the Julian Date, column (2) indicates the filter, columns (3) and (4) give the DoP and PA corrected for the interstellar medium and bias-noise, columns (5) and (6) give the observed DoP measurements after and before  the correction for the bias-noise, column (7) gives the observed PA and column (8) gives the interstellar degree of polarization. The PA of the interstellar polarization is also given in the header line of each object.}
\label{table}
\begin{tabular}{cccccccc}
\hline
 & & & IRAS Z02229+6208 & & &   PA$_{ISP}$= 99.1 $\pm$ 2.6\degree & \\
\hline
Julian Date & Filter & DoP(\%) & PA(\degree) & P$_{MAS}$(\%) & P$_{obs}$(\%) & PA$_{obs}$(\degree) & ISP(\%) \\
6166.9550 & B & 3.91 $\pm$  0.430 & 126.3 $\pm$  5.40 & 5.55 $\pm$  0.33 & 8.45 $\pm$ 0.33 & 116.60 $\pm$ 4.70 & 2.28 $\pm$ 0.28 \\
6166.9326 & V & 7.74 $\pm$  0.370 & 132.7 $\pm$  2.90 & 8.96 $\pm$  0.22 & 9.38 $\pm$ 0.22 & 125.50 $\pm$ 1.40 & 2.42 $\pm$ 0.30 \\
6166.9435 & R & 6.51 $\pm$  0.360 & 134.3 $\pm$  3.00 & 7.62 $\pm$  0.21 & 8.24 $\pm$ 0.21 & 125.90 $\pm$ 1.60 & 2.34 $\pm$ 0.29 \\
6166.9377 & I & 6.82 $\pm$  0.340 & 129.4 $\pm$  3.10 & 7.87 $\pm$  0.25 & 8.62 $\pm$ 0.25 & 123.60 $\pm$ 1.80 & 1.81 $\pm$ 0.23 \\
6188.9374 & B & 6.94 $\pm$  0.500 & 129.8 $\pm$  3.10 & 8.27 $\pm$  0.42 & 8.92 $\pm$ 0.42 & 122.80 $\pm$ 1.70 & 2.28 $\pm$ 0.28 \\
6188.9202 & R & 6.63 $\pm$  0.400 & 133.6 $\pm$  3.00 & 7.78 $\pm$  0.27 & 8.32 $\pm$ 0.27 & 125.50 $\pm$ 1.50 & 2.34 $\pm$ 0.29 \\
\hline
\end{tabular}
\end{table*}

\clearpage

\begin{table*}
\addtocounter{table}{-1}
\caption{continued}
\begin{tabular}{cccccccc}
\hline
  & & & IRAS 05040+4820 & & &   PA$_{ISP}$= 57.4 $\pm$ 28.8\degree & \\
 \hline
Julian Date & Filter & DoP(\%) & PA(\degree) & P$_{MAS}$(\%) & P$_{obs}$(\%) & PA$_{obs}$(\degree) & ISP(\%) \\
6261.8919 & U & 2.14 $\pm$  0.570 & 145.2 $\pm$  29.0 & 2.34 $\pm$  0.35 & 3.81 $\pm$ 0.35 & 145.40 $\pm$ 2.60 & 0.19 $\pm$ 0.45 \\
6263.9319 & U & 1.85 $\pm$  0.660 & 131.8 $\pm$  29.0 & 2.03 $\pm$  0.48 & 3.42 $\pm$ 0.48 & 130.90 $\pm$ 2.60 & 0.19 $\pm$ 0.45 \\
6261.8736 & B & 2.29 $\pm$  0.650 & 137.7 $\pm$  28.9 & 2.50 $\pm$  0.39 & 3.58 $\pm$ 0.39 & 138.50 $\pm$ 1.70 & 0.22 $\pm$ 0.52 \\
6263.9422 & B & 1.69 $\pm$  0.650 & 135.2 $\pm$  28.9 & 1.89 $\pm$  0.39 & 3.08 $\pm$ 0.39 & 136.60 $\pm$ 2.10 & 0.22 $\pm$ 0.52 \\
6261.8769 & V & 2.09 $\pm$  1.71 & 141.0 $\pm$  28.9 & 2.32 $\pm$  1.62 & 3.50 $\pm$ 1.62 & 141.60 $\pm$ 1.90 & 0.23 $\pm$ 0.55 \\
6263.9074 & V & 2.01 $\pm$  1.68 & 136.6 $\pm$  28.9 & 2.23 $\pm$  1.59 & 3.45 $\pm$ 1.59 & 137.70 $\pm$ 2.00 & 0.23 $\pm$ 0.55 \\
\hline
\end{tabular}
\end{table*}

\begin{table*}
\addtocounter{table}{-1}
\caption{continued}
\begin{tabular}{cccccccc}
\hline
  & & & IRAS 06072+0953 & &   &   PA$_{ISP}$= 78.3 $\pm$ 10.1\degree & \\
   \hline
Julian Date & Filter & DoP(\%) & PA(\degree) & P$_{MAS}$(\%) & P$_{obs}$(\%) & PA$_{obs}$(\degree) & ISP(\%) \\
6266.9234 & U & 0.890 $\pm$  0.330 & 152.9 $\pm$  12.0 & 1.26 $\pm$  0.25 & 2.48 $\pm$ 0.25 & 157.80 $\pm$ 6.50 & 0.42 $\pm$ 0.21 \\
6271.9043 & U & 0.630 $\pm$  0.430 & 158.5 $\pm$  12.6 & 1.03 $\pm$  0.38 & 2.04 $\pm$ 0.38 & 162.40 $\pm$ 7.50 & 0.42 $\pm$ 0.21 \\
6262.9289 & B & 0.170 $\pm$  0.410 & 137.6 $\pm$  11.1 & 0.57 $\pm$  0.33 & 1.14 $\pm$ 0.33 & 160.90 $\pm$ 4.70 & 0.48 $\pm$ 0.24 \\
6271.8883 & B & 0.570 $\pm$  0.360 & 127.3 $\pm$  11.0 & 0.79 $\pm$  0.27 & 1.56 $\pm$ 0.27 & 145.50 $\pm$ 4.30 & 0.48 $\pm$ 0.24 \\
6262.9509 & V & 0.300 $\pm$  0.350 & 140.3 $\pm$  10.7 & 0.72 $\pm$  0.24 & 1.41 $\pm$ 0.24 & 158.10 $\pm$ 3.60 & 0.50 $\pm$ 0.25 \\
6266.9377 & V & 0.530 $\pm$  0.390 & 167.9 $\pm$  10.4 & 1.03 $\pm$  0.30 & 1.94 $\pm$ 0.30 & 168.10 $\pm$ 2.70 & 0.50 $\pm$ 0.25 \\
\hline
\end{tabular}
\end{table*}

\begin{table*}
\addtocounter{table}{-1}
\caption{continued}
\begin{tabular}{cccccccc}
\hline
  & & & IRAS 07008+1050 & &   &   PA$_{ISP}$= 89.6 $\pm$ 19.8\degree &  \\
 \hline
Julian Date & Filter & DoP(\%) & PA(\degree) & P$_{MAS}$(\%) & P$_{obs}$(\%) & PA$_{obs}$(\degree) & ISP(\%) \\
5946.8776 & U & 1.22 $\pm$  1.51 & 2.300 $\pm$  20.1 & 1.79 $\pm$  0.39 & 3.23 $\pm$ 0.39 & 1.70 $\pm$ 3.30 & 0.58 $\pm$ 1.46 \\
5949.8090 & U & 1.14 $\pm$  1.50 & 11.80 $\pm$  20.1 & 1.69 $\pm$  0.34 & 3.05 $\pm$ 0.34 & 8.00 $\pm$ 3.20 & 0.58 $\pm$ 1.46 \\
5946.8861 & B & 0.990 $\pm$  1.73 & 42.60 $\pm$  20.2 & 1.24 $\pm$  0.47 & 2.39 $\pm$ 0.47 & 26.70 $\pm$ 4.10 & 0.66 $\pm$ 1.66 \\
5949.8042 & B & 0.850 $\pm$  1.73 & 34.60 $\pm$  20.2 & 1.25 $\pm$  0.48 & 2.39 $\pm$ 0.48 & 19.90 $\pm$ 4.00 & 0.66 $\pm$ 1.66 \\
5946.8843 & V & 0.520 $\pm$  1.77 & 29.30 $\pm$  19.9 & 1.07 $\pm$  0.21 & 1.95 $\pm$ 0.21 & 12.60 $\pm$ 2.10 & 0.70 $\pm$ 1.76 \\
5947.8286 & V & 0.640 $\pm$  1.77 & 39.30 $\pm$  20.0 & 1.04 $\pm$  0.21 & 1.92 $\pm$ 0.21 & 18.80 $\pm$ 2.30 & 0.70 $\pm$ 1.76 \\
\hline
\end{tabular}
\end{table*}

\begin{table*}
\addtocounter{table}{-1}
\caption{continued}
\begin{tabular}{cccccccc}
\hline
  & & & IRAS 08187-1905 & & &   PA$_{ISP}$= 103.2 $\pm$ 30.6\degree &  \\
 \hline
Julian Date & Filter & DoP(\%) & PA(\degree) & P$_{MAS}$(\%) & P$_{obs}$(\%) & PA$_{obs}$(\degree) & ISP(\%) \\
6262.9947 & U & 1.20 $\pm$  0.450 & 129.0 $\pm$  30.7 & 1.35 $\pm$  0.35 & 2.49 $\pm$ 0.35 & 125.10 $\pm$ 2.90 & 0.23 $\pm$ 0.29 \\
6271.9751 & U & 0.320 $\pm$  0.600 & 55.90 $\pm$  31.6 & 0.52 $\pm$  0.52 & 1.04 $\pm$ 0.52 & 64.60 $\pm$ 7.80 & 0.23 $\pm$ 0.29 \\
6264.9590 & B & 0.600 $\pm$  0.410 & 70.50 $\pm$  30.7 & 0.86 $\pm$  0.24 & 1.66 $\pm$ 0.24 & 72.40 $\pm$ 2.90 & 0.27 $\pm$ 0.33 \\
6271.9645 & B & 0.530 $\pm$  0.400 & 106.0 $\pm$  30.8 & 0.80 $\pm$  0.23 & 1.56 $\pm$ 0.23 & 105.10 $\pm$ 3.50 & 0.27 $\pm$ 0.33 \\
6262.9764 & V & $-$ $\pm$  $-$ & $-$ $\pm$  $-$ & 0.10 $\pm$  0.25 & 0.19 $\pm$ 0.25 & 88.00 $\pm$ 14.30 & 0.28 $\pm$ 0.35 \\
6264.9606 & V & 0.220 $\pm$  0.490 & 21.30 $\pm$  31.7 & 0.29 $\pm$  0.35 & 0.58 $\pm$ 0.35 & 54.20 $\pm$ 8.30 & 0.28 $\pm$ 0.35 \\
\hline
\end{tabular}
\end{table*}

\begin{table*}
\addtocounter{table}{-1}
\caption{continued}
\begin{tabular}{cccccccc}
\hline
  & & & IRAS F17495+0757 & & &   PA$_{ISP}$= 79.4 $\pm$ 6.2\degree &  \\
 \hline
Julian Date & Filter & DoP(\%) & PA(\degree) & P$_{MAS}$(\%) & P$_{obs}$(\%) & PA$_{obs}$(\degree) & ISP(\%) \\
6124.7437 & U & 0.460 $\pm$  0.460 & 88.20 $\pm$  21.5 & 0.90 $\pm$  0.44 & 1.80 $\pm$ 0.44 & 83.80 $\pm$ 20.60 & 0.45 $\pm$ 0.13 \\
6123.7623 & B & 0.380 $\pm$  0.360 & 98.70 $\pm$  7.50 & 0.85 $\pm$  0.33 & 1.67 $\pm$ 0.33 & 87.60 $\pm$ 4.20 & 0.52 $\pm$ 0.15 \\
6124.7325 & B & 0.650 $\pm$  0.440 & 97.80 $\pm$  7.00 & 1.11 $\pm$  0.41 & 2.12 $\pm$ 0.41 & 89.70 $\pm$ 3.40 & 0.52 $\pm$ 0.15 \\
6123.7648 & V & 0.170 $\pm$  0.270 & 73.30 $\pm$  7.40 & 0.72 $\pm$  0.22 & 1.41 $\pm$ 0.22 & 78.00 $\pm$ 4.10 & 0.55 $\pm$ 0.15 \\
6124.7380 & V & 0.110 $\pm$  0.270 & 64.90 $\pm$  8.00 & 0.65 $\pm$  0.23 & 1.29 $\pm$ 0.23 & 77.00 $\pm$ 5.10 & 0.55 $\pm$ 0.15 \\
6123.7818 & R & 0.0700 $\pm$  1.41 & 142.5 $\pm$  8.40 & 0.49 $\pm$  1.40 & 0.98 $\pm$ 1.40 & 82.80 $\pm$ 5.70 & 0.53 $\pm$ 0.15 \\
\hline
\end{tabular}
\end{table*}

\begin{table*}
\addtocounter{table}{-1}
\caption{continued}
\begin{tabular}{cccccccc}
\hline
  & & & IRAS 19090+3829 & & &   PA$_{ISP}$= 59.9 $\pm$ 8.9\degree &  \\
 \hline
Julian Date & Filter & DoP(\%) & PA(\degree) & P$_{MAS}$(\%) & P$_{obs}$(\%) & PA$_{obs}$(\degree) & ISP(\%) \\
6166.6787 & B & 0.0700 $\pm$  0.590 & 148.7 $\pm$  13.3 & 0.95 $\pm$  0.34 & 1.89 $\pm$ 0.34 & 58.30 $\pm$ 9.90 & 0.91 $\pm$ 0.48 \\
6166.6852 & V & 0.260 $\pm$  0.920 & 143.7 $\pm$  10.0 & 0.72 $\pm$  0.77 & 1.42 $\pm$ 0.77 & 62.20 $\pm$ 4.60 & 0.97 $\pm$ 0.51 \\
6166.6559 & R & 0.510 $\pm$  0.590 & 127.1 $\pm$  9.70 & 0.69 $\pm$  0.32 & 1.35 $\pm$ 0.32 & 76.10 $\pm$ 3.90 & 0.94 $\pm$ 0.49 \\
6166.6617 & I & 1.16 $\pm$  0.460 & 101.8 $\pm$  9.30 & 1.43 $\pm$  0.26 & 2.56 $\pm$ 0.26 & 86.70 $\pm$ 2.50 & 0.73 $\pm$ 0.38 \\
\hline
  & & & IRAS 19114+0002 & & &   PA$_{ISP}$= 92.00 $\pm$ 12.0\degree &  \\
 \hline
Julian Date & Filter & DoP(\%) & PA(\degree) & P$_{MAS}$(\%) & P$_{obs}$(\%) & PA$_{obs}$(\degree) & ISP(\%) \\
\hline
\end{tabular}
\end{table*}

\begin{table*}
\addtocounter{table}{-1}
\caption{continued}
\begin{tabular}{cccccccc}
\hline
  & & & IRAS 19386+0155 & & &   PA$_{ISP}$= 82.2 $\pm$ 29.0\degree &  \\
 \hline
Julian Date & Filter & DoP(\%) & PA(\degree) & P$_{MAS}$(\%) & P$_{obs}$(\%) & PA$_{obs}$(\degree) & ISP(\%) \\
6165.7463 & U & 5.08 $\pm$  0.360 & 146.2 $\pm$  29.3 & 5.18 $\pm$  0.30 & 7.52 $\pm$ 0.30 & 146.80 $\pm$ 3.70 & 0.15 $\pm$ 0.19 \\
6165.7294 & I & 0.630 $\pm$  0.310 & 109.9 $\pm$  29.4 & 0.76 $\pm$  0.26 & 1.50 $\pm$ 0.26 & 107.80 $\pm$ 4.80 & 0.13 $\pm$ 0.17 \\
6188.6614 & U & 5.94 $\pm$  0.310 & 151.6 $\pm$  29.3 & 6.05 $\pm$  0.25 & 8.48 $\pm$ 0.25 & 152.10 $\pm$ 3.80 & 0.15 $\pm$ 0.19 \\
6188.6753 & B & 3.82 $\pm$  0.320 & 158.8 $\pm$  29.1 & 3.97 $\pm$  0.24 & 4.75 $\pm$ 0.24 & 159.30 $\pm$ 1.40 & 0.17 $\pm$ 0.21 \\
6188.6809 & V & 1.40 $\pm$  0.340 & 146.6 $\pm$  29.1 & 1.52 $\pm$  0.25 & 2.65 $\pm$ 0.25 & 149.20 $\pm$ 2.30 & 0.18 $\pm$ 0.23 \\
\hline
  & & & IRAS 19475+3119 & & &   PA$_{ISP}$= 80.5 $\pm$ 10.4\degree &  \\
\hline
\end{tabular}
\end{table*}

\begin{table*}
\addtocounter{table}{-1}
\caption{continued}
\begin{tabular}{cccccccc}
\hline
  & & & IRAS 20056+1834& & &   PA$_{ISP}$= 98.1 $\pm$ 6.4\degree &  \\
 \hline
Julian Date & Filter & DoP(\%) & PA(\degree) & P$_{MAS}$(\%) & P$_{obs}$(\%) & PA$_{obs}$(\degree) & ISP(\%) \\
6122.9072 & I & 12.0 $\pm$  0.420 & 138.4 $\pm$  6.60 & 12.14 $\pm$  0.41 & 12.71 $\pm$ 0.41 & 138.90 $\pm$ 1.90 & 0.21 $\pm$ 0.08 \\
6122.8913 & B & 13.4 $\pm$  0.290 & 139.8 $\pm$  6.50 & 13.60 $\pm$  0.27 & 13.81 $\pm$ 0.27 & 140.30 $\pm$ 1.20 & 0.27 $\pm$ 0.10 \\
6122.8994 & V & 12.5 $\pm$  0.250 & 141.2 $\pm$  6.50 & 12.71 $\pm$  0.23 & 12.93 $\pm$ 0.23 & 141.80 $\pm$ 1.20 & 0.29 $\pm$ 0.10 \\
6122.8790 & R & 12.1 $\pm$  0.260 & 144.6 $\pm$  6.50 & 12.30 $\pm$  0.24 & 12.61 $\pm$ 0.24 & 145.10 $\pm$ 1.40 & 0.28 $\pm$ 0.10 \\
6192.7346 & B & 16.3 $\pm$  0.280 & 139.6 $\pm$  6.50 & 16.50 $\pm$  0.26 & 16.65 $\pm$ 0.26 & 140.00 $\pm$ 1.10 & 0.27 $\pm$ 0.10 \\
6192.7063 & V & 14.9 $\pm$  0.260 & 139.4 $\pm$  6.50 & 15.10 $\pm$  0.24 & 15.29 $\pm$ 0.24 & 139.90 $\pm$ 1.20 & 0.29 $\pm$ 0.10 \\
\hline
\end{tabular}
\end{table*}

\begin{table*}
\addtocounter{table}{-1}
\caption{continued}
\begin{tabular}{cccccccc}
\hline
  & & & IRAS 22223+4327 & &  &   PA$_{ISP}$= 61.9 $\pm$ 12.5\degree & \\
 \hline
Julian Date & Filter & DoP(\%) & PA(\degree) & P$_{MAS}$(\%) & P$_{obs}$(\%) & PA$_{obs}$(\degree) & ISP(\%) \\
6164.7945 & U & $-$ $\pm$  $-$ & $-$ $\pm$  $-$ & 0.73 $\pm$  0.38 & 1.46 $\pm$ 0.38 & 133.50 $\pm$ 17.20 & 2.13 $\pm$ 1.12 \\
6164.7990 & B & $-$ $\pm$  $-$ & $-$ $\pm$  $-$ & 0.14 $\pm$  3.35 & 0.29 $\pm$ 3.35 & 36.80 $\pm$ 18.40 & 2.42 $\pm$ 1.27 \\
6164.8016 & V & $-$ $\pm$  $-$ & $-$ $\pm$  $-$ & 0.81 $\pm$  0.25 & 1.58 $\pm$ 0.25 & 28.40 $\pm$ 3.30 & 2.57 $\pm$ 1.35 \\
6164.8066 & R & $-$ $\pm$  $-$ & $-$ $\pm$  $-$ & 0.87 $\pm$  0.20 & 1.66 $\pm$ 0.20 & 28.70 $\pm$ 2.70 & 2.49 $\pm$ 1.31 \\
6164.7891 & I & $-$ $\pm$  $-$ & $-$ $\pm$  $-$ & 0.20 $\pm$  0.25 & 0.40 $\pm$ 0.25 & 55.40 $\pm$ 20.50 & 1.93 $\pm$ 1.01 \\
6188.8285 & U & $-$ $\pm$  $-$ & $-$ $\pm$  $-$ & 1.31 $\pm$  0.47 & 2.51 $\pm$ 0.47 & 87.30 $\pm$ 4.10 & 2.13 $\pm$ 1.12 \\
\hline

\end{tabular}
\end{table*}

\end{document}